\documentclass[a4paper,11pt]{article}

\usepackage{jcappub}            
\usepackage{amsmath,amssymb}
\usepackage{graphicx}
\usepackage{subcaption}
\usepackage{booktabs}
\usepackage{graphicx}
\usepackage{subcaption}
\usepackage{hyperref}

\numberwithin{equation}{section}

\newcommand{\mpl}{M_{{}_{\rm Pl}}}
\newcommand{\rdm}{\mathrm{d}}
\newcommand{\hh}{\mathcal{H}}
\newcommand{\BD}{\mathrm{BD}}
\newcommand{\nBD}{\mathrm{nBD}}
\newcommand{\Ob}{\mathrm{obs}}
\newcommand{\varphis}{\varphi_{{}_S}}
\newcommand{\chis}{\chi_{{}_S}}
\newcommand{\dk}[1]{\overset{\scriptscriptstyle\circ}{#1}}
\newcommand{\ddk}[1]{\overset{\scriptscriptstyle\circ\circ}{#1}}
\newcommand{\dns}{\delta^{(n_{\rm s})}_{{}_{\nBD}}}
\newcommand{\das}{\delta^{(\alpha_{\rm s})}_{{}_{\nBD}}}
\newcommand{\Bs}{\mathcal{B}_{{}_S}}
\newcommand{\panel}[2]{%
  \begin{subfigure}[t]{0.48\textwidth}
    \centering
    \includegraphics[width=\linewidth]{#1}
    \caption{#2}
  \end{subfigure}}

\title{ACT Excited: Reconciling Starobinsky and Quadratic Chaotic Inflationary Models with ACT Results via Excited States}

\author[a]{Amjad Ashoorioon,}
\author[a]{Abdolreza Yousefi-Sostani}

\affiliation[a]{School of Physics, Institute for Research in Fundamental Sciences (IPM),\\
P.O. Box 19395-5531, Tehran, Iran}

\emailAdd{amjad@ipm.ir}
\emailAdd{yousefiae@ipm.ir}

\abstract{Data Release 6 of the Atacama Cosmology Telescope (ACT DR6), combined with Planck, CMB lensing and DESI BAO data, favours a higher scalar spectral index $n_{\rm s}$ than Planck and a mildly positive running $\alpha_{\rm s}$, putting the Starobinsky model in tension with the data, while the quadratic chaotic (QC) model is already excluded by its large tensor-to-scalar ratio $r$. We show that relaxing the Bunch--Davies (BD) assumption for the initial state of the scalar perturbations can restore agreement without modifying the inflaton potential. Parametrising the Bogoliubov coefficients by an excitation parameter $\chi_{{}_S}$ and a relative phase $2\varphi_{{}_S}$, we derive the shifts of $n_{\rm s}$ and $\alpha_{\rm s}$ induced by their scale dependence and map the regions of the $(\varphi_{{}_S},\chi_{{}_S})$ plane compatible with the $1\sigma$ ACT DR6 ranges, the bound $r<0.038$, and the requirement that the scale of new physics $\Lambda$ exceed the Hubble scale $H$. For the Starobinsky model, a scale-dependent $\chi_{{}_S}$ or $\varphi_{{}_S}$ works almost everywhere in this plane: only small neighbourhoods of $\varphi_{{}_S}=0,\pi$ with $\chi_{{}_S}\gtrsim1.22$ are excluded, and a scale-dependent phase fails at $\varphi_{{}_S}=0,\pi/2,\pi$ and at $\chi_{{}_S}=0$, where the required ${\rm d}\varphi_{{}_S}/{\rm d}\ln k$ and ${\rm d}^2\varphi_{{}_S}/{\rm d}(\ln k)^2$ diverge. For the QC model, the excited state must also suppress $r$, which requires $\chi_{{}_S}\gtrsim0.63$ and $\Lambda\lesssim6H$. A scale-dependent $\chi_{{}_S}$ must satisfy $|{\rm d}\chi_{{}_S}/{\rm d}\ln k|\gtrsim0.0021$ in both models, and in the QC model we show analytically that $0.0021\lesssim{\rm d}\chi_{{}_S}/{\rm d}\ln k\lesssim0.0058$ and $0.0008\lesssim{\rm d}^2\chi_{{}_S}/{\rm d}(\ln k)^2\lesssim0.0062$. If the tensor modes remain in the BD state, the scenario also predicts a violation of the consistency relation $r=-8n_{{}_T}$, testable once the tensor power spectrum and its tilt are measured.}


\begin{document}
\maketitle
\flushbottom

\section{Introduction}
\label{sec:intro}

The inflationary paradigm, in particular its single-field slow-roll realisations
\cite{Starobinsky:1980te, Guth:1980zm, Linde:1981mu}, provides a simple and robust framework
for resolving several long-standing problems of the hot Big Bang model, including the flatness
and horizon problems, and for explaining the origin of large-scale structure
\cite{Baumann:2009ds, Achucarro:2022qrl, Vazquez:2018qdg}. Observations from missions such as
Planck \cite{Planck:2018jri, Planck:2018nkj} have placed stringent constraints on key
inflationary observables, namely the scalar spectral index $n_{\rm s}$ and the tensor-to-scalar ratio
$r$, and thereby restricted or excluded many models. The Starobinsky model
\cite{Starobinsky:1980te, Starobinsky:1983zz}, for instance, was found to be in excellent
agreement with Planck and earlier data, lying well inside the allowed region of the $n_{\rm s}$--$r$
plane.

\begin{figure}[t]
  \centering
  \includegraphics[width=0.8\linewidth]{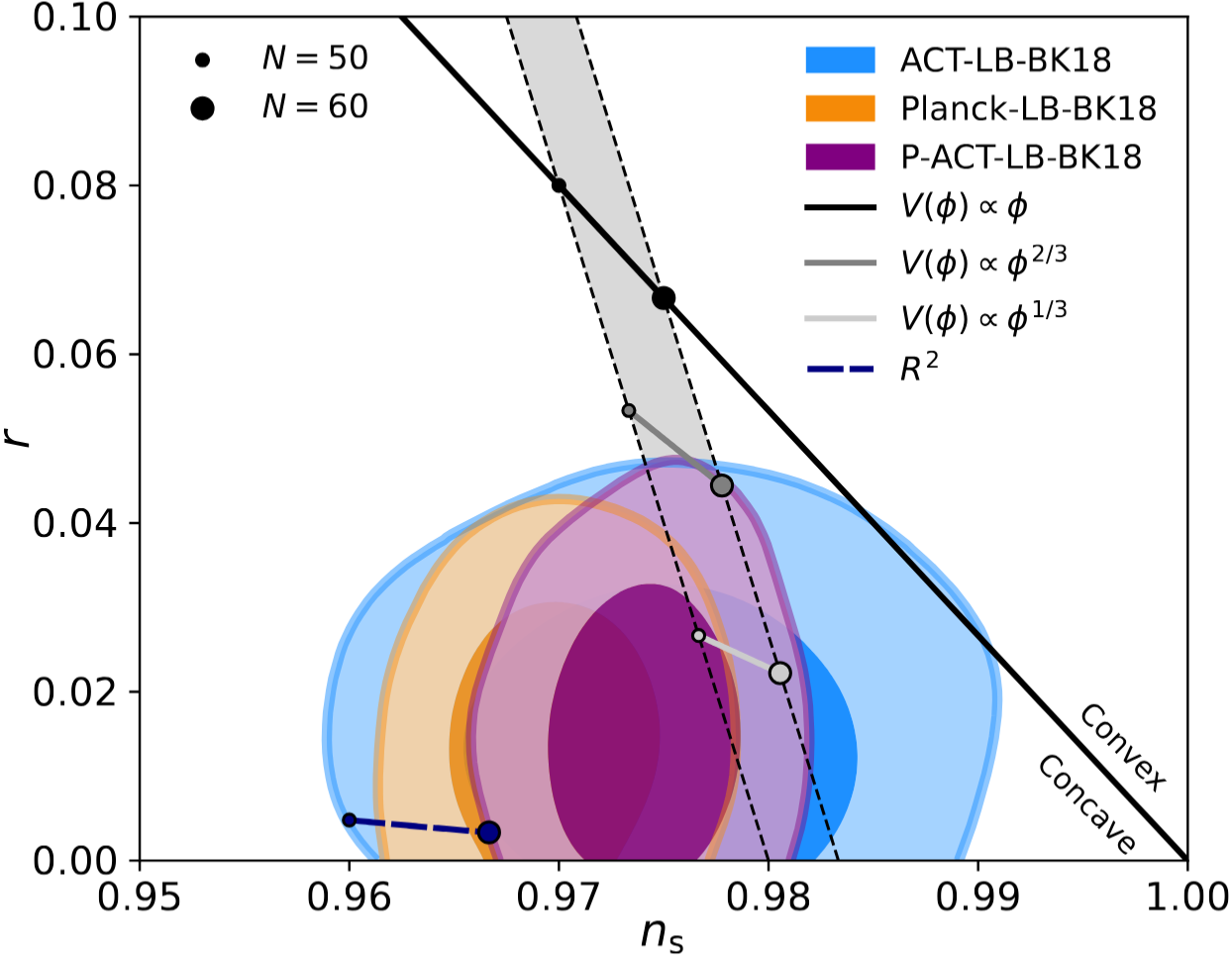}
  \caption{Constraints on the scalar and tensor primordial power spectra at the pivot scale
  $k_{*}=0.05\ \mathrm{Mpc}^{-1}$ in the $n_{\rm s}$--$r$ plane, taken from
  ref.~\cite{AtacamaCosmologyTelescope:2025nti}. The upper limit on $r$ is determined mainly by
  the BICEP/Keck data \cite{BICEP:2021xfz, BICEPKeck:2024stm}, while the constraint on $n_{\rm s}$
  comes from Planck (orange), ACT (blue), or the combined Planck--ACT analysis (P-ACT, purple).
  In all cases, the joint data sets include CMB lensing and BAO measurements.}
  \label{fig:Planck_ACT}
\end{figure}

The recent Data Release 6 (DR6) of the Atacama Cosmology Telescope (ACT)
\cite{AtacamaCosmologyTelescope:2025blo, AtacamaCosmologyTelescope:2025nti}, however, prefers
a higher value of $n_{\rm s}$ and thereby puts several models, including Starobinsky inflation, under
tension (see table~\ref{tab:obs} and figure~\ref{fig:Planck_ACT}). For $N_e=60$ e-folds, the Starobinsky model predicts $n_{\rm s}\simeq1-2/N_e\simeq0.9667$, about $2.2\sigma$ below the combined ACT DR6 value $n_{\rm s}=0.9743\pm0.0034$, and a small negative running, $\alpha_{\rm s}\simeq-2/N_e^2\simeq-5.6\times10^{-4}$, whereas ACT DR6 mildly favours a positive running, $\alpha_{\rm s}=0.0062\pm0.0052$. The QC model, $V\propto\phi^2$, has the same $n_{\rm s}$ and $\alpha_{\rm s}$, but its tensor-to-scalar ratio, $r\simeq8/N_e\simeq0.13$, is far above the current upper bound $r<0.038$ \cite{BICEP:2021xfz, BICEPKeck:2024stm}. Such tensions are not unique
to ACT DR6: analyses of other data sets have also posed challenges for inflationary models
\cite{Balkenhol:2025wms}. Part of the shift towards higher $n_{\rm s}$ arises when the CMB data are combined with the DESI BAO measurements, between which there are marginally significant differences.
It is not yet clear whether this amounts to a genuine challenge for
the inflationary paradigm itself, since the quoted numerical values refer to the simplest
versions of the models.

From a theoretical viewpoint, inflation is a flexible framework. Without resorting to exotic or
ad hoc assumptions, one can study corrections that modify the predicted values of the
inflationary observables. A standard and central assumption is that the quantum state of the
cosmological (scalar and tensor) perturbations is the Bunch--Davies (BD) vacuum
\cite{Bunch:1978yq}. Although there are good motivations for this choice, it is neither
necessary nor self-evident, and one may consider other initial states, or corrections to the BD
state, especially for the scalar modes.

\begin{table}[t]
  \centering
  \begin{tabular}{lcc}
    \toprule
    Experiment & $n_{\rm s}$ & $\alpha_{\rm s}$ \\
    \midrule
    ACT DR6 & $0.9743 \pm 0.0034$ & $\phantom{-}0.0062 \pm 0.0052$ \\
    Planck 2018 & $0.9651 \pm 0.0044$ & $-0.004 \pm 0.013$ \\
    \bottomrule
  \end{tabular}
  \caption{Observed values of the scalar spectral index $n_{\rm s}$ and its running $\alpha_{\rm s}$, taken
  from refs.~\cite{Planck:2018jri, AtacamaCosmologyTelescope:2025blo, AtacamaCosmologyTelescope:2025nti}. The ACT DR6 values are those of the combination of ACT DR6 with Planck, CMB lensing and DESI BAO data (P-ACT-LB); the value of $n_{\rm s}$ is obtained in $\Lambda$CDM, while $\alpha_{\rm s}$ is obtained when the running is allowed to vary.}
  \label{tab:obs}
\end{table}
There are both theoretical and phenomenological reasons to take non-BD initial states seriously. It has been repeatedly shown that the choice of the initial state can have significant consequences for the evolution of
early-universe models and their viability (see, e.g., refs.~\cite{Chen:2006nt, Holman:2007na, Ashoorioon:2010xg, Ashoorioon:2013eia, Ashoorioon:2014nta, Ashoorioon:2018sqb}). Our knowledge of the high-energy physics underlying inflation is incomplete, so the BD vacuum is not uniquely selected by first principles \cite{Kaloper:2002cs, Kowalczyk:2024ech}. If inflation lasted only slightly longer than the minimum number of e-folds required to solve the standard cosmological problems, observable modes could retain imprints of Planck-scale physics, known as trans-Planckian signatures, through their excitation during inflation \cite{Brandenberger:2000wr, Kempf:2000ac, Easther:2001fi, Easther:2001fz, Hui:2001ce, Ashoorioon:2011eg, Schalm:2004qk, Ashoorioon:2004vm, Ashoorioon:2004wd, Ashoorioon:2005ep, Powell:2006yg, Bahrami:2015bva}. In that case, it is natural for some modes to start in excited states, provided that backreaction limits are respected \cite{Tanaka:2000jw, Greene:2004np, Holman:2007na}. Independently, if inflation lasted only just long enough to solve the problems of standard big bang cosmology, the pre-inflationary phase could influence the initial conditions
of observable modes and excite some of them, making a non-standard initial state a natural possibility \cite{Sarangi:2006yy, Cicoli:2014bja, Bahrami:2015bva, Melcher:2023kpd}; the past incompleteness of inflationary space-times \cite{Borde:2001nh} indeed implies that some other phase must have preceded inflation. Non-BD states also provide a broader framework for capturing possible signatures of heavy fields and non-standard interactions \cite{Jackson:2010cw, Yin:2023jlv}; alternatively, they could be signatures of a modified dispersion relation for the modes during inflation \cite{Ashoorioon:2017toq}.

An excited initial state modifies the primordial power spectra through the Bogoliubov coefficients that relate it to the BD vacuum. On super-horizon scales, the scalar power spectrum is multiplied by a factor $\gamma_{{}_S}=|\alpha^{{}_S}_k-\beta^{{}_S}_k|^2$, and similarly for the tensor modes \cite{Ashoorioon:2013eia}. If the excitation is the same for all observable modes, $\gamma_{{}_S}$ only rescales the amplitude of the spectrum and leaves $n_{\rm s}$ and $\alpha_{\rm s}$ unchanged. If, however, only the scalar modes are excited, the tensor-to-scalar ratio is suppressed by $1/\gamma_{{}_S}$; this mechanism was used in refs.~\cite{Ashoorioon:2013eia} to reconcile chaotic models like the quadratic chaotic model with the Planck data. Scale-dependent Bogoliubov coefficients induced running in $n_{\rm s}$ to reconcile BICEP2 \cite{BICEP2:2014owc}  with the Planck 2013 \cite{Planck:2013jfk} results \cite{Ashoorioon:2014nta}. Such a dependence arises naturally when the modes are excited by new physics at a fixed physical scale, or by a pre-inflationary phase that affects different modes differently, and is known to produce a running of the spectral index and to modify the consistency relation \cite{Ashoorioon:2005ep, Broy:2016zik}. The excitation cannot be arbitrarily large: the energy density of the excited quanta must not spoil slow-roll inflation \cite{Tanaka:2000jw, Greene:2004np, Holman:2007na}. This backreaction condition bounds the Bogoliubov coefficient $\beta^{{}_S}_k$ in terms of the slow-roll parameters and the scale of new physics $\Lambda$, and requiring $\Lambda>H$, so that the effective field theory description remains valid, restricts the allowed excitations.

In this paper we ask whether a scale-dependent non-BD initial state of the scalar perturbations can supply the shifts of $n_{\rm s}$ and $\alpha_{\rm s}$ needed to reconcile the Starobinsky and QC models with ACT DR6, while keeping $r$ below the observational bound and the backreaction under control. We parametrise the Bogoliubov coefficients by an excitation parameter $\chi_{{}_S}$ and a relative phase $2\varphi_{{}_S}$, derive the shifts of $n_{\rm s}$ and $\alpha_{\rm s}$ produced by their first and second derivatives with respect to $\ln k$, and consider two cases, in which either the phase or the excitation parameter carries the scale dependence. For each model, we determine the region of the $(\varphi_{{}_S},\chi_{{}_S})$ plane allowed by the bound on $r$ and by the condition $\Lambda>H$, and the scale dependence required within this region to reproduce the ACT DR6 values \footnote{In doing so, we use the definition of the slow-roll parameter $\eta$ that enters the backreaction bound \cite{Holman:2007na}. For the QC model this $\eta$ vanishes at leading order in $1/N_e$, which makes the condition $\Lambda>H$ considerably more restrictive than estimated in earlier work.}.

We find that, for the Starobinsky model, the required shifts can be generated almost everywhere in the $(\varphi_{{}_S},\chi_{{}_S})$ plane, with only small neighbourhoods of $\varphi_{{}_S}=0,\pi$ excluded and with $\Lambda$ well above $H$. For the QC model, the excited state must also suppress $r$, which requires $\chi_{{}_S}\gtrsim0.63$ and limits the scale of new physics to $\Lambda\lesssim6H$. When the scale dependence is carried by $\chi_{{}_S}$, we show that the relevant equations depend on the position in the $(\varphi_{{}_S},\chi_{{}_S})$ plane only through a single combination. This allows us to obtain the shapes of the allowed regions analytically and, for the QC model, narrow analytic bounds on the required running of $\chi_{{}_S}$.

The paper is organised as follows. In section~\ref{sec:excited} we
review the effect of excited initial states on the primordial power spectra, introduce the
Bogoliubov parametrisation and the backreaction constraint, and derive the resulting shifts in
$n_{\rm s}$ and $\alpha_{\rm s}$. In section~\ref{sec:models} we introduce the two models, determine the regions of the $(\varphi_{{}_S},\chi_{{}_S})$ plane allowed by the bound on $r$ and by the validity of the effective field theory, and compute the scale dependence of $\varphi_{{}_S}$ or $\chi_{{}_S}$ required to reproduce the ACT DR6 values of $n_{\rm s}$ and $\alpha_{\rm s}$. We summarise our results and discuss their observational implications in
section~\ref{sec:conclusions}.

\section{Excited initial states}
\label{sec:excited}

\subsection{Power spectra for a generic initial state}
\label{sec:powerspectra}

We start from the simplest inflationary set-up, a single minimally coupled scalar field
$\phi$ with potential $V(\phi)$, described by the action
\begin{equation}\label{eq:action}
  S = \int \rdm^4x \sqrt{-g}\left[\frac{\mpl^2}{2}R - \frac{1}{2}g^{\mu\nu}\partial_\mu\phi\,\partial_\nu\phi - V(\phi)\right],
\end{equation}
where $\mpl$ is the reduced Planck mass. The perturbed metric contains scalar perturbations
$\Phi$\footnote{The perturbed Einstein equations imply that the two Bardeen potentials are
equal.} and tensor perturbations $h_{ij}$ \footnote{$h_{ij}$ is a symmetric, traceless and
divergence-free tensor.},
\begin{equation}\label{eq:metric}
  \rdm s^2 = a^2(\tau)\left[-e^{2\Phi}\rdm\tau^2 + \left(e^{-2\Phi}\delta_{ij} + h_{ij}\right)\rdm x^i \rdm x^j\right],
\end{equation}
where $\tau$ is conformal time and a prime denotes a derivative with respect to $\tau$. The
inflaton is decomposed around its background value as \footnote{Unperturbed quantities are denoted
by an overbar.}
\begin{equation}\label{eq:inflaton-pert}
  \phi = \bar\phi + \delta\phi\,.
\end{equation}
The scalar perturbations are described by the Mukhanov--Sasaki variable
\begin{equation}\label{eq:MS-var}
  u(\tau,\mathbf{x}) = -\frac{a^2\bar\phi'}{a'}\left(\frac{a'}{a}\frac{\delta\phi}{\bar\phi'} + \Phi\right),
\end{equation}
whose Fourier modes, to leading order in the slow-roll parameters, solve the
Mukhanov--Sasaki equation with the general solution
\begin{equation}\label{eq:MS-solution}
  u^{{}_S}(k,\tau) \simeq \frac{\sqrt{\pi|\tau|}}{2}\left[\alpha^{{}_S}_k H^{(1)}_{{}_{3/2}}(k|\tau|) + \beta^{{}_S}_k H^{(2)}_{{}_{3/2}}(k|\tau|)\right].
\end{equation}
Here, following ref.~\cite{Holman:2007na}, the slow-roll parameters are defined through $\dot H=-\epsilon H^2$ and $\ddot H=2\epsilon\eta H^3$, where a dot denotes a derivative with respect to cosmic time. In terms of conformal time, with $\hh\equiv a'/a=aH$, they read
\begin{equation}\label{eq:slow-roll}
  \epsilon \equiv -\frac{\dot H}{H^2}= 1 - \frac{\hh'}{\hh^2}\,, \qquad \eta \equiv \frac{\ddot H}{2\epsilon H^3}= \epsilon - \frac{\epsilon'}{2\hh\epsilon}=\frac{\hh''-4\hh\hh'+2\hh^3}{2\hh\left(\hh^2-\hh'\right)}\,.
\end{equation}
With these definitions, the BD spectral index is $n_{{}_{{\rm s},\BD}}-1=2\eta-4\epsilon$ to leading order in slow roll.
The coefficients $\alpha^{{}_S}_k$ and $\beta^{{}_S}_k$ are Bogoliubov coefficients, which obey the
normalisation (Wronskian) condition $|\alpha^{{}_S}_k|^2 - |\beta^{{}_S}_k|^2 = 1$. The BD vacuum
corresponds to $\alpha^{{}_S}_k=1$ and $\beta^{{}_S}_k=0$, while a generic excited (non-BD) state has
different values. The second Bogoliubov coefficients are bounded by the requirement that the energy density in the excited quanta should not derail
slow-roll inflation \cite{Holman:2007na,Ashoorioon:2013eia},
\begin{equation}\label{eq:beta-bound}
|\beta^{{}_{S,T}}_k|\lesssim \sqrt{\epsilon|\eta|}\,\frac{H\mpl}{\Lambda^2_{{}_{S,T}}}\,,
\end{equation}
where $\Lambda_{{}_{S,T}}$ is the scale of new physics at which the scalar (tensor) modes are excited \footnote{One should note that the $\eta$ parameter defined in eq. \eqref{eq:beta-bound} is the same as the one in \eqref{eq:slow-roll}, and is different from what is usually identified as $\eta_{{}_V}\equiv \mpl^2 \frac{V_{{}_{\phi\phi}}}{V}$, where $V_{{}_{\phi\phi}}$ denotes the second derivative of $V$ with respect to $\phi$. For the Starobinsky model the parameter $\eta$ of eq.~\eqref{eq:slow-roll} is $\eta\simeq-1/N_e$, and we use $|\eta|=1/N_e$. For the QC model, the parameter of eq.~\eqref{eq:slow-roll} vanishes at the leading order in $1/N_e$; at next order $\eta\simeq\epsilon_V^2/3\simeq1/(12N_e^2)\simeq2.3\times10^{-5}$ for $N_e=60$. In ref.~\cite{Ashoorioon:2013eia}, for the QC model, $\eta$ instead was mistakenly identified to be $\eta_{{}_V}$, and of the same order as the first slow-roll parameter, $|\eta|\simeq\epsilon\simeq1/(2N_e)$. As we will see, $|\eta|\simeq1/(12N_e^2)$ would lower $\Lambda^{\rm max}_{{}_S}/H$ by a factor $(6N_e)^{1/4}\simeq4.4$ for the QC model. }.
As shown in ref.~\cite{Ashoorioon:2013eia}, the scalar power spectrum is then
modified as
\begin{equation}\label{eq:PS}
  P_{{}_S} = \gamma_{{}_S}\, P_{{}_{S,\BD}}\,,
\end{equation}
with
\begin{equation}\label{eq:gammaS}
  \gamma_{{}_S} = \left|\alpha^{{}_S}_k - \beta^{{}_S}_k\right|^2_{k/\hh\to 0}\,.
\end{equation}
Consequently, the scalar spectral tilt
\begin{equation}\label{eq:ns}
  n_{\rm s} \equiv 1 + \frac{\rdm\ln P_{{}_S}}{\rdm\ln k}
\end{equation}
and its running
\begin{equation}\label{eq:alphas}
  \alpha_{\rm s} \equiv \frac{\rdm n_{\rm s}}{\rdm\ln k}
\end{equation}
are modified as well.

A similar analysis can be carried out for the tensor perturbations $h_{ij}$, whose mode functions are likewise expressed in terms of Hankel functions
\begin{equation}\label{p-sol-ds}
h_k(\tau)\simeq\frac{\sqrt{\pi|\tau|}}{2}\left[\alpha^{{}_T}_k~  H_{3/2}^{(1)}(k|\tau|)+\beta^{{}_T}_k~ H_{3/2}^{(2)}(k|\tau|)\right]\,,
\end{equation}
where $h_k(\tau)$ denotes the Fourier mode of the amplitude of either polarization of the gravitational waves. The tensor power spectrum can be treated in the same way,
\begin{equation}\label{eq:PT}
  P_{{}_T} = \gamma_{{}_T}\, P_{{}_{{T,\BD}}}\,, \qquad \gamma_{{}_T} = \left|\alpha^{{}_T}_k - \beta^{{}_T}_k\right|^2_{k/\hh\to 0}\,,
\end{equation}
so that for a non-BD state the tensor-to-scalar ratio becomes
\begin{equation}\label{eq:r}
  r \equiv \frac{P_{{}_T}}{P_{{}_S}} = \gamma\, r_{{}_{\BD}} = 16\,\gamma\,\epsilon\,,
  \qquad
  \gamma \equiv \frac{\gamma_{{}_T}}{\gamma_{{}_S}} = \left|\frac{\alpha^{{}_T}_k - \beta^{{}_T}_k}{\alpha^{{}_S}_k - \beta^{{}_S}_k}\right|^2 .
\end{equation}
This modifies the Lyth bound and the consistency relation \cite{Hui:2001ce, Ashoorioon:2004wd, Broy:2016zik}. In all of these quantities, the
dependence on the initial state enters only through the combinations $|\alpha_k-\beta_k|$.

The factor $\gamma$ depends on the initial states of both the scalar and the tensor modes.
Different mechanisms can excite the scalar and tensor modes separately, so in general
$\alpha^{{}_S}_k\neq\alpha^{{}_T}_k$ and $\beta^{{}_S}_k\neq\beta^{{}_T}_k$, and $\gamma$ is not
constrained a priori other than by observations. Since observations demand a suppressed $r$,
values $\gamma<1$ are of particular interest. If, for example, one assumes $\gamma_{{}_T}= 1$, then
$\gamma_{{}_S}>1$ is required to suppress $r$.

\subsection{Parametrisation of non-BD states and backreaction}
\label{sec:param}

To analyse non-BD initial states, we parametrise the Bogoliubov coefficients as
\begin{subequations}\label{eq:param}
\begin{align}
  \alpha^{{}_S}_k &= e^{i\varphi_{{}_S}}\cosh\chi_{{}_S}\,, & \beta^{{}_S}_k &= e^{-i\varphi_{{}_S}}\sinh\chi_{{}_S}\,, \label{eq:param-S}\\
  \alpha^{{}_T}_k &= e^{i\varphi_{{}_T}}\cosh\chi_{{}_T}\,, & \beta^{{}_T}_k &= e^{-i\varphi_{{}_T}}\sinh\chi_{{}_T}\,, \label{eq:param-T}
\end{align}
\end{subequations}
where $\varphi_{{}_{S,T}}$ is the relative phase between $\alpha^{{}_{S,T}}_k$ and $\beta^{{}_{S,T}}_k$. With
this parametrisation, $\gamma_{{}_{S,T}}$ lies in the interval $[e^{-2\chi_{{}_{S,T}}},\,e^{2\chi_{{}_{S,T}}}]$ and
is given by
\begin{align}
  \gamma_{{}_S} &= e^{2\chi_{{}_S}}\sin^2\varphi_{{}_S} + e^{-2\chi_{{}_S}}\cos^2\varphi_{{}_S}\,, \label{eq:gammaS-chi}\\
  \gamma_{{}_T} &= e^{2\chi_{{}_T}}\sin^2\varphi_{{}_T} + e^{-2\chi_{{}_T}}\cos^2\varphi_{{}_T}\,. \label{eq:gammaT-chi}
\end{align}
Using $|\beta^{{}_{S,T}}_k|=\sinh\chi_{{}_{S,T}}$ and eliminating $\sqrt{\epsilon}\,\mpl$ in
eq.~\eqref{eq:beta-bound} through the normalisation of the scalar power spectrum,
$A_{{}_S}=\gamma_{{}_S}H^2/(8\pi^2\epsilon\mpl^2)$, i.e.\ $\sqrt{\epsilon}\,\mpl=\mathcal{C}\sqrt{\gamma_{{}_S}}\,H$,
the backreaction of the excited modes on the background is small for both scalar and tensor
modes, provided \cite{Ashoorioon:2013eia}
\begin{equation}\label{eq:backreaction}
  \Lambda_{{}_{S,T}} \lesssim \Lambda^{{\rm max}}_{{}_{S,T}}(\varphi_{{}_S},\chi_{{}_{S,T}})\equiv \frac{\mathcal{C}^{1/2} H}{\sinh\chi_{{}_{S,T}}^{{}_{1/2}}}\sqrt[4]{\gamma_{{}_S}\,|\eta|}\,,
\end{equation}
where $A_{{}_S}$ is the measured amplitude of the scalar power spectrum $P_{{}_S}$, $\Lambda^{{\rm max}}_{{}_{S,T}}$ designates the maximum scale of new physics allowed in the scalar or the tensor sector, and the parameter $\mathcal{C}$ is defined as
\begin{equation}\label{mC}
\mathcal{C}\equiv\frac{1}{(8\pi^2 A_{{}_S})^{1/2}}\simeq 2456.46\,,
\end{equation}
where we have used $A_{{}_S}=2.0989\times10^{-9}$ \cite{Planck:2018jri}.
One should note that $\Lambda^{\rm max}_{{}_{T}}$ depends on the occupation number in the scalar perturbations, $\chi_{{}_S}$ too.

Two regimes can be considered depending on the values of $\chi_{{}_{S, T}}$:

\paragraph{$\bullet ~\mathbf{\chi_{{}_{S,T}}\ll 1}:$} Expanding to first order, we obtain
\begin{equation}\label{eq:gamma-small}
  \gamma_{{}_{S,T}}\simeq 1 - 2\chi_{{}_{S,T}}\cos 2\varphi_{{}_{S,T}}\,.
\end{equation}
Since $|\cos2\varphi_{{}_{S, T}}|\le1$, we have $\gamma_{{}_{S, T}}\simeq 1$, indicating that the scalar and/or tensor modes are close to the BD state. In this regime eq.~\eqref{eq:backreaction} gives
$\Lambda^{{\rm max}}_{{}_{S,T}}/H\simeq \mathcal{C}^{1/2}|\eta|^{1/4}\chi_{{}_{S,T}}^{-1/2}\gg1$, irrespective of the values of $\varphi_{{}_{S,T}}$, so the scale of new physics can lie far above $H$.

\paragraph{$\bullet ~\mathbf{\chi_{{}_{S,T}}\gtrsim 1:}$} Away from the limit $\sin\varphi_{{}_{S,T}}\to0$, we have
\begin{equation}\label{eq:gamma-large}
  \gamma_{{}_{S,T}}\simeq e^{2\chi_{{}_{S,T}}}\sin^2\varphi_{{}_{S,T}}\,.
\end{equation}
In this regime, the maximum separation between the new-physics scale $\Lambda_{{}_{S}}^{\rm max}$ and the Hubble scale $H$ is achieved for $\varphi_{{}_S}\simeq \frac{\pi}{2}$. For a given $\epsilon$, the Hubble parameter is reduced from its BD value as
\begin{equation}\label{eq:H-large}
  H=\sqrt{8\pi^2A_{{}_S}}\,\sqrt{\frac{\epsilon}{\gamma_{{}_S}}}\,\mpl\simeq 4.071\times 10^{-4}\,\frac{\sqrt{\epsilon}\,e^{-\chi_{{}_S}}}{\sin\varphi_{{}_S}}\,\mpl\,.
\end{equation}
For $\varphi_{{}_S}\simeq\pi/2$, and assuming $\chi_{{}_T}\lesssim \chi_{{}_S}$, the most recent determination of the scalar power-spectrum amplitude by the Planck experiment, $A_{{}_S}=2.0989_{-0.034}^{+0.031}\times 10^{-9}$ \cite{Planck:2018jri}, yields
\begin{equation}\label{lambda-over-H-chi-large}
\frac{\Lambda}{H}\leq \frac{\Lambda^{{\rm max}}_{{}_S}(\varphis\simeq \frac{\pi}{2},\chis\gg 1)}{H}\approx \sqrt{2\,\mathcal{C}}\,|\eta|^{1/4}\simeq 70.0922\, {|\eta| }^{1/4}\,,
\end{equation}
and more generally $\Lambda^{{\rm max}}_{{}_S}/H\simeq 70.0922\,|\eta|^{1/4}\sin^{1/2}\varphi_{{}_S}$ for $\chi_{{}_S}\gg1$ and $\varphi_{{}_S}$ not too close to $0$ or $\pi$. Here $\eta$ is the second slow-roll parameter of eq.~\eqref{eq:slow-roll}, whose values for the two models are given in eq.~\eqref{eq:eta-values}. In this case, assuming BD initial conditions for the tensor perturbations, the tensor-to-scalar ratio can be suppressed well below its BD value by a factor of $e^{-2\chi_{{}_S}}$. Finally, we note that the backreaction bound provides only an upper limit on the scale of new physics $\Lambda_{{}_{S}}$.

In this paper, we investigate the problem from a more general perspective, without restricting ourselves to any specific region of parameter space. We therefore make no a priori assumptions about $\chi$ or $\varphi$. Since the
ACT DR6 data primarily constrain the scalar sector, while the effects of non-BD tensor modes may become detectable only in future observations, we focus on the scalar sector in what follows.

\subsection{Shifts in $n_{\rm s}$ and $\alpha_{\rm s}$}
\label{sec:shifts}

From eq.~\eqref{eq:PS}, the shifts of the spectral tilt and of its running with respect to the
BD values are
\begin{equation}\label{eq:ns-shift}
  \dns{} \equiv n_{{}_{\rm s}} - n_{{}_{\mathrm{s},\BD}} = \frac{\partial\ln\gamma_{{}_S}}{\partial\ln k} = \frac{\dk{\gamma}_{{}_S}}{\gamma_{{}_S}}\,,
  \qquad
\das{} \equiv \alpha_{\rm s} - \alpha_{{{}_{\rm s,\BD}}} = \frac{\rdm}{\rdm\ln k}\!\left(\frac{\dk{\gamma}_{{}_S}}{\gamma_{{}_S}}\right),
\end{equation}
where a circle denotes a derivative with respect to $\ln k$. A non-vanishing shift therefore
requires $\chi_{{}_S}$ and/or $\varphi_{{}_S}$ to depend on $k$. Allowing both to be scale-dependent, and
defining
\begin{equation}\label{eq:B}
  \Bs \equiv e^{2\chi_{{}_S}}\sin^2\varphi_{{}_S} - e^{-2\chi_{{}_S}}\cos^2\varphi_{{}_S}\,,
\end{equation}
we find
\begin{equation}\label{eq:ns-shift-general}
  \dns{} = \frac{2}{\gamma_{{}_S}}\Big(\sinh2\chi_{{}_S}\,\sin2\varphi_{{}_S}\;\dk{\varphi}_{{}_S} + \Bs\,\dk{\chi}_{{}_S}\Big),
\end{equation}
and
\begin{equation}\label{eq:alphas-shift-general}
\begin{split}
  \das{} =& \frac{2}{\gamma_{{}_S}}\Big(\sinh2\chi_{{}_S}\,\sin2\varphi_{{}_S}\;\ddk{\varphi}_{{}_S} + \Bs\,\ddk{\chi}_{{}_S}\Big)
  - \frac{4\sinh2\chi_{{}_S}\,\Bs}{\gamma_{{}_S}^2}\,\big(\dk{\varphi}_{{}_S}\big)^2\\
  &+ \frac{4\sin^2 2\varphi_{{}_S}}{\gamma_{{}_S}^2}\,\big(\dk{\chi}_{{}_S}\big)^2
  + \frac{8\sin2\varphi_{{}_S}}{\gamma_{{}_S}^2}\,\dk{\chi}_{{}_S}\,\dk{\varphi}_{{}_S}\,.
\end{split}
\end{equation}
When both $\dk\varphi_{{}_S}$ and $\dk\chi_{{}_S}$ are non-zero, the relative importance of the two terms in
eq.~\eqref{eq:ns-shift-general} depends on $\varphi_{{}_S}$ and $\chi_{{}_S}$. To keep the analysis
transparent, we instead consider the two separate cases in which only one of the two parameters
is scale dependent, and determine the allowed ranges of $\varphi_{{}_S}$ and $\chi_{{}_S}$ in each case.
Accordingly, we assume that the running arises from the variation of either $\varphi_{{}_S}$ or
$\chi_{{}_S}$ alone and neglect the mixed term $\dk\chi_{{}_S}\dk\varphi_{{}_S}$ in
eq.~\eqref{eq:alphas-shift-general}. For each case one must check whether there exist values of
$\varphi_{{}_S}$ and $\chi_{{}_S}$ for which the observed $n_{\rm s}$ and $\alpha_{\rm s}$ are reproduced, and whether
the regions allowed by the two requirements overlap reasonably.

\section{Allowed ranges of $\chi_{{}_S}$ and $\varphi_{{}_S}$ in two inflationary models}
\label{sec:models}

\subsection{The Starobinsky and quadratic chaotic models}
\label{sec:models-intro}

We study two well-known models: (i) the Starobinsky model and (ii) the quadratic chaotic (QC)
model. The Starobinsky potential is
\begin{equation}\label{eq:V-St}
  V_{{}_{\rm St}} = \frac{3}{4}\mpl^2M^2\left(1 - e^{-\sqrt{2/3}\,\phi/\mpl}\right)^2,
\end{equation}
and the QC potential is
\begin{equation}\label{eq:V-QC}
  V_{{}_{\rm QC}} = \frac{1}{2}m^2\phi^2\,.
\end{equation}
For $N_e$ e-folds of inflation after horizon exit of the pivot scale, the BD predictions of the
two models are
\begin{align}
  n_{{}_{S,\BD}}^{{}_{\rm St}} &= 1-\frac{2}{N_e}\,, & r_{{}_{\BD}}^{{}_{\rm St}} &= \frac{12}{N_e^2}\,, & \alpha_{{}_{S,\BD}}^{{}_{\rm St}} &= -\frac{2}{N_e^2}\,, \label{eq:St-pred}\\
  n_{{}_{S,\BD}}^{{}_{\rm QC}} &= 1-\frac{2}{N_e}\,, & r_{{}_\BD}^{{}_{\rm QC}} &= \frac{8}{N_e}\,, & \alpha_{{}_{S,\BD}}^{{}_{\rm QC}} &= -\frac{2}{N_e^2}\,, \label{eq:QC-pred}
\end{align}
whose numerical values for $N_e=60$ are collected in table~\ref{tab:BD}. In terms of the slow-roll parameters \eqref{eq:slow-roll}, at horizon exit of the pivot scale
\begin{equation}\label{eq:eta-values}
\epsilon^{{}_{\rm St}}\simeq\frac{3}{4N_e^2}\,,\quad \eta^{{}_{\rm St}}\simeq-\frac{1}{N_e}\,;\qquad
\epsilon^{{}_{\rm QC}}\simeq\frac{1}{2N_e}\,,\quad \eta^{{}_{\rm QC}}\simeq\frac{1}{12N_e^2}\,.
\end{equation}
For the QC model, $\eta$ vanishes at leading order in $1/N_e$, since there $\epsilon'/(2\hh\epsilon)\simeq\epsilon$, and the value quoted is the next-order result $\eta\simeq\epsilon^2/3$. We have checked both values against a numerical solution of the background equations, which for $N_e=60$ gives $\eta^{{}_{\rm St}}\simeq-0.0159$ and $\eta^{{}_{\rm QC}}\simeq2.31\times10^{-5}$. In the numerical estimates below we therefore use $|\eta|=1/N_e$ for the Starobinsky model and $|\eta|=1/(12N_e^2)$ for the QC model, with $N_e=60$. The value of $r$ for the
Starobinsky model is well below the current bound $r<r_{{}_{\rm max}}$ (95\% CL), whereas $n_{\rm s}$ and
$\alpha_s$ are under tension with ACT DR6. The QC model has the same $n_{\rm s}$ and $\alpha_s$, but a
much larger $r$. If the BD vacuum were the only possible initial state, the QC model would
therefore have to be regarded as ruled out. As we show below, allowing for non-BD initial
states changes this conclusion.
\begin{table}[t]
  \centering
  \begin{tabular}{lccc}
    \toprule
    Model ($N_e=60$) & $n_{s,\BD}$ & $r_\BD$ & $\alpha_{s,\BD}$ \\
    \midrule
    Starobinsky & $0.9667$ & $0.0033$ & $-0.00056$ \\
    Quadratic Chaotic (QC) & $0.9667$ & $0.1333$ & $-0.00056$ \\
    \bottomrule
  \end{tabular}
  \caption{BD predictions of the Starobinsky and QC models for $N_e=60$.}
  \label{tab:BD}
\end{table}

The shifts $\dns{}$ and $\das{}$ are required to equal the
difference between the observed and the BD values,
\begin{equation}\label{eq:delta-obs}
  \dns{} = n_{\rm s}^{\Ob} - n_{{\rm s}, {}_{\BD}}\,, \qquad
  \das{} = \alpha_{\rm s}^{\Ob} - \alpha_{{\rm s},{}_{\BD}}\,.
\end{equation}
Using the ACT DR6 values of table~\ref{tab:obs} together with table~\ref{tab:BD}, and since both models share the same BD values of $n_{\rm s}$ and $\alpha_{\rm s}$, the tension is
resolved within the $1\sigma$ ranges of ACT DR6 in both models if
$\dns{}\in[0.00423,\,0.01103]$ and $\das{}\in[0.00156,\,0.01196]$.
For the ACT mean values and $N_e=60$ these shifts are
\begin{equation}\label{eq:delta-mean}
  \dns{} = 0.00763\,, \qquad \das{} = 0.00676\,.
\end{equation}

\subsection{Constraints from the bounds on $r$ and the validity of the EFT}
\label{sec:r-constraint}

The allowed values of $\varphi_{{}_S}$ and $\chi_{{}_S}$ follow from eqs.~\eqref{eq:ns-shift-general} and
\eqref{eq:alphas-shift-general}, but not all of them are permitted. First, assuming $\gamma_{{}_T}\simeq1$ we have
$r=r_{{}_\BD}/\gamma_{{}_S}$, and we only retain the values of $\varphi_{{}_S}$ and $\chi_{{}_S}$ that satisfy
$r<r_{{}_{\rm max}}=0.038$, i.e.
\begin{equation}\label{eq:r0.038}
R\equiv  \frac{r_{{}_\BD}}{r_{{}_{\rm max}}} < e^{2\chi_{{}_S}}\sin^2\varphi_{{}_S} + e^{-2\chi_{{}_S}}\cos^2\varphi_{{}_S}\,.
\end{equation}
This equation guarantees that the predictions for both the QC and Starobinsky with non-BD initial conditions satisfy the latest BICEP-Planck constraint on the tensor-to-scalar ratio. The second condition is that the EFT theory should hold during inflation. Namely, the cutoff of the EFT theory, in both the tensor and scalar sectors, should satisfy
$\Lambda^{\rm max}_{{}_{S,T}}(\varphi_{{}_{S}},\chi_{{}_{S,T}})>H$. Assuming that $\chi_{{}_T}\leq \chi_{{}_S}$, the most stringent bound is imposed by
\begin{equation}
\frac{\Lambda^{{\rm max}}_{{}_S}(\varphi_{{}_{S}},\chi_{{}_{S}})}{H}>1\,.
\label{EFT-Validity}
\end{equation}
\begin{figure}[t]
  \centering
  \begin{subfigure}[t]{0.48\textwidth}
    \centering
    \includegraphics[width=\linewidth]{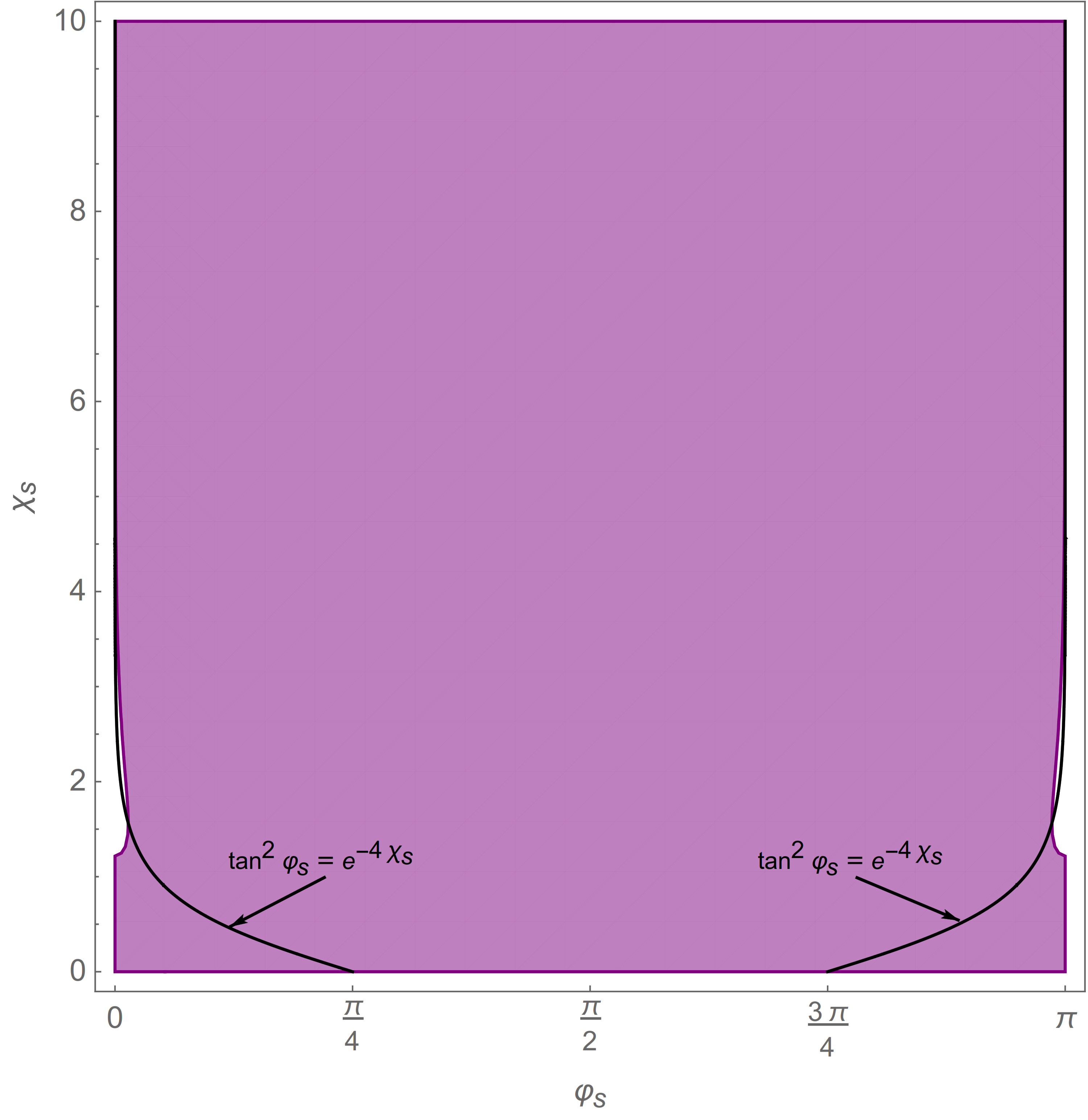}
    \caption{Starobinsky, $0\leq \varphi_{{}_S}\leq \pi$}
  \end{subfigure}
  \hfill
  \begin{subfigure}[t]{0.48\textwidth}
    \centering
    \includegraphics[width=\linewidth]{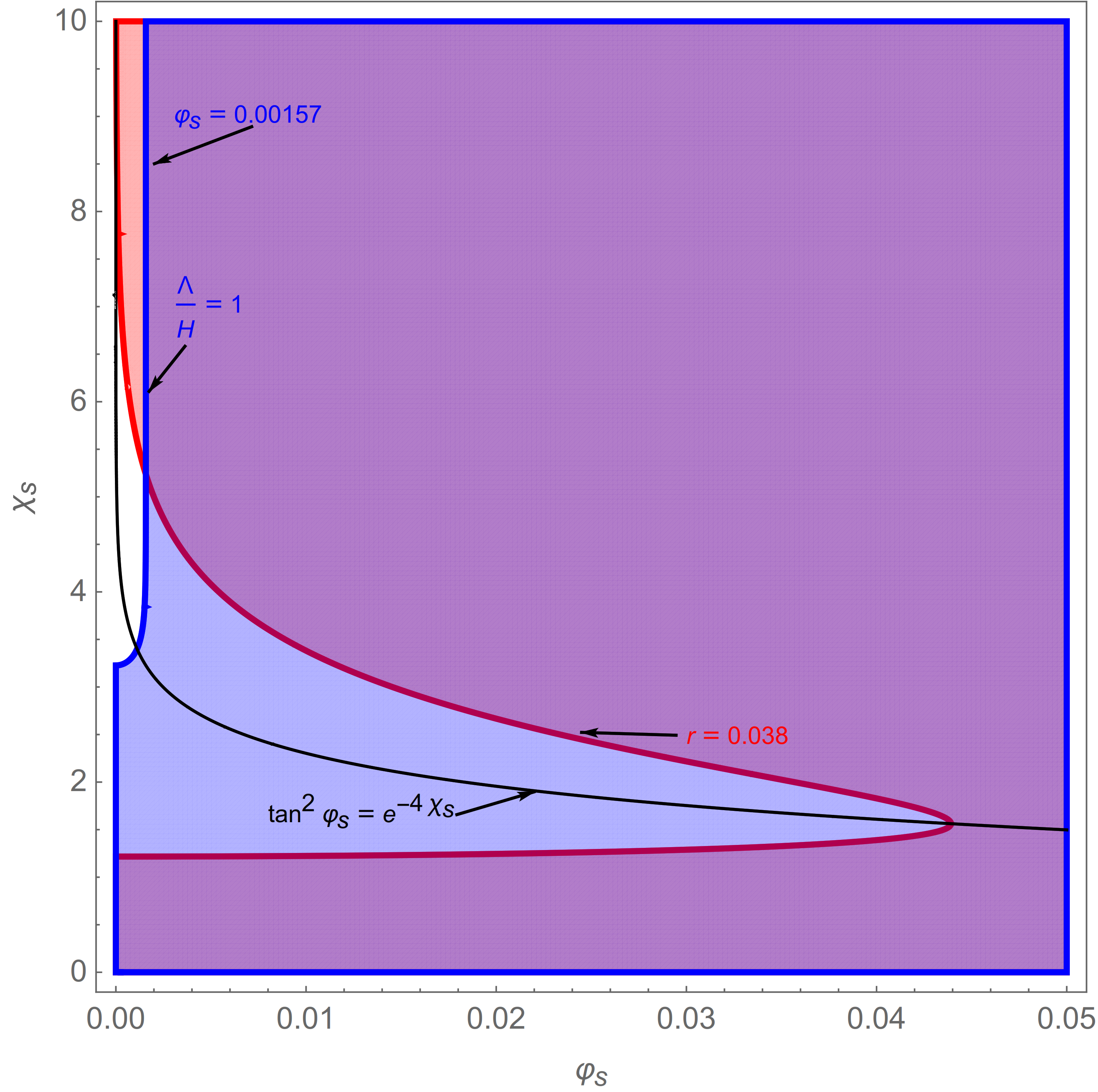}
    \caption{Starobinsky, $0\leq \varphi_{{}_S}\leq 0.05$}
  \end{subfigure}
  \caption{The left plot shows the allowed region in the $(\varphi_{{}_S},\chi_{{}_S})$ plane implied by the bound
  \eqref{eq:r0.038} for the Starobinsky model in the range $0\leq \varphi_{{}_S}\leq \pi$. The purple region is the region of parameter space in which both $r<r_{{}_{\rm max}}$ and $\Lambda_{{}_S}^{{\rm max}}/H\geq1$ are satisfied. In the right plot, we have zoomed in on the interval $0\leq \varphi_{{}_S}\leq 0.05$ to show how different bounds confine the boundary of the region close to $\varphi_{{}_S}\approx 0$. A mirror image of the left plot with respect to $\varphi_{{}_S}=\frac{\pi}{2}$ exists for the boundary close to $\varphi_{{}_S}\approx \pi$. Part of the curve $\Bs=0$, $\tan^2\varphi_{{}_S}=e^{-4\chi_{{}_S}}$, passes through the allowed region in the Starobinsky model.}
    \label{domain-starobinsky-fig}
\end{figure}

For the Starobinsky Inflation, the tensor-to-scalar ratio, $r$, already respects the latest experimental bounds on $r$. Therefore for $\varphis$ away from $0$ and $\pi$, any value of $\chis$ is allowed. Close to $\varphis=0$ and $\pi$, the parameter space starts to feel the tension from $r<r_{{}_{\rm max}}$. For $\varphis=0,\ \pi$ exactly, $\gamma_{{}_S}=e^{-2\chis}$ and $r_{{}_{\rm BD}}$, instead of being suppressed, gets an enhancement factor of $e^{2\chis}$. This means that for $\varphis=0,\ \pi$, $\chis\geq\chi_{{}_{S,1}}\equiv -\frac{1}{2}\ln R \simeq 1.21681$ should be excluded from the allowed region of parameter space. For larger $\chis$, not only $\varphis=0$ (or $\varphis=\pi$) is removed, but slightly larger values of $\varphis$ (or of $\pi-\varphis$) are excluded as well. The boundary of the excluded region is the curve $\gamma_{{}_S}=r_{{}_{\rm BD}}/r_{{}_{\rm max}}$, i.e.
\begin{equation}\label{eq:r-boundary}
\sin^2\varphis=\frac{R\,e^{2\chis}-1}{e^{4\chis}-1}\,.
\end{equation}
The maximum exclusion in $\varphis$ occurs where this curve satisfies $\partial\gamma_{{}_S}/\partial\chi_{{}_S}=2\Bs=0$, i.e.\ on the curve $\tan\varphis=e^{-2\chis}$, on which $\gamma_{{}_S}=1/\cosh2\chis=\sin2\varphis$. This gives
\begin{equation}\label{chis2-phis2-star}
\chi_{{}_{S,2}}=\frac{1}{2}\operatorname{arccosh} R\simeq1.56242\,,\quad
\varphi_{{}_{S,2}}=\frac{1}{2}\arcsin R \simeq 0.04392
\end{equation}
(or $\pi-\varphi_{{}_{S,2}}$ close to $\varphis=\pi$). For $\chis\gtrsim\chi_{{}_{S,2}}$, eq.~\eqref{eq:r-boundary} quickly approaches $\sin^2\varphis \simeq 0.08772\, e^{-2\chis}$, and the boundary follows this curve until it meets the curve $\Lambda^{{\rm max}}_{{}_{S}}=H$, i.e.\ $\sinh\chis=\mathcal{C}\sqrt{\gamma_{{}_S}|\eta|}$, at the point
\begin{align}
\chi_{{}_{S,3}} &=
\operatorname{arcsinh}\left(\mathcal{C}\sqrt{|\eta|\,R}\right)
\simeq 5.23567\,,\\
\varphi_{{}_{S,3}} &=
\arcsin\sqrt{
\frac{R\,e^{2\chi_{{}_{S,3}}}-1}{e^{4\chi_{{}_{S,3}}}-1}
}\simeq 0.00157636\,.
\label{chis3-phis3-star}
\end{align}
\begin{figure}[t]
  \centering
  \begin{subfigure}[t]{0.48\textwidth}
    \centering
    \includegraphics[width=\linewidth]{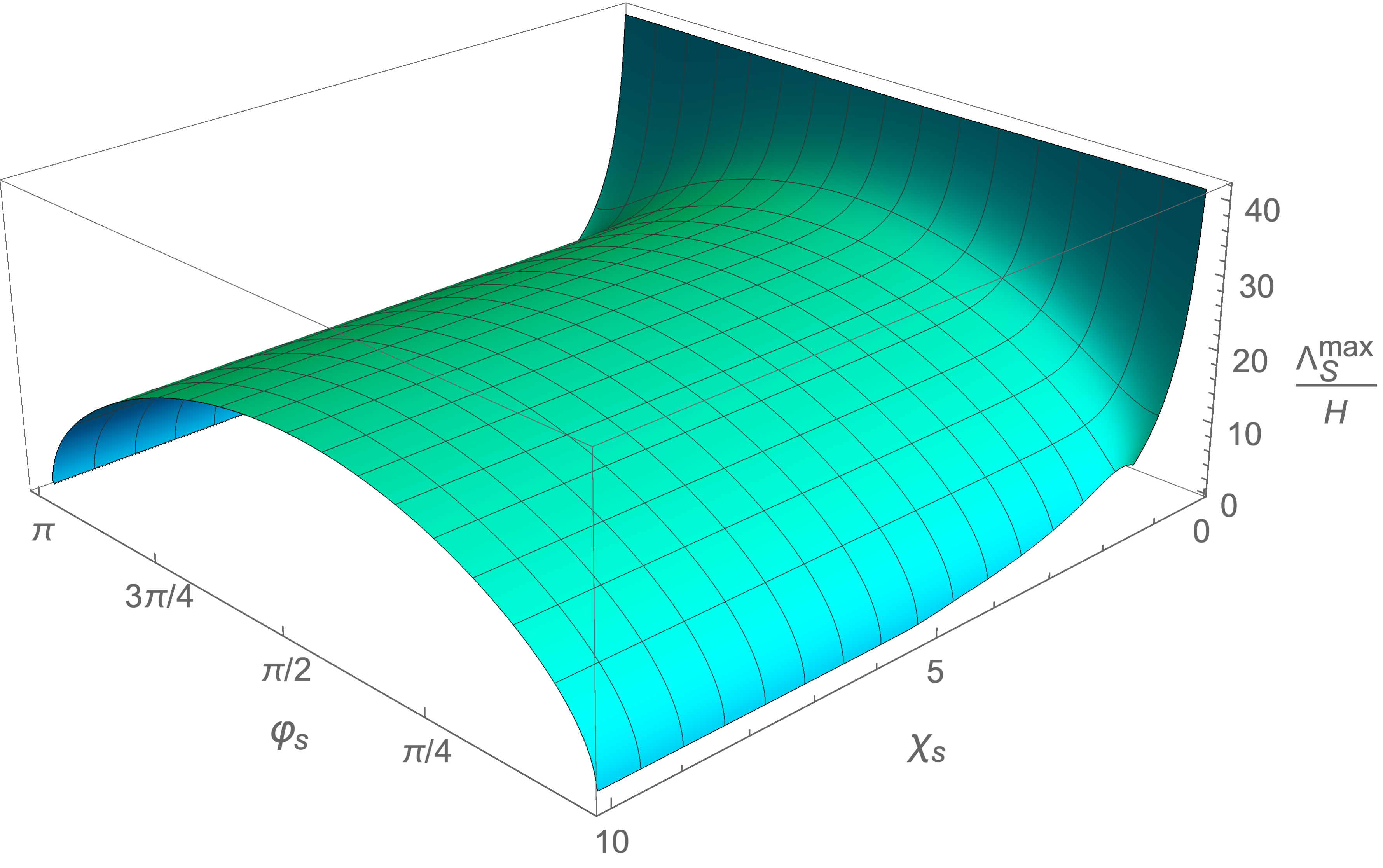}
    \caption{Starobinsky}
    \label{Lambda-over-H-fig-Starobinsky}
  \end{subfigure}
  \hfill
  \begin{subfigure}[t]{0.48\textwidth}
    \centering
    \includegraphics[width=\linewidth]{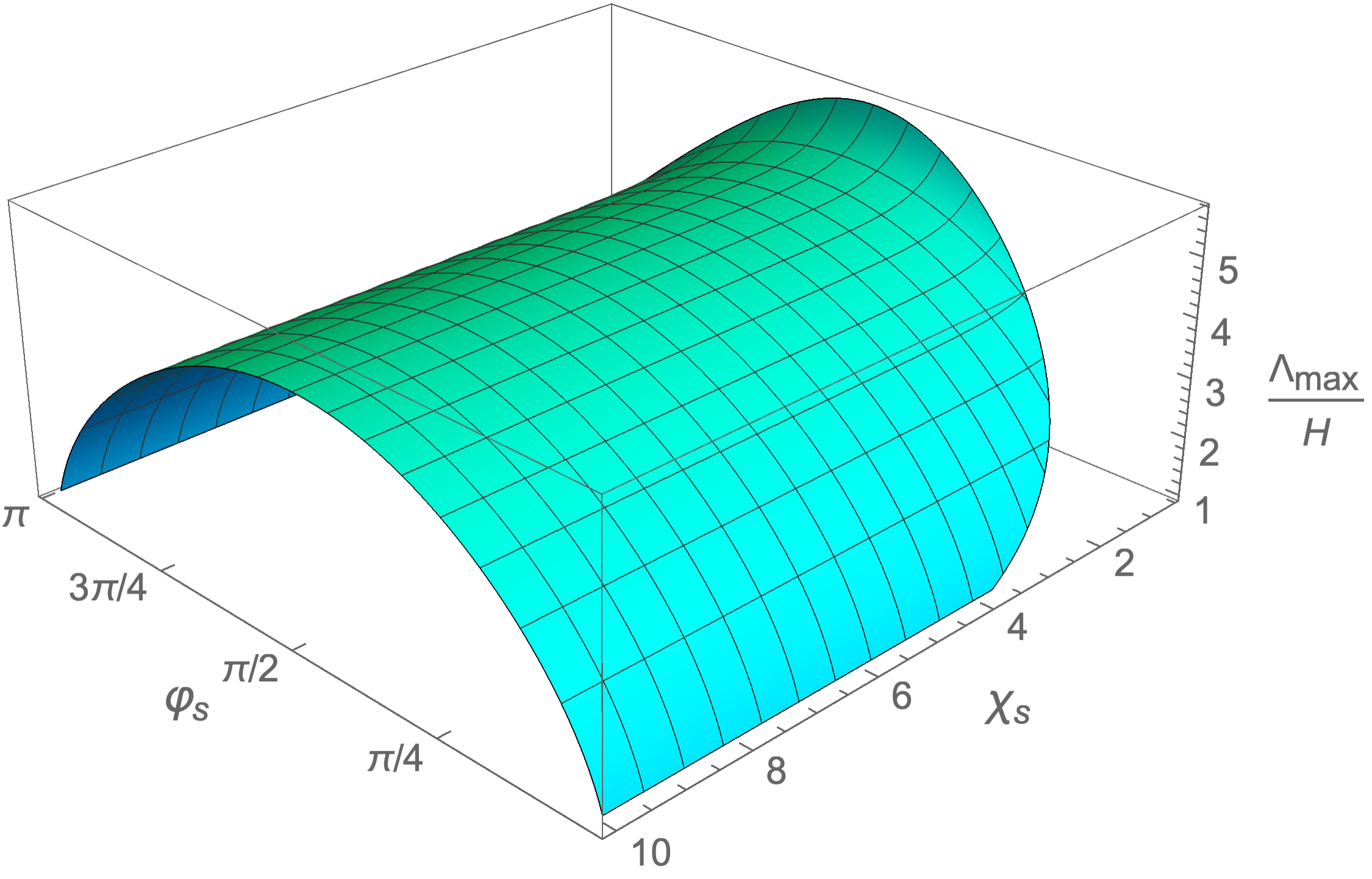}
    \caption{QC}
    \label{Lambda-over-H-fig-QC}
  \end{subfigure}
 \caption{Left and right plots, respectively, plot the behavior of $\frac{\Lambda^{{\rm max}}_{{}_S}(\varphi_{{}_{S}},\chi_{{}_{S}})}{H}$ in viable domains of $\varphis$ and $\chis$ for each model. In the QC model, the regime where $\chis\ll1$, in which $\Lambda^{{\rm max}}_{{}_S}/H$ grows like $\chis^{-1/2}$, is not accessible.}
    \label{Lambda-over-H-fig}
\end{figure}
For $\chis>\chi_{{}_{S,3}}$ the lower edge of the allowed region is set by $\Lambda^{{\rm max}}_{{}_{S}}=H$. Using $\gamma_{{}_S}\simeq e^{2\chis}\sin^2\varphis$ and $\sinh\chis\simeq e^{\chis}/2$, this edge becomes independent of $\chis$ and $\varphis$ approaches
\begin{equation}\label{phis-infty}
\varphi_{{}_{S,\infty}}=\arcsin\left(\frac{1}{2\,\mathcal{C}\,|\eta|^{1/2}}\right)\simeq0.0015767\simeq\varphi_{{}_{S,3}}\,.
\end{equation}
The plots demonstrating the allowed domain for $\chis$ and $\varphis$ are displayed in figure~\ref{domain-starobinsky-fig}. The plot of $\frac{\Lambda^{{\rm max}}_{S}(\varphi_{{}_{S}},\chi_{{}_{S}})}{H}$ for the Starobinsky model is shown in figure~\ref{Lambda-over-H-fig-Starobinsky}. For $\chis\gg 1$, and values of $\varphis$ not too close to $0$ or $\pi$,
$\frac{\Lambda^{{\rm max}}_{S}(\varphi_{{}_{S}},\chi_{{}_{S}})}{H}$ approaches $25.1844 \sin^{1/2}\varphis$, which takes its maximum value at $\varphis=\frac{\pi}{2}$. Along the lower edge of the allowed region with $\chi_{{}_{S,2}}\lesssim\chis\lesssim\chi_{{}_{S,3}}$, we have $\gamma_{{}_S}=R$ and
$\sin \varphis \approx \varphis \approx 0.296174\, e^{-\chis}$ (or $\pi-\varphis \approx 0.296174\, e^{-\chis}$ close to $\varphis=\pi$), so that $\frac{\Lambda^{{\rm max}}_{{}_S}}{H}\approx\sqrt{2\mathcal{C}}\left(|\eta|\,R\right)^{1/4}e^{-\frac{\chis}{2}}\approx 13.7058\, e^{-\frac{\chis}{2}}$, which indeed reaches unity at $\chis=\chi_{{}_{S,3}}$. On the other hand, for $\chis\ll 1$ we have $\gamma_{{}_S}\simeq1$ and $\frac{\Lambda^{{\rm max}}_{S}}{H}\simeq\mathcal{C}^{1/2}|\eta|^{1/4}\chis^{-1/2}\simeq17.81\,\chis^{-1/2}$, so the scale of new physics can become arbitrarily large as $\chis\to0$, almost independently of $\varphis$.
\begin{figure}[t]
  \centering
  \begin{subfigure}[t]{0.48\textwidth}
    \centering
    \includegraphics[width=\linewidth]{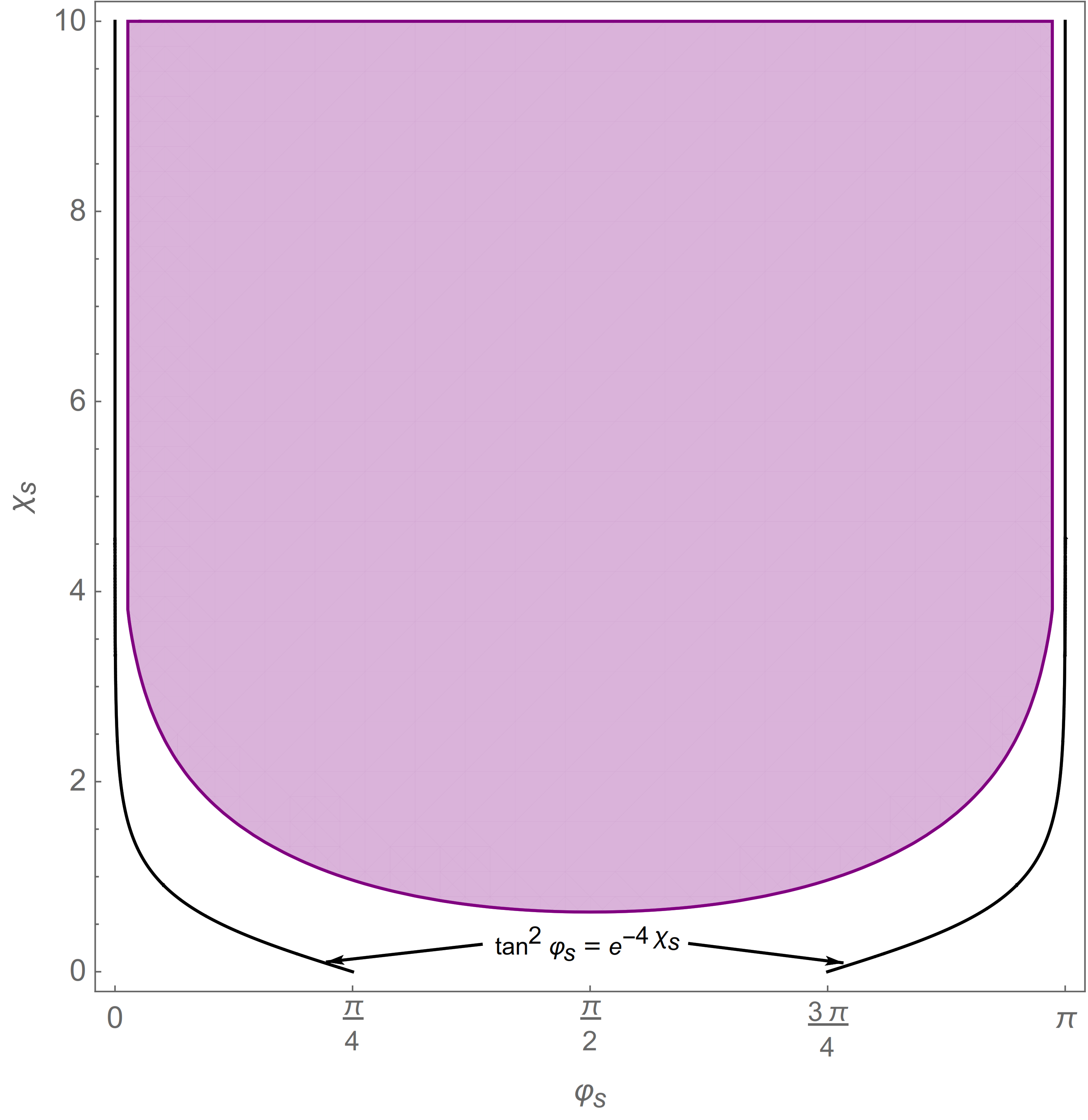}
    \caption{QC, $0\leq\varphi_{{}_S}\leq \pi$}
    \label{domain-QC-0-varphis-pi}
  \end{subfigure}
  \hfill
  \begin{subfigure}[t]{0.48\textwidth}
    \centering
    \includegraphics[width=\linewidth]{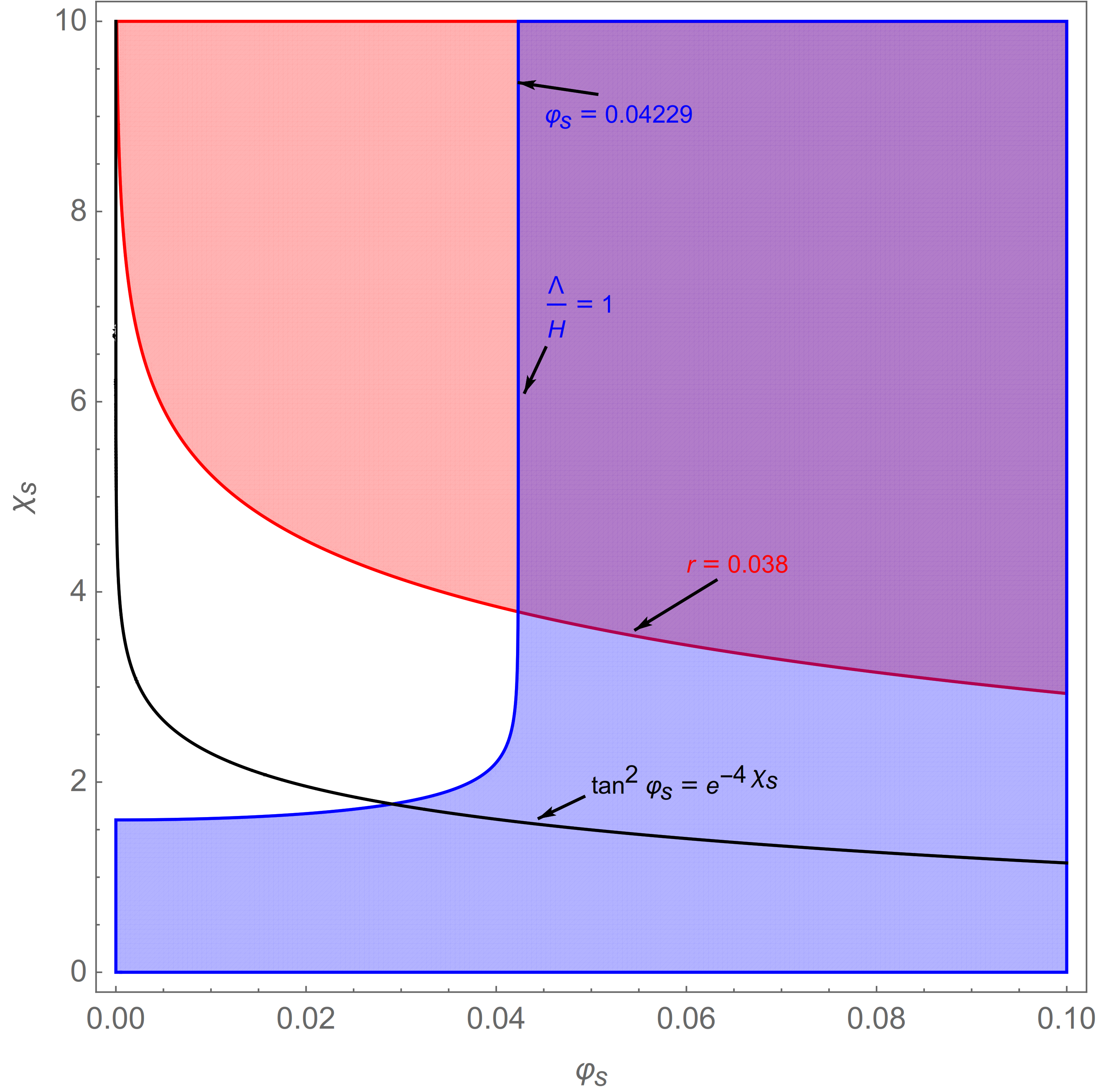}
    \caption{QC, $0\leq\varphi_{{}_S}\leq 0.05$}
    \label{domain-QC-0-varphis-0.05}
  \end{subfigure}
  \caption{The left plot shows the allowed region in the $(\varphi_{{}_S},\chi_{{}_S})$ plane implied by the bound
  \eqref{eq:r0.038} for the QC model in the range $0\leq \varphi_{{}_S}\leq \pi$. The purple region is the region of parameter space in which both $r<r_{{}_{\rm max}}$ and $\frac{{\Lambda_{{}_S}}^{\rm max}}{H}\geq1$ are satisfied. In the right plot, we have zoomed in on the interval $0\leq \varphi_{{}_S}\leq 0.05$ to show how different bounds confine the boundary of the region close to $\varphi_{{}_S}\approx 0$. A mirror image of the left plot with respect to $\varphi_{{}_S}=\frac{\pi}{2}$ exists for the boundary close to $\varphi_{{}_S}\approx \pi$. The curve $\Bs=0$, $\tan^2\varphi_{{}_S}=e^{-4\chi_{{}_S}}$, never passes through the allowed region in the QC model.}
  \label{fig:domain-bound-phi2}
\end{figure}

 For the QC model, $R\simeq 3.5088$, so for any value of $\chis$ and $\varphis$ eq.~\eqref{eq:r0.038} requires $\gamma_{{}_S}> 3.5088$. Since $\gamma_{{}_S}\leq e^{2\chis}$, with equality at $\varphi_{{}_S}=\pi/2$, even for $\varphi_{{}_S}=\pi/2$ this demands that $\chi_{{}_S}$ be larger than $\chi_{{}_{S,1}}=\frac{1}{2}\ln R\simeq 0.6276$.
 In contrast with the Starobinsky inflation, as $\varphis$ goes toward $0$ or $\pi$, the minimum allowed value of $\chis$ increases monotonically and there is no equivalent of $\chi_{{}_{S,2}}$ in the QC model (the curve $\Bs=0$, on which $\gamma_{{}_S}=\sin2\varphis\leq1$, lies entirely in the excluded region). For $\chis\gtrsim 1$ the minimum of $\chis$ follows $\sin^2\varphis\approx 3.5088\, e^{-2\chis}$, until it reaches the points close to $\varphis= 0$ and $\varphis=\pi$ where the constraint from the validity of the EFT becomes important, at
 \begin{align}
\chi_{{}_{S,3}} &=
\operatorname{arcsinh}\left(\mathcal{C}\sqrt{|\eta|\,R}\right)
\simeq 3.79097\,,\\
\varphi_{{}_{S,3}} &=
\arcsin\sqrt{
\frac{R\,e^{2\chi_{{}_{S,3}}}-1}{e^{4\chi_{{}_{S,3}}}-1}
}\simeq 0.0422940\,.
\label{chis3-phis3-QC}
\end{align}
 For $\chis>\chi_{{}_{S,3}}$, $\varphis$ saturates at $\varphi_{{}_{S,\infty}}=\arcsin\left(1/(2\mathcal{C}|\eta|^{1/2})\right)\simeq0.0423187\simeq\varphi_{{}_{S,3}}$. We have displayed the allowed domain for $\chis$ and $\varphis$ in the QC model in figure~\ref{fig:domain-bound-phi2}. The plot of $\frac{\Lambda^{{\rm max}}_{{}_S}(\varphi_{{}_{S}},\chi_{{}_{S}})}{H}$ for the QC model is shown in figure~\ref{Lambda-over-H-fig-QC}. For $\chis\gg 1$, and values of $\varphis\approx \frac{\pi}{2}$, $\frac{\Lambda^{{\rm max}}_{{}_S}(\varphi_{{}_{S}},\chi_{{}_{S}})}{H}$ approaches $\sqrt{2\mathcal{C}}\,|\eta|^{1/4}\simeq 4.86$. Since $\chis>0.6276$ in the QC model, the regime $\chis\ll1$, in which $\Lambda^{{\rm max}}_{{}_S}/H$ grows like $\chis^{-1/2}$, is not accessible; the largest value, $\Lambda^{{\rm max}}_{{}_S}/H\simeq5.75$, is attained at $\varphis=\pi/2$, $\chis=\chi_{{}_{S,1}}$.

\subsection{Scale dependence of $\chi_{{}_S}$ and $\varphi_{{}_S}$}
\label{sec:scale-dep}

For the allowed values of $\varphi_{{}_S}$ and $\chi_{{}_S}$ we must also determine the required scale
dependence, i.e.\ $\dk\varphi_{{}_S}$, $\dk\chi_{{}_S}$, $\ddk\varphi_{{}_S}$ and $\ddk\chi_{{}_S}$. As explained in
section~\ref{sec:shifts}, we assume that $\dns{}$ is generated either by
$\dk\varphi_{{}_S}$ or by $\dk\chi_{{}_S}$, and that $\das{}$ is generated either by
$(\dk\varphi_{{}_S},\ddk\varphi_{{}_S})$ or by $(\dk\chi_{{}_S},\ddk\chi_{{}_S})$. This provides us with two options.

\begin{figure}[t]
  \centering
  \panel{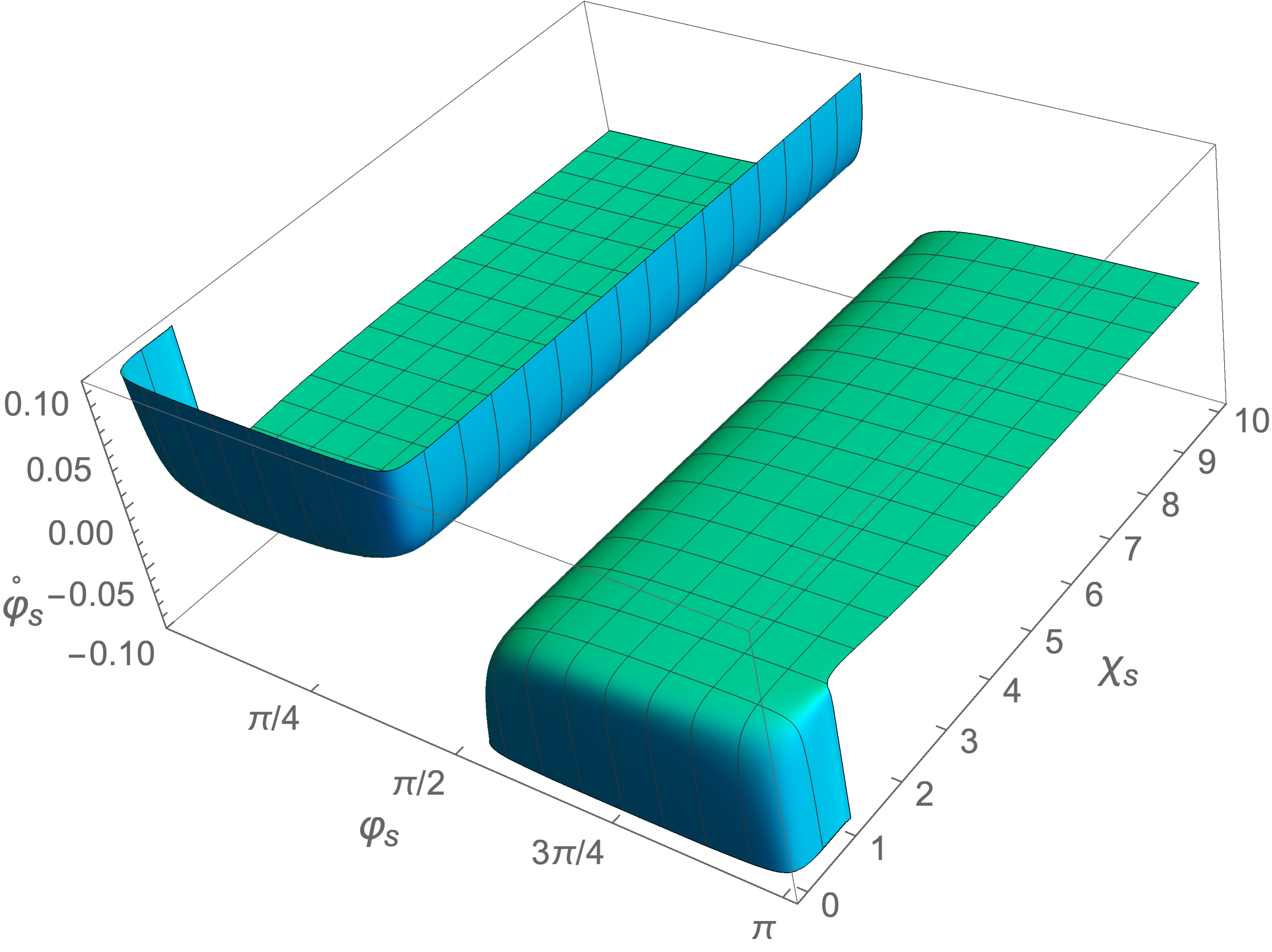}{Starobinsky}\hfill\panel{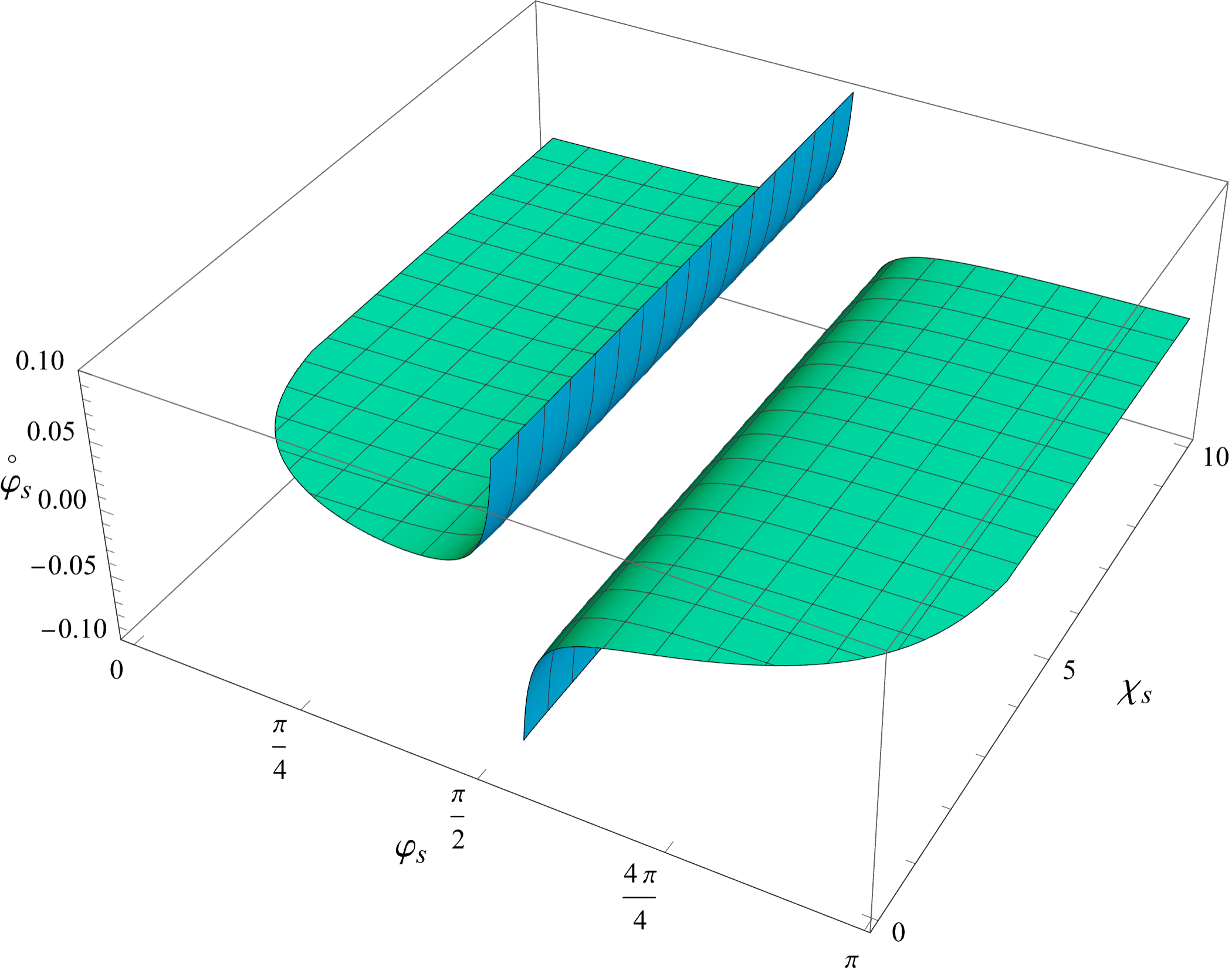}{QC}
  \caption{Allowed values for $\dk\varphi_{{}_S}$ as a function of $(\varphi_{{}_S},\chi_{{}_S})$, when $\dk\chi_{{}_S}=0$. In this plot, we have assumed that $n_{\rm s}$ has taken
  the mean value of $n_{\rm s}^{\Ob}$ from ACT DR6.  The regions clipped are from requiring $r<r_{{}_{\rm max}}$. }
  \label{fig:phidot-mean}
\end{figure}

\begin{figure}[t]
  \centering
  \panel{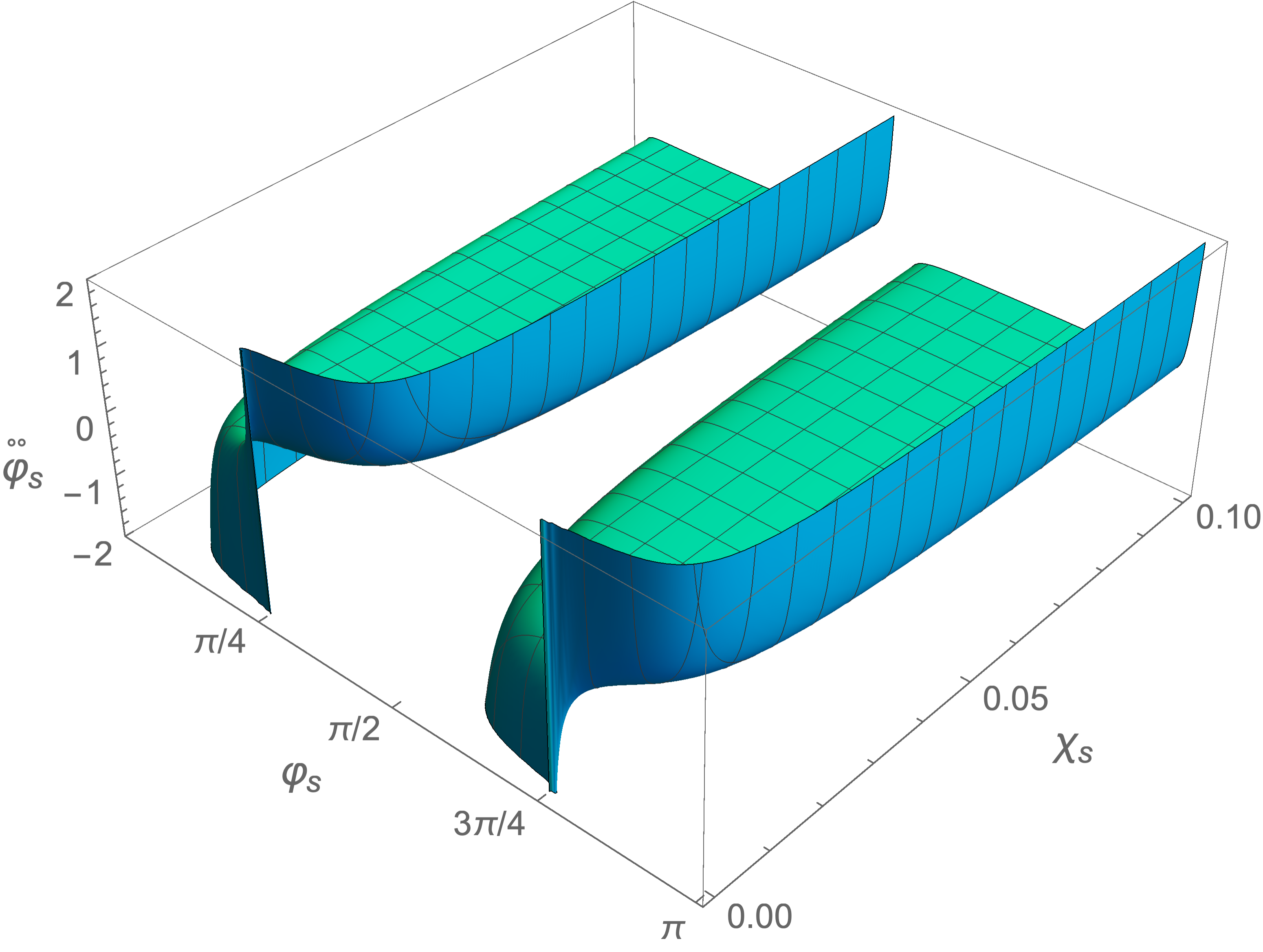}{Starobinsky, $0\leq \chi_{{}_S}\leq 0.1$}\hfill\panel{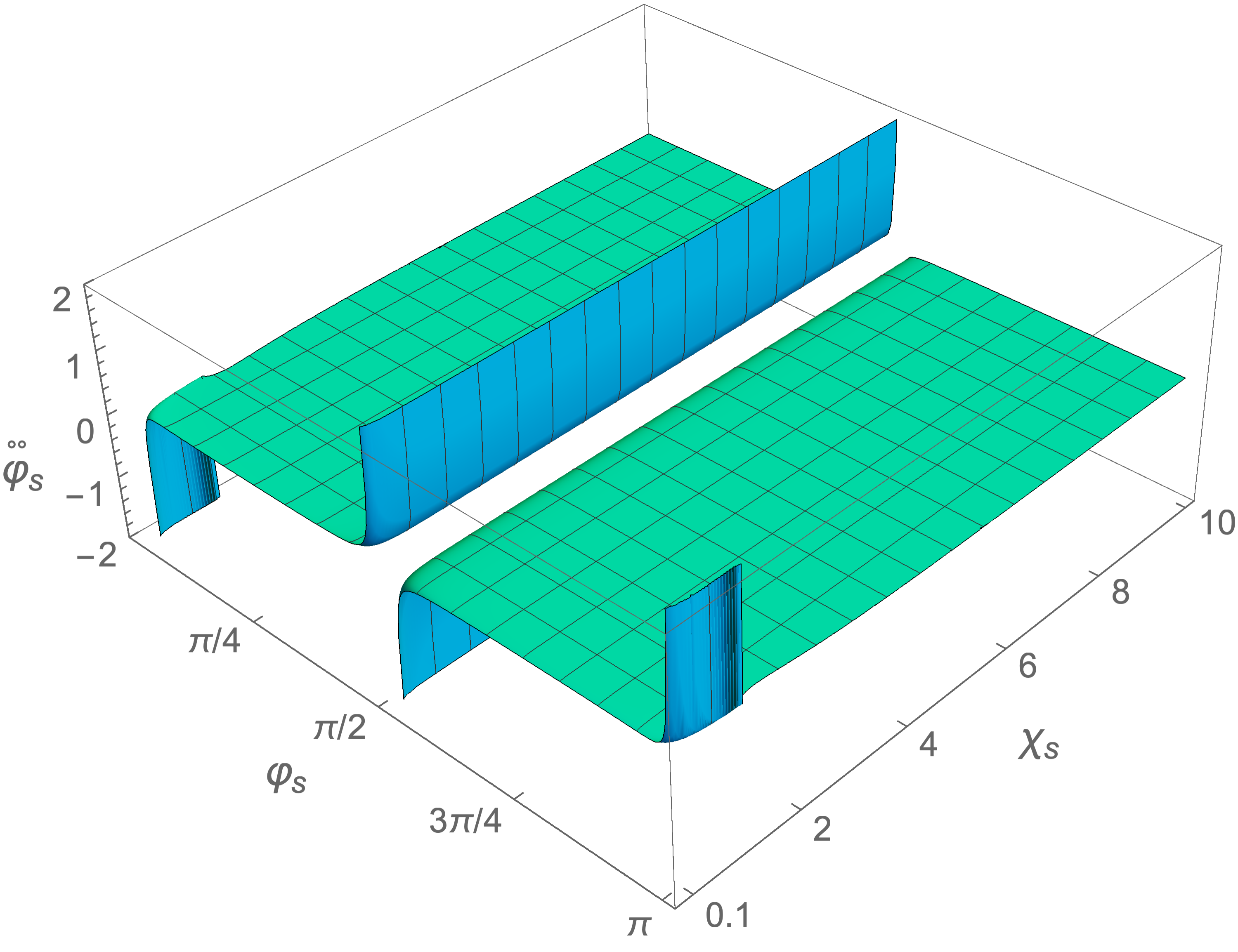}{Starobinsky, $0.1\leq \chi_{{}_S}\leq 10$}
  \caption{Plot of  $\ddk\varphi_{{}_S}$ as a function of $(\varphi_{{}_S},\chi_{{}_S})$ for scale-independent $\chi_{{}_S}$ ($\dk\chi_{{}_S}=\ddk\chi_{{}_S}=0$), assuming the mean values of $n_{\rm s}^{\Ob}$ and $\alpha_{{}_S}^{\Ob}$
  from the ACT DR6, for the Starobinsky model in different regions of $\chi_{{}_S}$. For the Starobinsky model, $|\ddk\varphi_{{}_S}|$ could become large, at
 $\varphi_{{}_S}=0, \frac{\pi}{2}$, $\pi$, and around $\chi_{{}_S}=0$. Parts of the domain along $\varphi_{{}_S}=0$ and $\varphi_{{}_S}=\pi$ are snipped by the requirement of $r<r_{{}_{\rm max}}$, Eq. \eqref{eq:r0.038}. }
  \label{fig:phiddot-mean-Starobinsky}
\end{figure}
\begin{figure}[t]
  \centering
  \panel{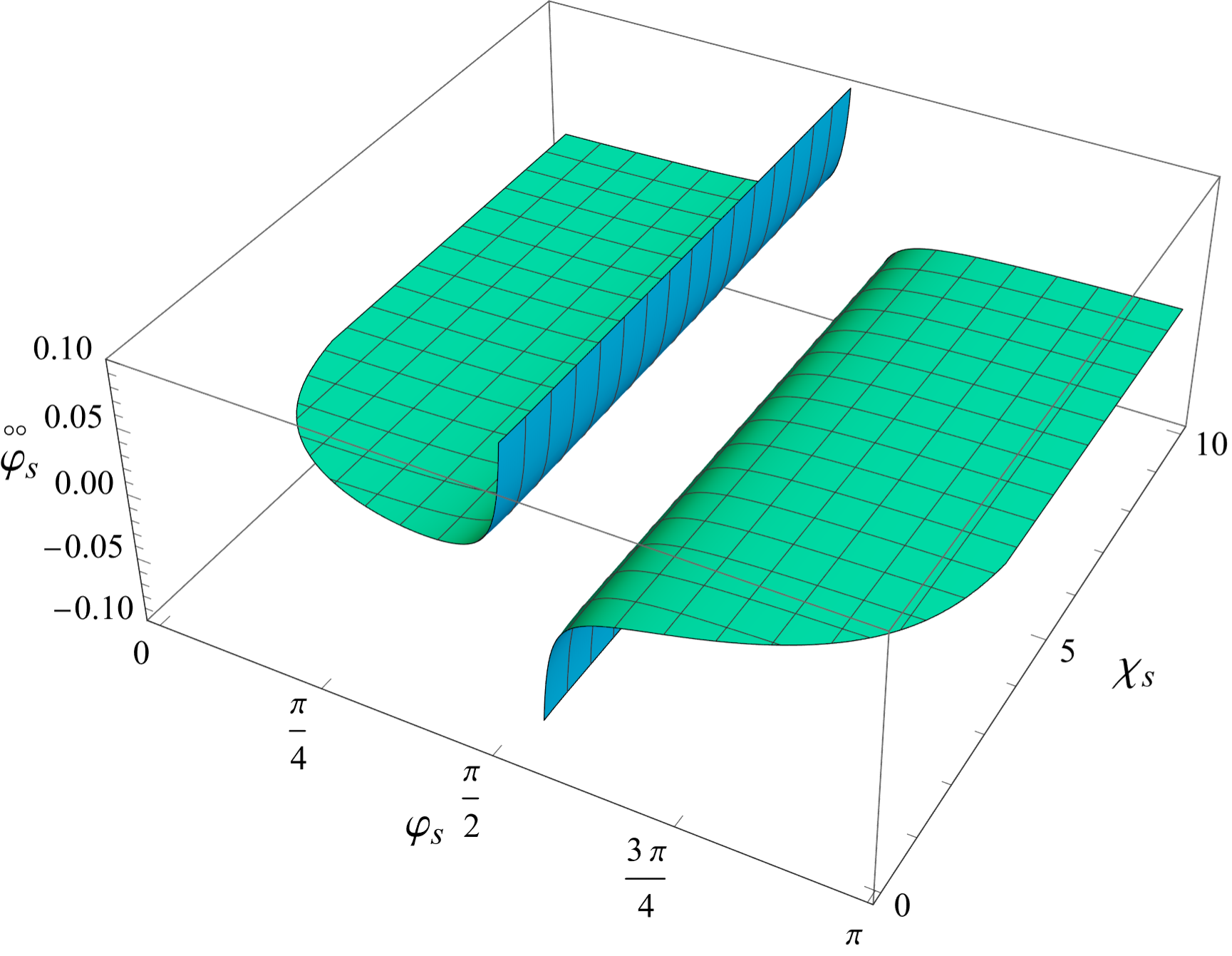}{QC}
  \caption{Plot of  $\ddk\varphi_{{}_S}$ as a function of $(\varphi_{{}_S},\chi_{{}_S})$ for scale-independent $\chi_{{}_S}$, assuming the mean values of $n_{\rm s}^{\Ob}$ and $\alpha_{{}_S}^{\Ob}$ from the ACT DR6, for the QC model. For the QC model, regions where $|\ddk\varphi_{{}_S}|$ could potentially become large, at
 $\varphi_{{}_S}=0, \pi$, and around $\chi_{{}_S}=0$ are trimmed by the bound that exists on the tensor-to-scalar ratio, Eq. \eqref{eq:r0.038}, and the validity of EFT, $\Lambda^{{\rm max}}_{S}\geq H$. For $\varphi_{{}_S}\simeq \frac{\pi}{2}$,  $|\ddk\varphi_{{}_S}|$ must take large values.}
  \label{fig:phiddot-mean-QC}
\end{figure}

\begin{figure}[p]
  \centering
  \panel{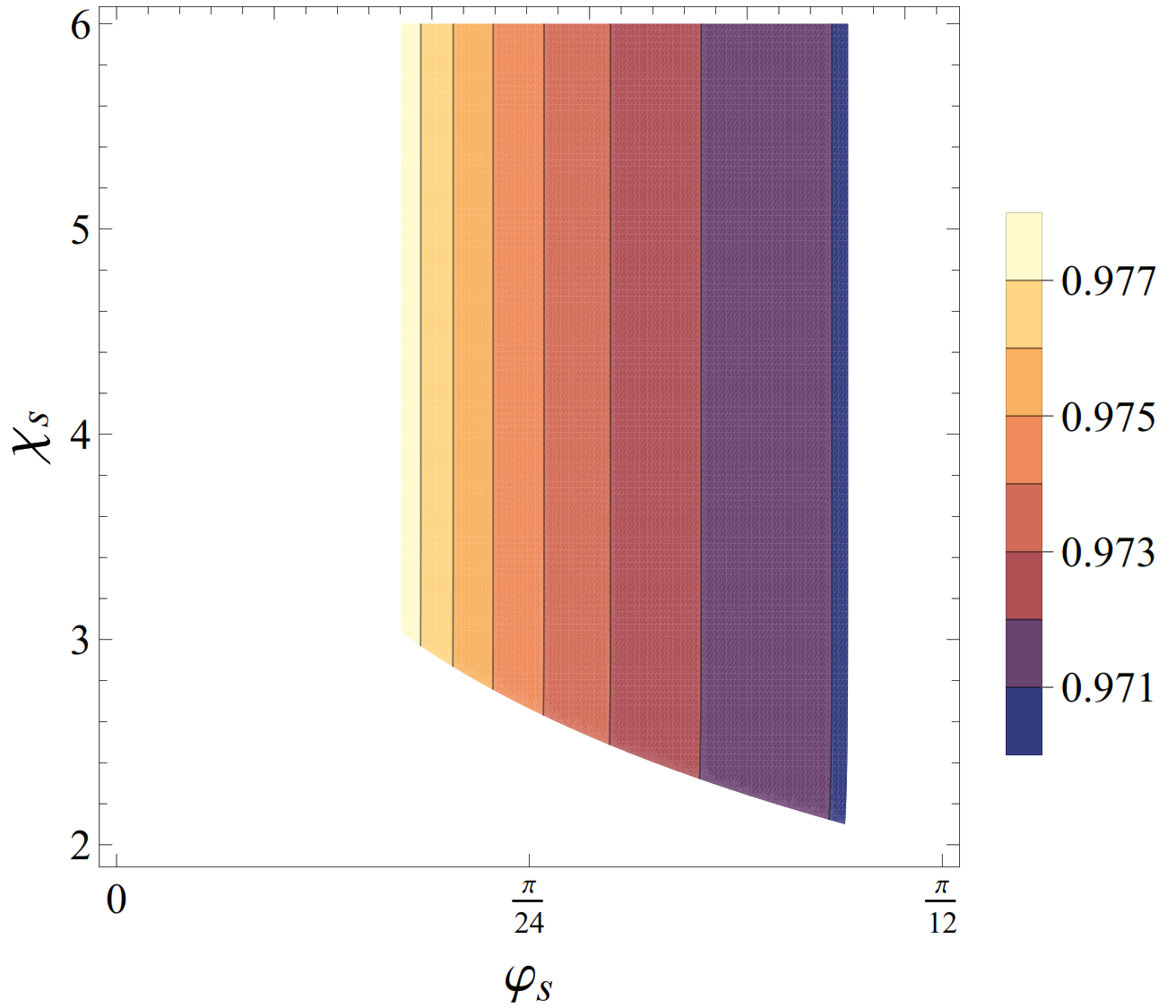}{$\dk\varphi_{{}_S}=0.0005$}\hfill\panel{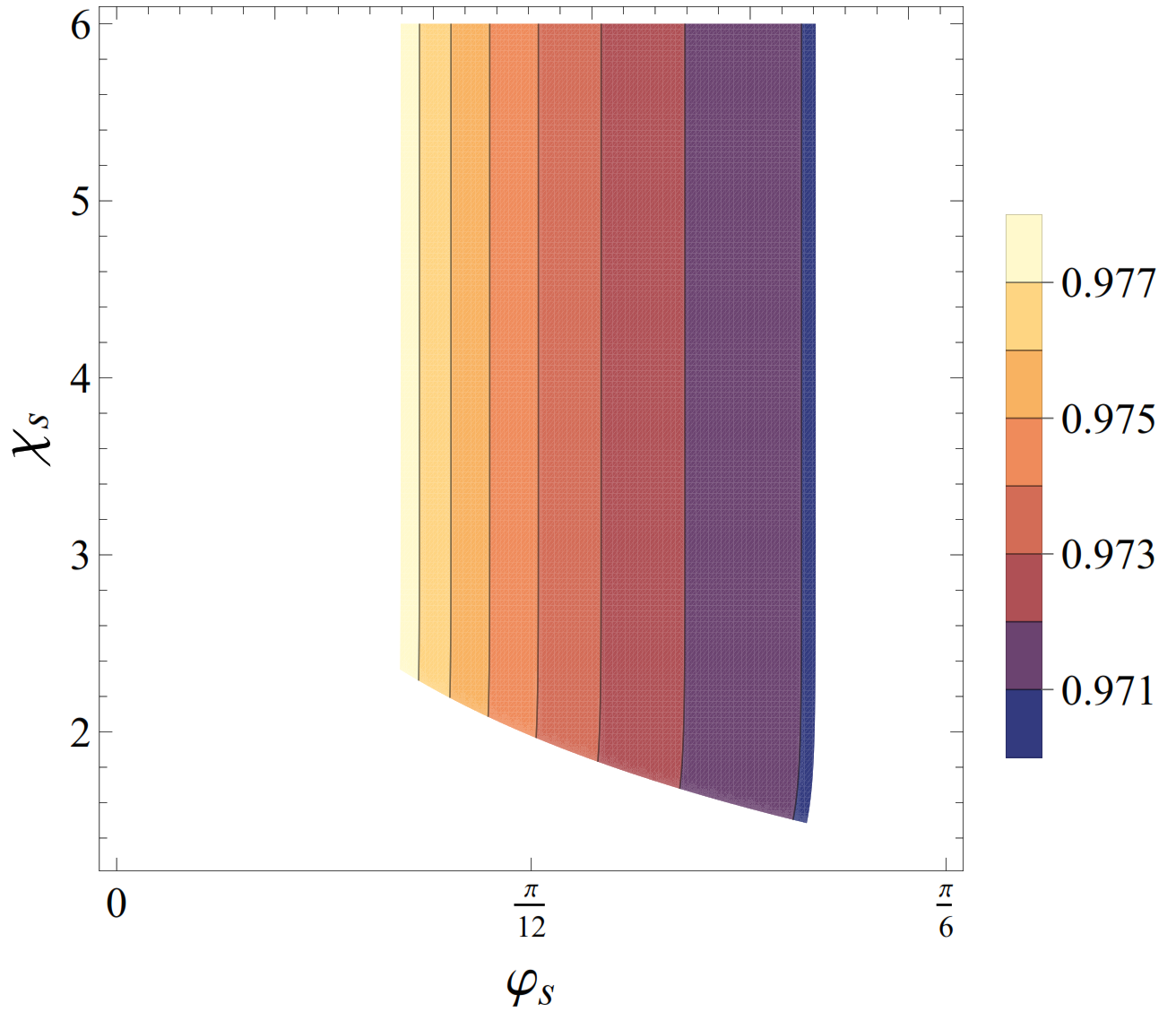}{$\dk\varphi_{{}_S}=0.001$}\\[2mm]
  \panel{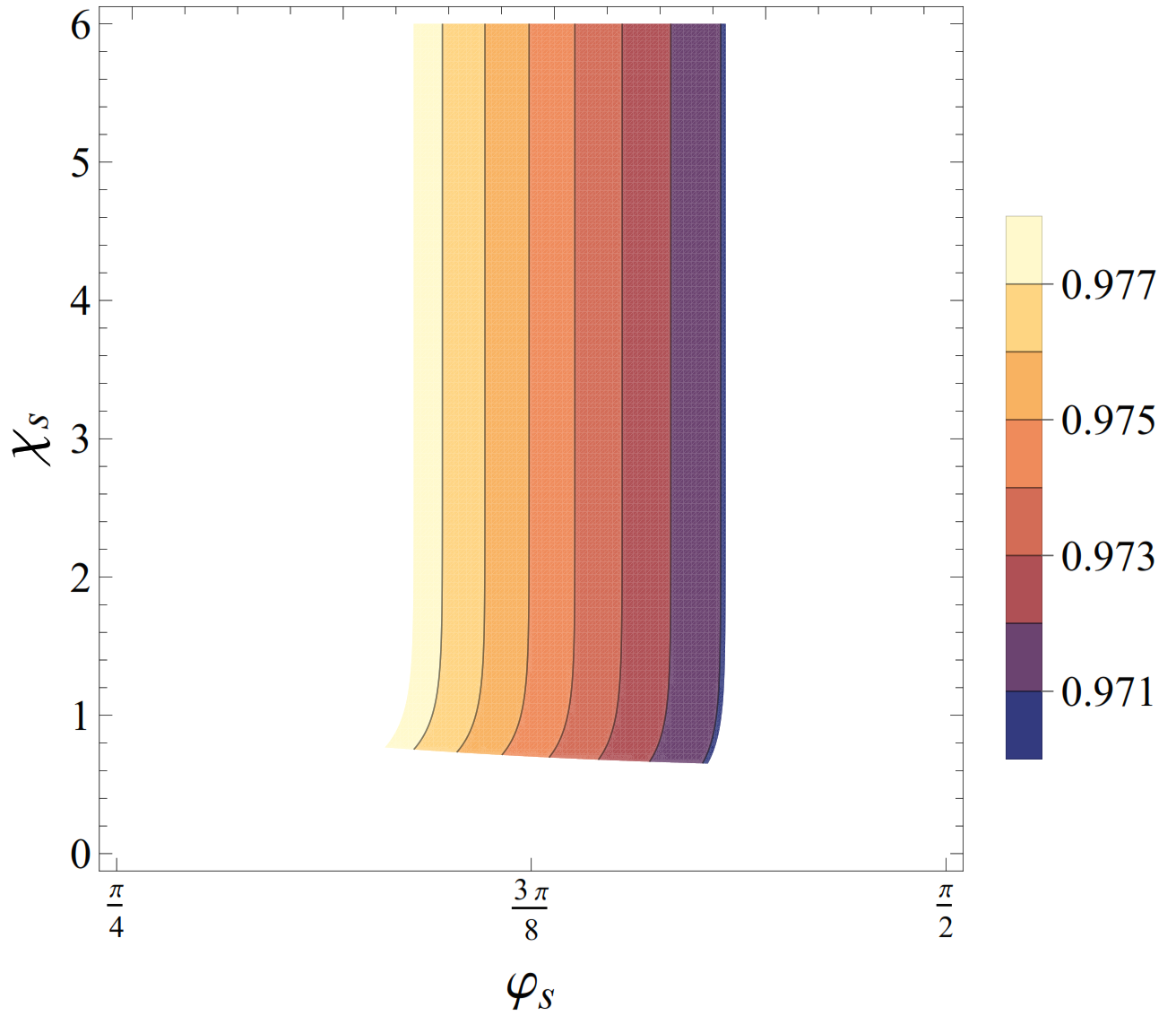}{$\dk\varphi_{{}_S}=0.01$}\hfill\panel{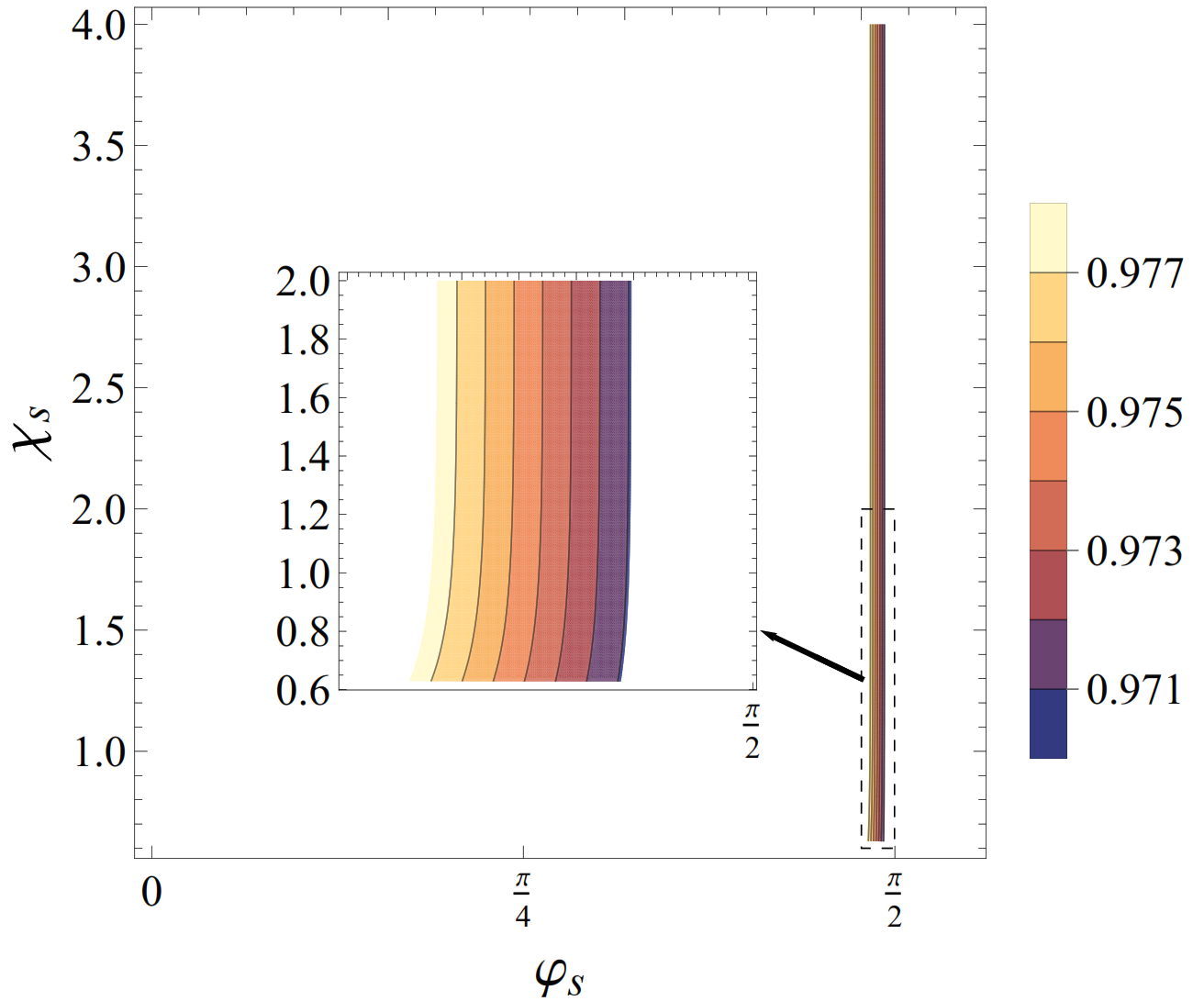}{$\dk\varphi_{{}_S}=0.1$}\\[2mm]
  \panel{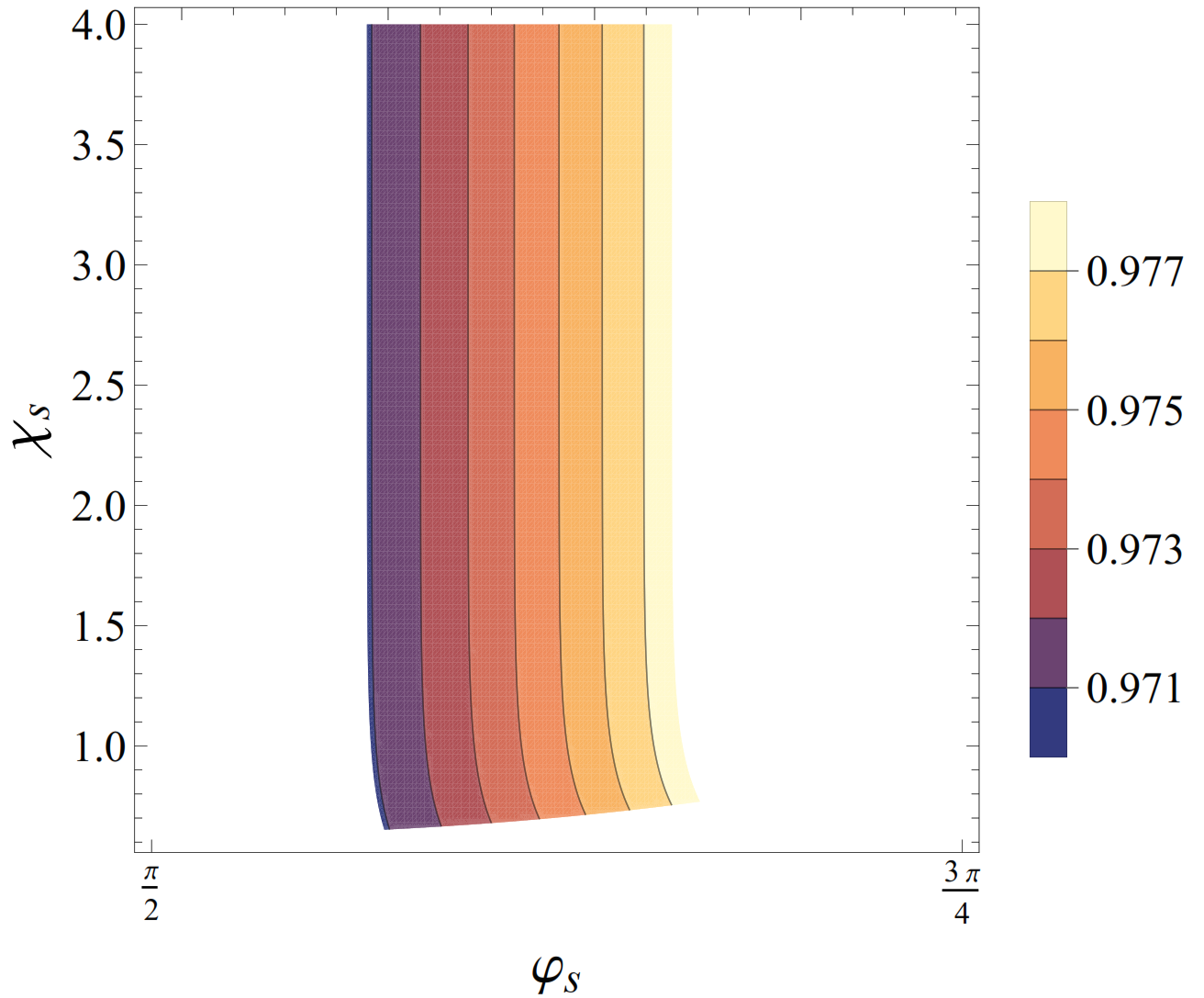}{$\dk\varphi_{{}_S}=-0.01$}\hfill\panel{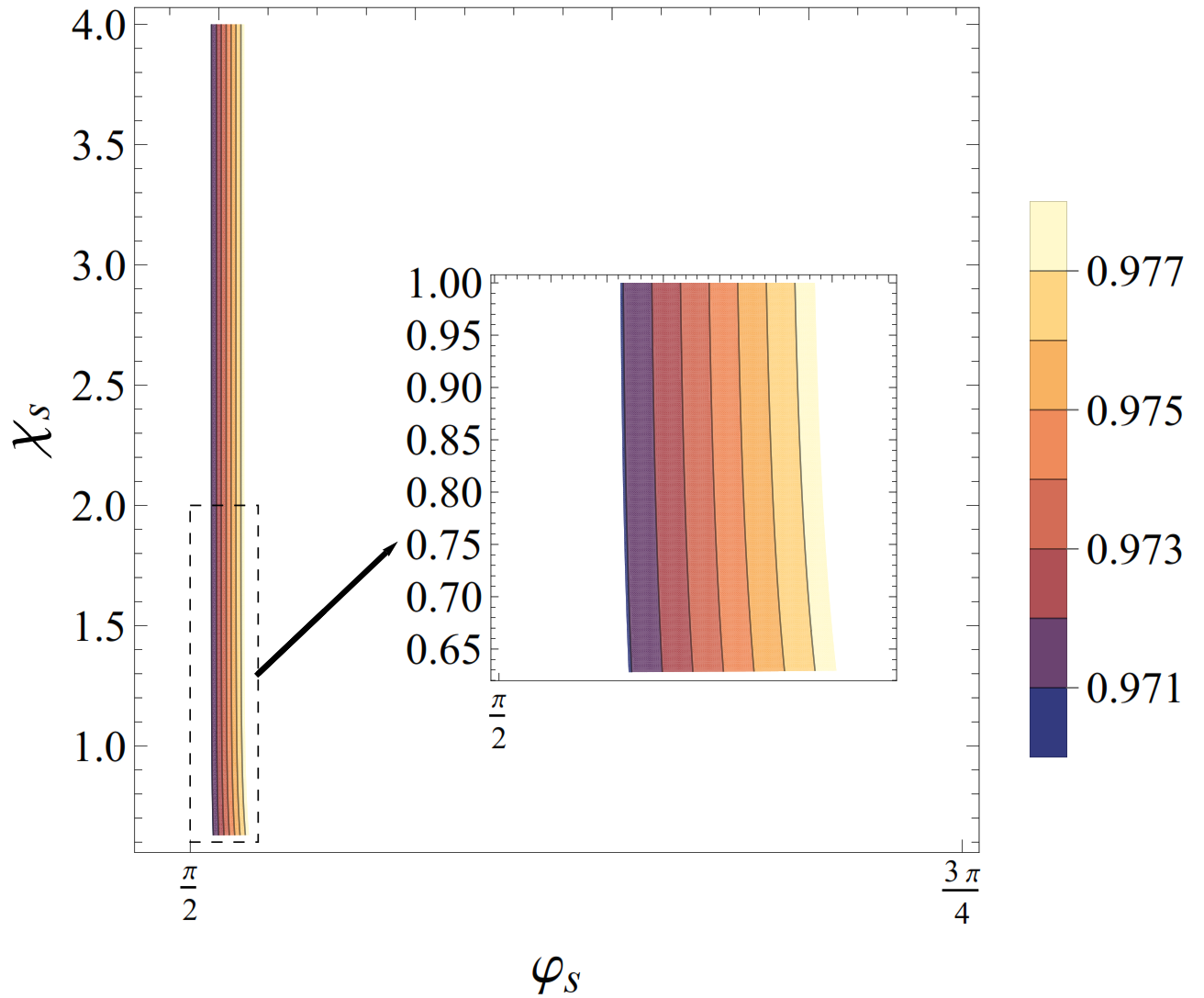}{$\dk\varphi_{{}_S}=-0.1$}
  \caption{QC model: allowed region in $(\varphi_{{}_S},\chi_{{}_S})$ for fixed $\dk\varphi_{{}_S}$.}
  \label{fig:A-QC-phidot}
\end{figure}

\paragraph{Case A: $\dk\varphi_{{}_S}\neq0$, $\ddk\varphi_{{}_S}\neq0$, $k$-independent $\chi_{{}_S}$:} Inverting
eqs.~\eqref{eq:ns-shift-general} and \eqref{eq:alphas-shift-general} gives
\begin{align}
  \dk\varphi_{{}_S} &= \frac{\dns{}}{2}\,\frac{\gamma_{{}_S}}{\sinh2\chi_{{}_S}\,\sin2\varphi_{{}_S}}\,, \label{eq:phidot}\\
  \ddk\varphi_{{}_S} &= \frac{\das{}}{2}\,\frac{\gamma_{{}_S}}{\sinh2\chi_{{}_S}\,\sin2\varphi_{{}_S}}
  + \big(\dns{}\big)^2\,\frac{\Bs\,\gamma_{{}_S}}{2\sinh^2 2\chi_{{}_S}\,\sin^3 2\varphi_{{}_S}}\,. \label{eq:phiddot}
\end{align}
\begin{figure}[h]
  \centering
  \panel{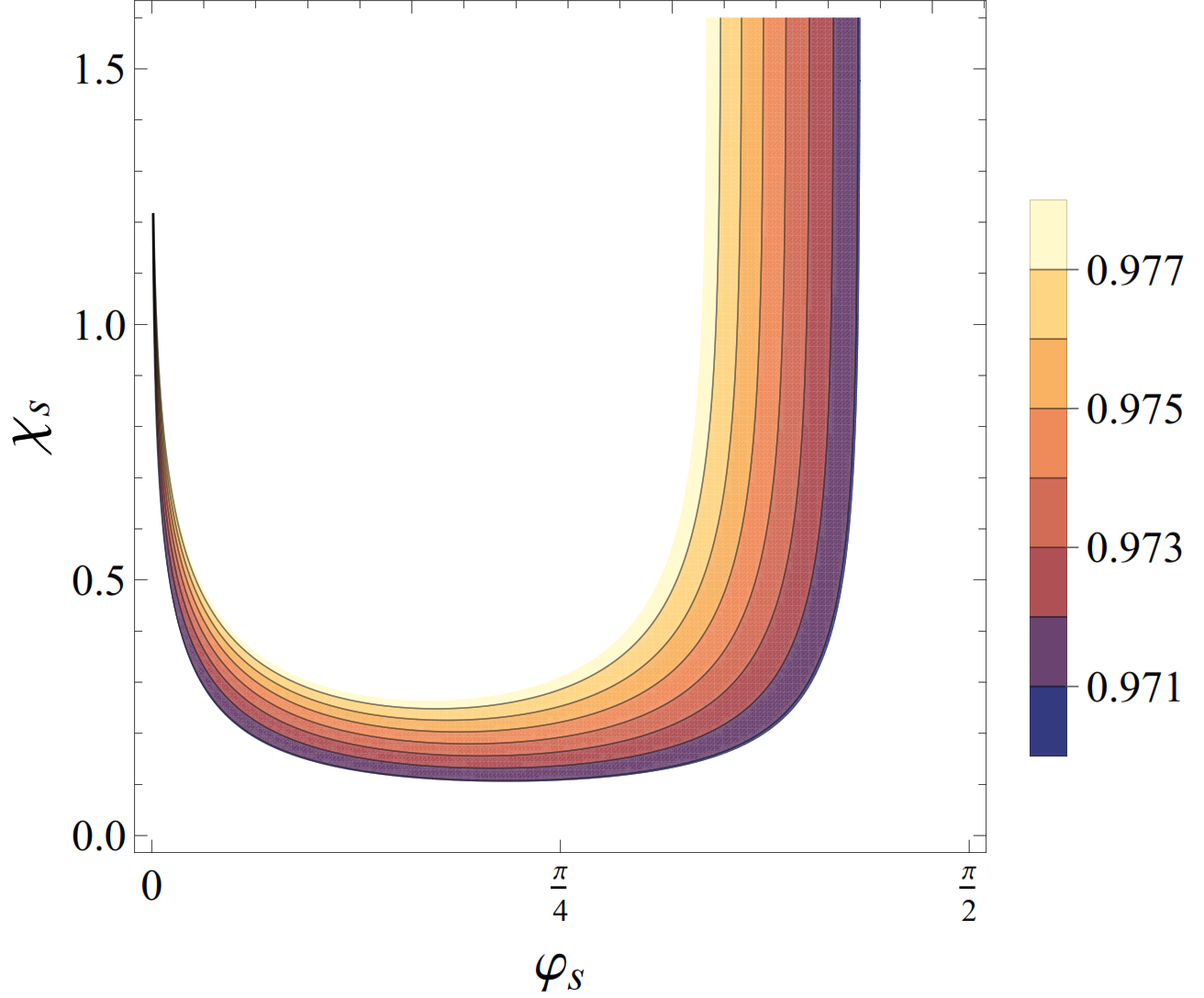}{$\dk\varphi_{{}_S}=0.01$}\hfill\panel{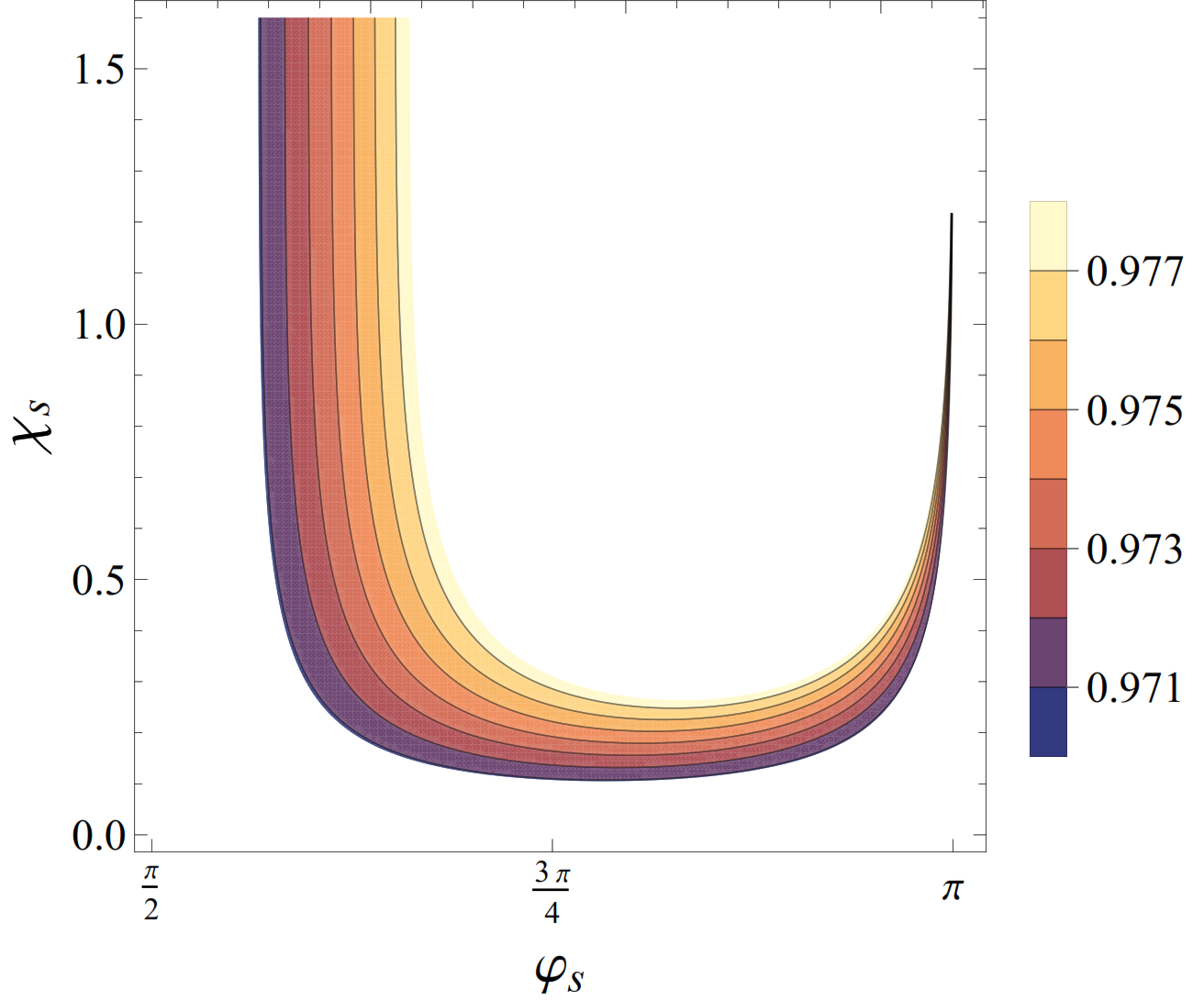}{$\dk\varphi_{{}_S}=-0.01$}
  \caption{Starobinsky model: allowed region in $(\varphi_{{}_S},\chi_{{}_S})$ for fixed $\dk\varphi_{{}_S}$, varying $n_{\rm s}^{\Ob}$ in the 1$\sigma$ region of the ACT DR6 results. For $\dk\varphi_{{}_S}>0$, $0<\varphi_{{}_S}<\frac{\pi}{2}$, whereas if  $\dk\varphi_{{}_S}<0$, $\frac{\pi}{2}<\varphi_{{}_S}<\pi$.}
  \label{fig:A-St-phidot}
\end{figure}
We plot the values of $\dk\varphi_{{}_S}$ as a function of $(\varphi_{{}_S},\chi_{{}_S})$ for the Starobinsky and
QC models in figure~\ref{fig:phidot-mean}, using the ACT DR6 mean value of $n_{\rm s}^{\Ob}$. For $ \dk\varphi_{{}_S}\neq0$, $ \ddk\varphi_{{}_S}\neq0$, the plots of $\ddk\varphi_{{}_S}$ for different values of $\varphi_{{}_S}$ and $\chi_{{}_S}$  are presented in figure \ref{fig:phiddot-mean-Starobinsky} for the Starobinsky model in two intervals $0\leq \chi_{{}_S}\leq 0.1$, and  $0.1\leq \chi_{{}_S}\leq 10$, and in figure \ref{fig:phiddot-mean-QC} for the QC model, assuming $n_{\rm s}^{\Ob}$ and $\alpha_{\rm s}^{\Ob}$ take the ACT DR6 mean values. Changes in the values of $n_{\rm s}^{\Ob}$ and $ \alpha_{\rm s}^{\Ob} $ within the ACT DR6 1$\sigma$ region will not change the overall shape of the plots, but will slightly change the magnitude of $\dk\varphi_{{}_S}$ and $\ddk\varphi_{{}_S}$. It is also useful to present the contour plots in the $(\varphi_{{}_S},\chi_{{}_S})$ plane, letting $n_{\rm s}^{\Ob}$ take all viable values within the $1\sigma$ region of ACT DR6 results for the scalar spectral index. The corresponding plots are shown in
figures~\ref{fig:A-St-phidot} and \ref{fig:A-QC-phidot}, respectively, for the Starobinsky and QC models.

Let us examine Eqs.~\eqref{eq:phidot} and \eqref{eq:phiddot} in several limits.  Both $\dk\varphi_{{}_S}$ and $\ddk\varphi_{{}_S}$ contain
the factor $1/\sin2\varphi_{{}_S}$, so they diverge wherever
$\sin2\varphi_{{}_S}\to0$, i.e.\ at $\varphi_{{}_S}=0,\pi/2,\pi$. This is simply the statement that
$\gamma_{{}_S}$ is stationary with respect to $\varphi_{{}_S}$ at these points
($\partial\gamma_{{}_S}/\partial\varphi_{{}_S}=2\sinh2\chi_{{}_S}\sin2\varphi_{{}_S}$), so that a finite $\dk\varphi_{{}_S}$
cannot generate a shift of $n_{\rm s}$ there.

For $\chi_{{}_S}\to\infty$ at fixed $\sin\varphi_{{}_S}\neq0$ we have
$\gamma_{{}_S}\simeq e^{2\chi_{{}_S}}\sin^2\varphi_{{}_S}$, $\Bs\simeq\gamma_{{}_S}$ and
$\sinh2\chi_{{}_S}\simeq e^{2\chi_{{}_S}}/2$, and the $\chi_{{}_S}$ dependence drops out:
\begin{equation}\label{eq:phi-limit-large}
\begin{split}
  \dk\varphi_{{}_S} &\simeq \frac{\dns{}}{2}\,\tan\varphi_{{}_S}
  \left(1+\frac{e^{-4\chi_{{}_S}}}{\sin^2\varphi_{{}_S}}\right),\\
  \ddk\varphi_{{}_S} &\simeq \frac{\das{}}{2}\,\tan\varphi_{{}_S}
  + \frac{\big(\dns{}\big)^2}{4}\,\tan\varphi_{{}_S}\,\sec^2\varphi_{{}_S}
  + \mathcal{O}\!\left(e^{-4\chi_{{}_S}}\right).
\end{split}
\end{equation}
This is what one obtains directly from $\ln\gamma_{{}_S}\simeq2\chi_{{}_S}+2\ln\sin\varphi_{{}_S}$, namely
$\dns{}=2\cot\varphi_{{}_S}\,\dk\varphi_{{}_S}$ and
$\das{}=2\cot\varphi_{{}_S}\,\ddk\varphi_{{}_S}-2\csc^2\varphi_{{}_S}\,(\dk\varphi_{{}_S})^2$.
At large $\chi_{{}_S}$ the allowed region is therefore independent of $\chi_{{}_S}$: for a given value of
$\dk\varphi_{{}_S}$ it is the vertical band
\begin{equation}
\tan\varphi_{{}_S}\simeq\frac{2\,\dk\varphi_{{}_S}}{\dns{}}\,,\label{varphis-varphisdot}
\end{equation}
whose width reflects the allowed range of $n_{\rm s}^{\Ob}$. Positive (negative) $\dk\varphi_{{}_S}$
selects $\varphi_{{}_S}<\pi/2$ ($\varphi_{{}_S}>\pi/2$). As $|\dk\varphi_{{}_S}|$ increases,
$\varphi_{{}_S}$ gets closer to $\pi/2$. From eq.~\eqref{varphis-varphisdot}, at fixed $\dk\varphi_{{}_S}$ a change $\Delta\dns{}$ of the required shift translates into a band of width
\begin{equation}
\Delta\varphis\simeq\frac{1}{2}\sin2\varphis\,\frac{\Delta\dns{}}{\dns{}}=\frac{2|\dk\varphi_{{}_S}|\,\Delta\dns{}}{(\dns{})^2+4(\dk\varphi_{{}_S})^2}\,.
\end{equation}
In particular, for $|\dk\varphi_{{}_S}|\gg\dns{}/2$ the band shrinks like $\Delta\dns{}/(2|\dk\varphi_{{}_S}|)$, as could be verified in plots of Fig. \eqref{fig:A-QC-phidot}. One should note that for the ACT DR6 mean values,
$\dk\varphi_{{}_S}\simeq0.0038\tan\varphi_{{}_S}$.

The above must be combined with the bound
\eqref{eq:r0.038}. In the Starobinsky model the neighbourhood of $\varphi_{{}_S}=0,\pi$ is retained
for $\chi_{{}_S}<1.2168$ (at $\varphi_{{}_S}=0,\pi$ exactly). For these values of $\chis$, as $\varphis\to 0$, i.e.\ for $v\equiv e^{4\chi_{{}_S}}\tan^2\varphi_{{}_S}\ll1$, eqs.~\eqref{eq:phidot} and \eqref{eq:phiddot} give
\begin{equation}\label{eq:phi-limit-small}
\dk\varphi_{{}_S}\simeq\frac{\dns{}}{2\left(e^{4\chis}-1\right)\varphis}\,,\qquad
\ddk\varphi_{{}_S}\simeq\frac{\das{}}{2\left(e^{4\chis}-1\right)\varphis}-\frac{\big(\dns{}\big)^2}{4\left(e^{4\chis}-1\right)^2\varphis^3}\,,
\end{equation}
\begin{figure}[t]
  \centering
  \panel{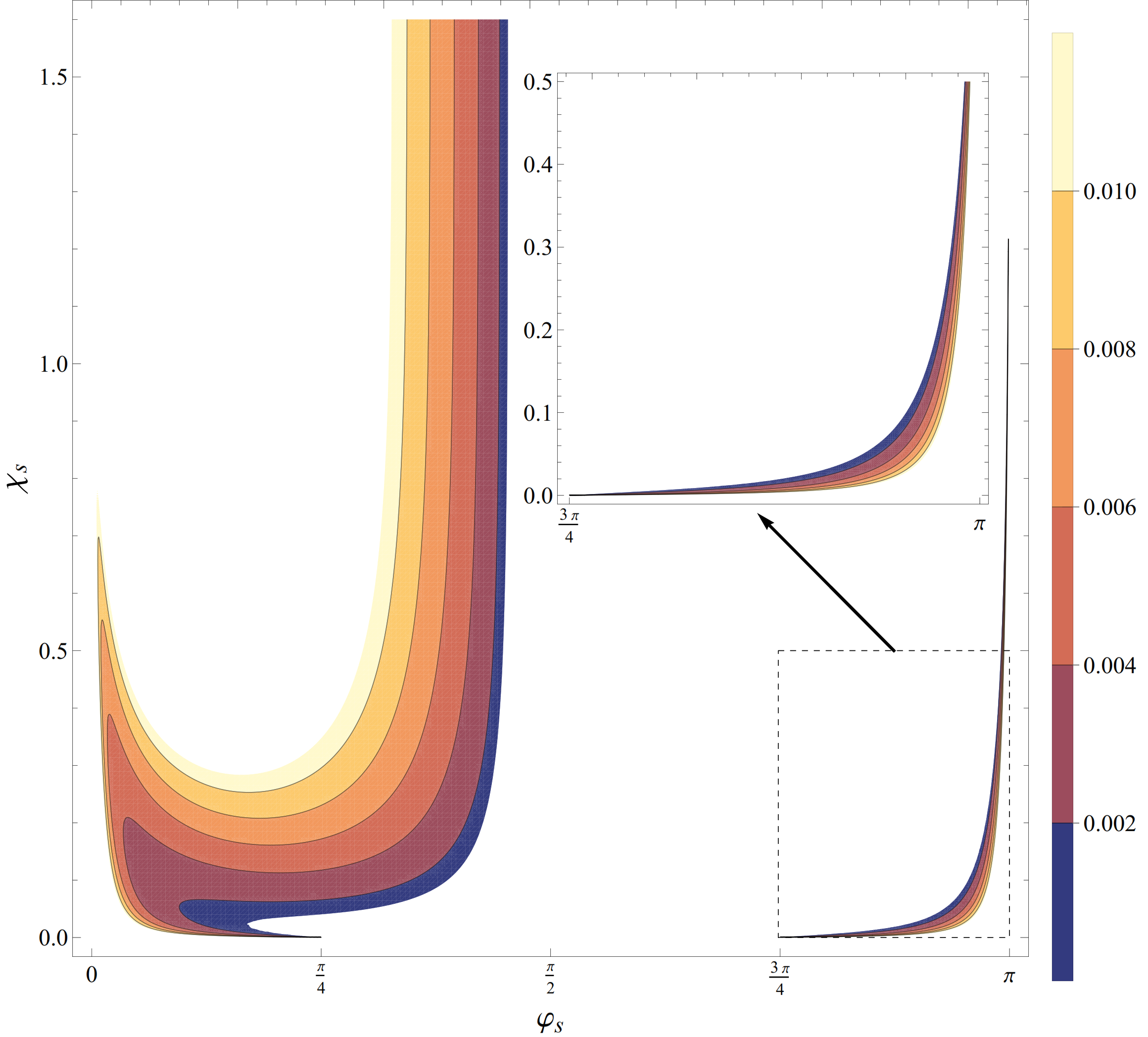}{$\ddk\varphi_{{}_S}=0.01$}\hfill\panel{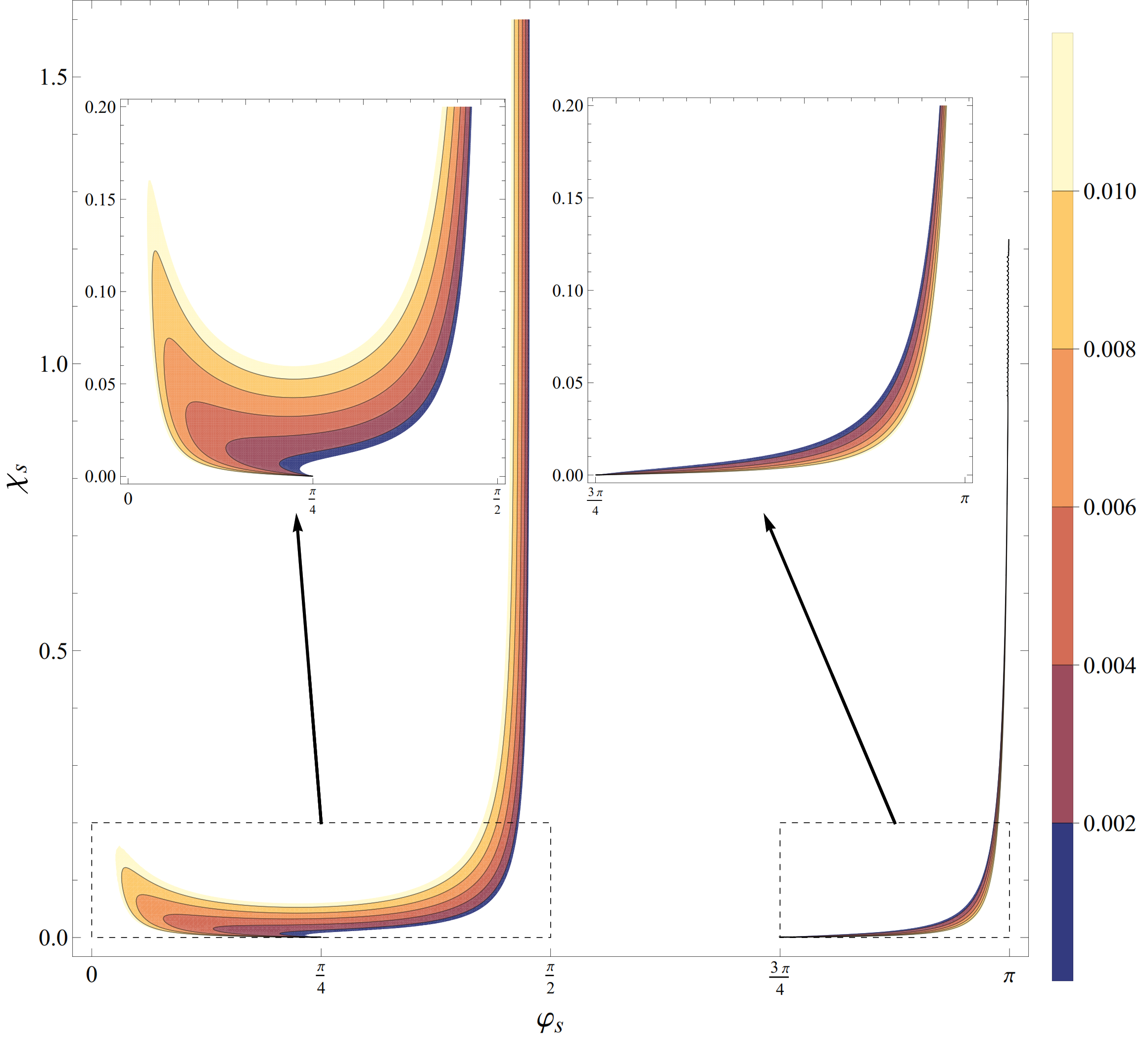}{$\ddk\varphi_{{}_S}=0.05$}\\[2mm]
  \panel{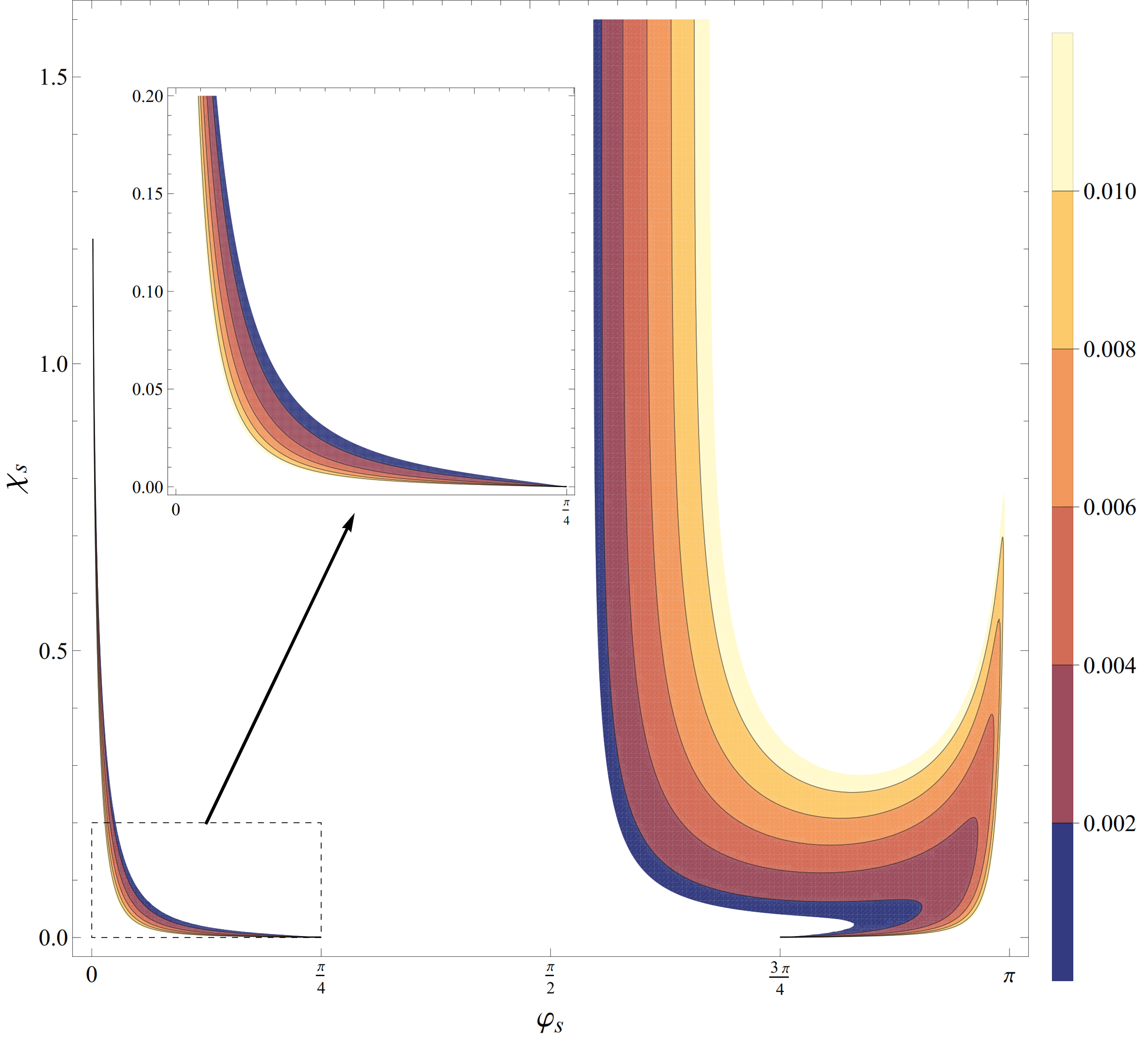}{$\ddk\varphi_{{}_S}=-0.01$}\hfill\panel{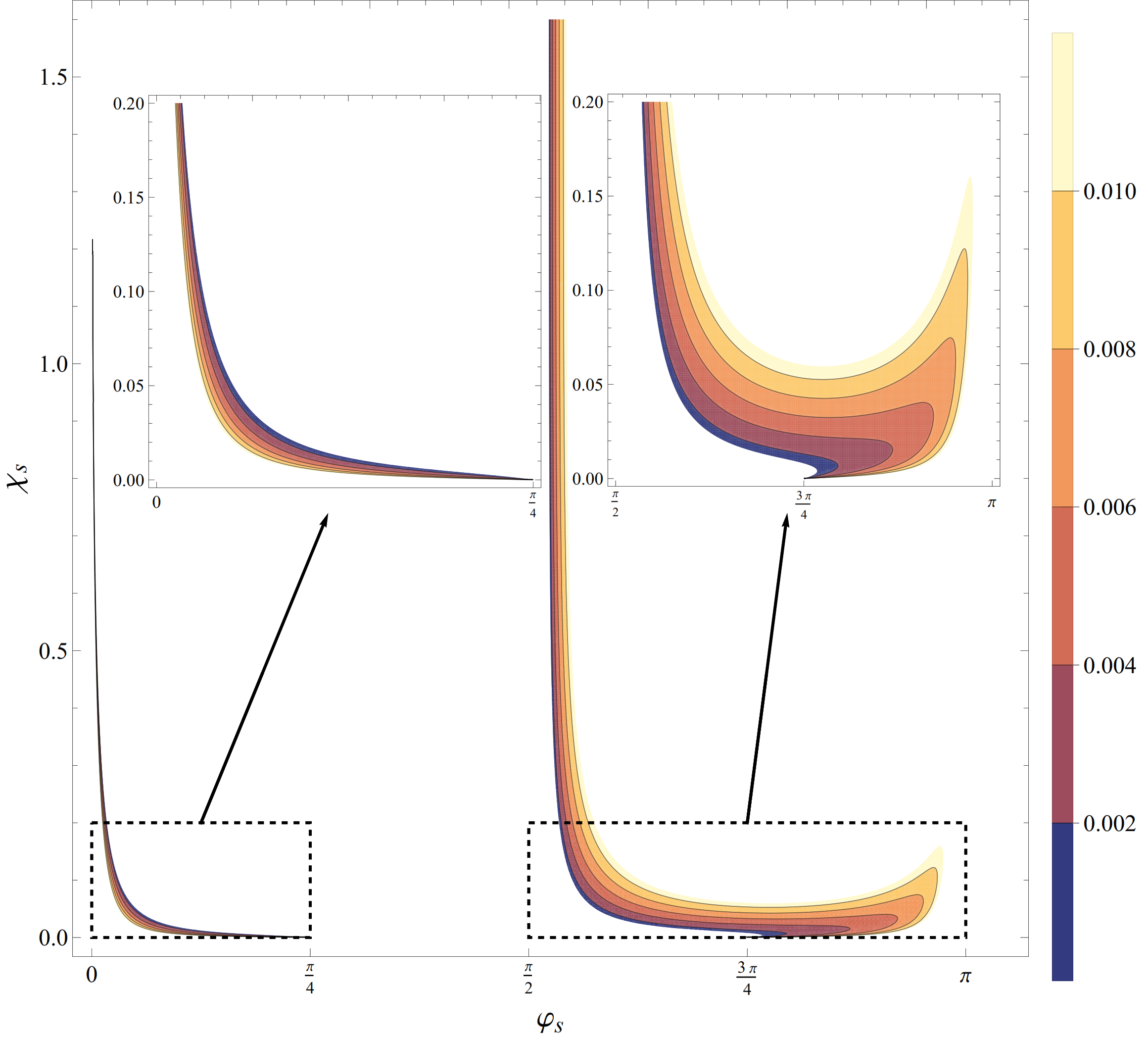}{$\ddk\varphi_{{}_S}=-0.05$}
  \caption{Starobinsky model: allowed region in $(\varphi_{{}_S},\chi_{{}_S})$ for fixed $\ddk\varphi_{{}_S}$.}
  \label{fig:A-St-phiddot}
\end{figure}

\begin{figure}[htbp]
  \centering
  \panel{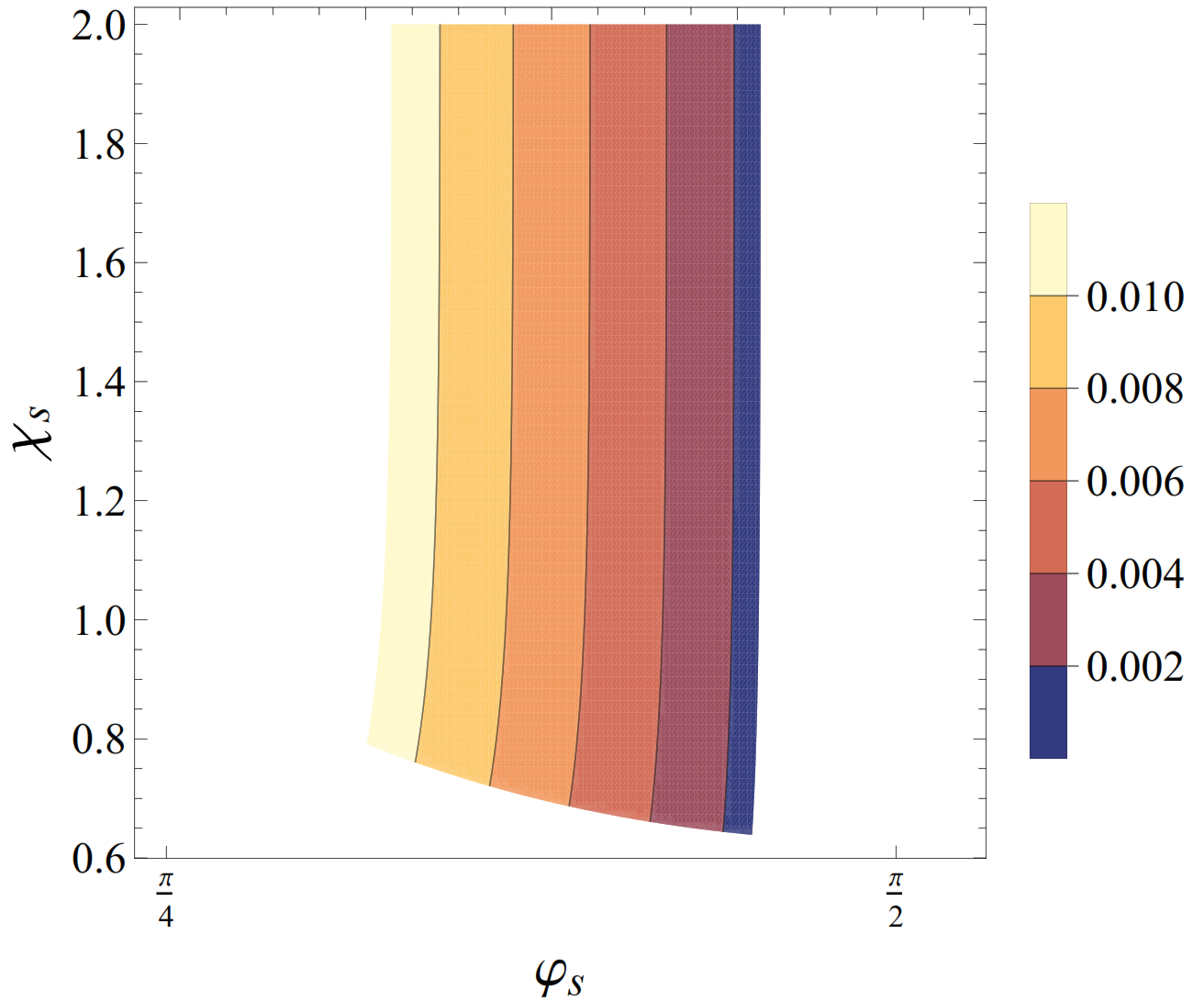}{$\ddk\varphi_{{}_S}=0.01$}\hfill\panel{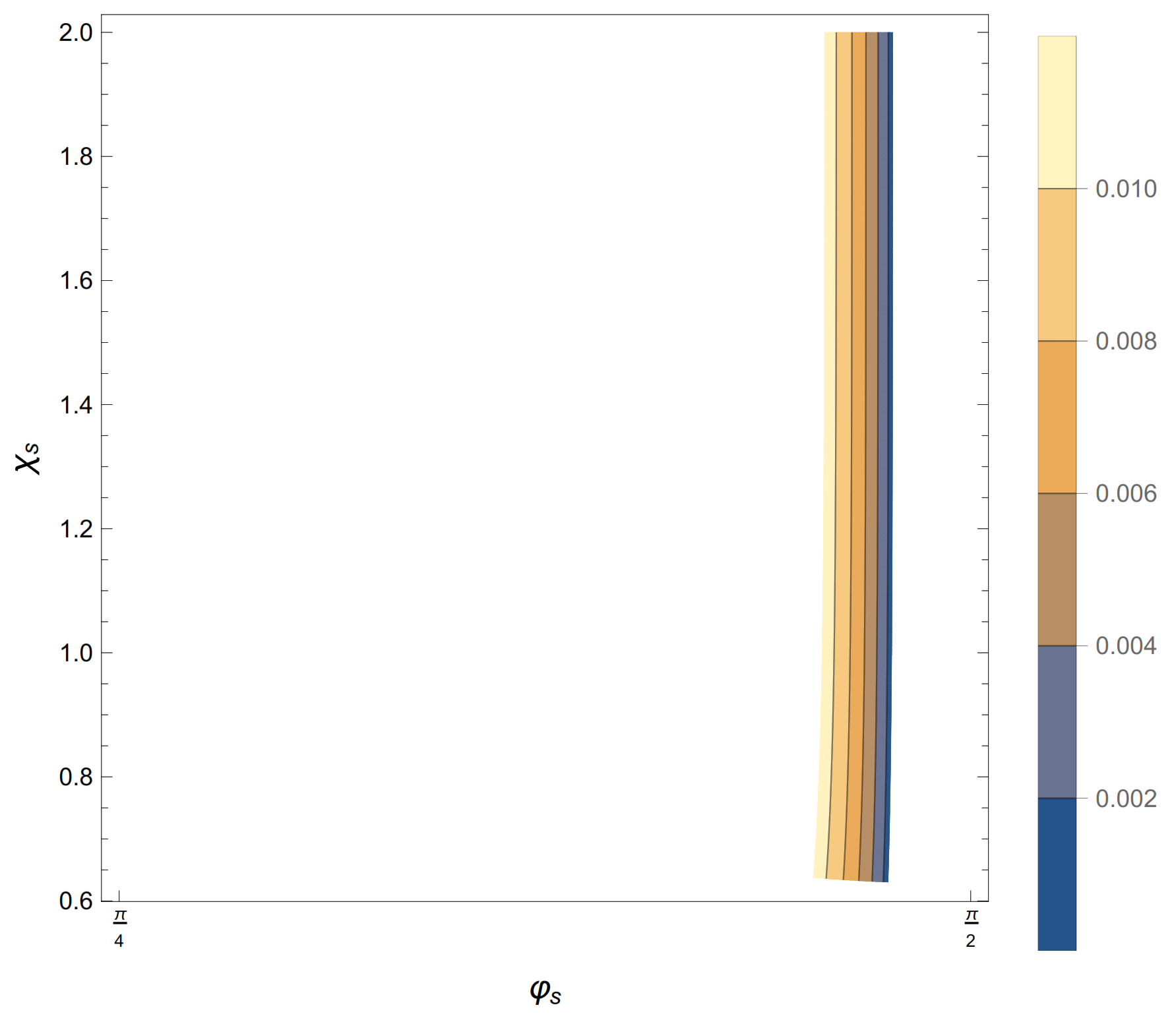}{$\ddk\varphi_{{}_S}=0.05$}\\[2mm]
  \panel{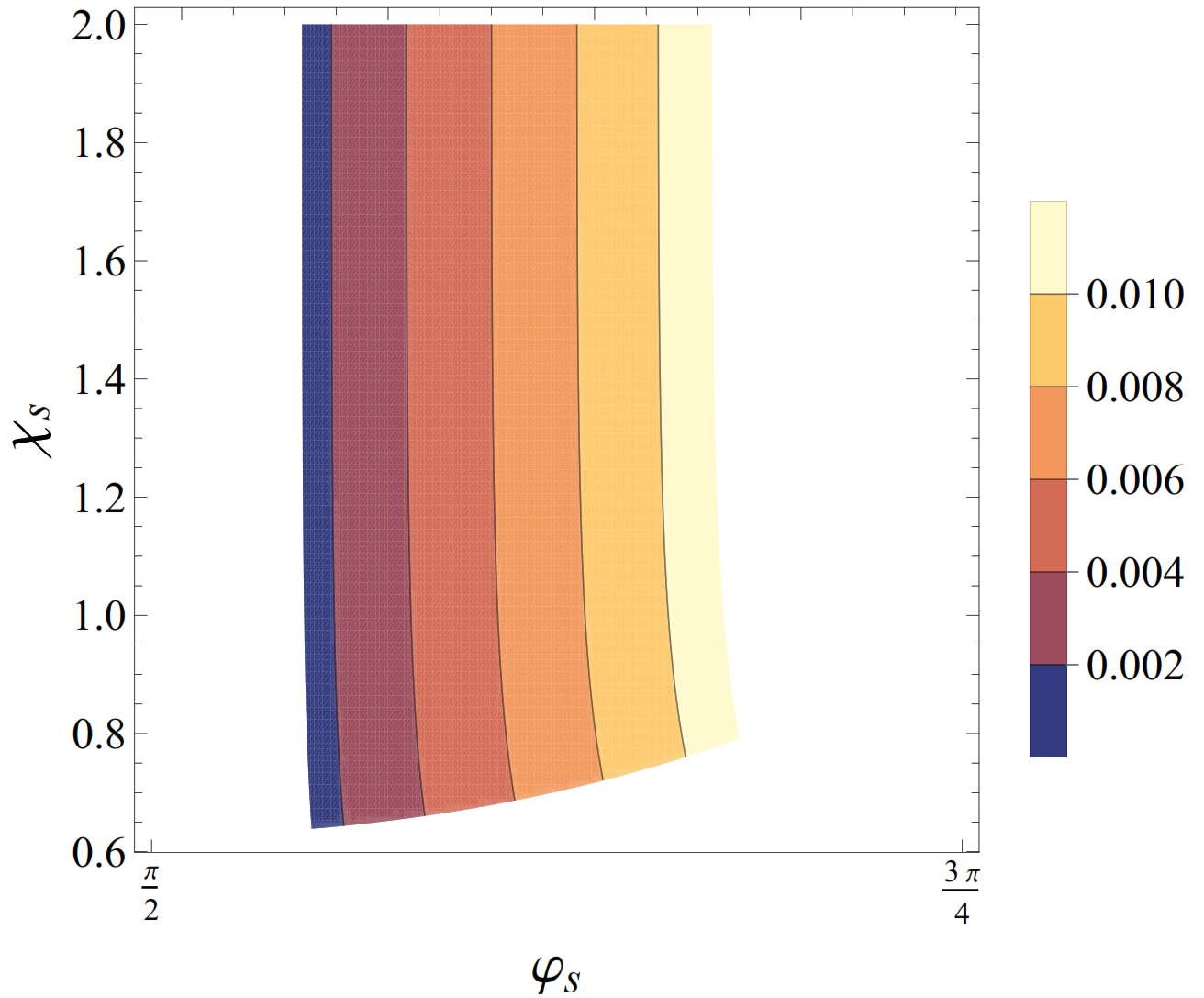}{$\ddk\varphi_{{}_S}=-0.01$}\hfill\panel{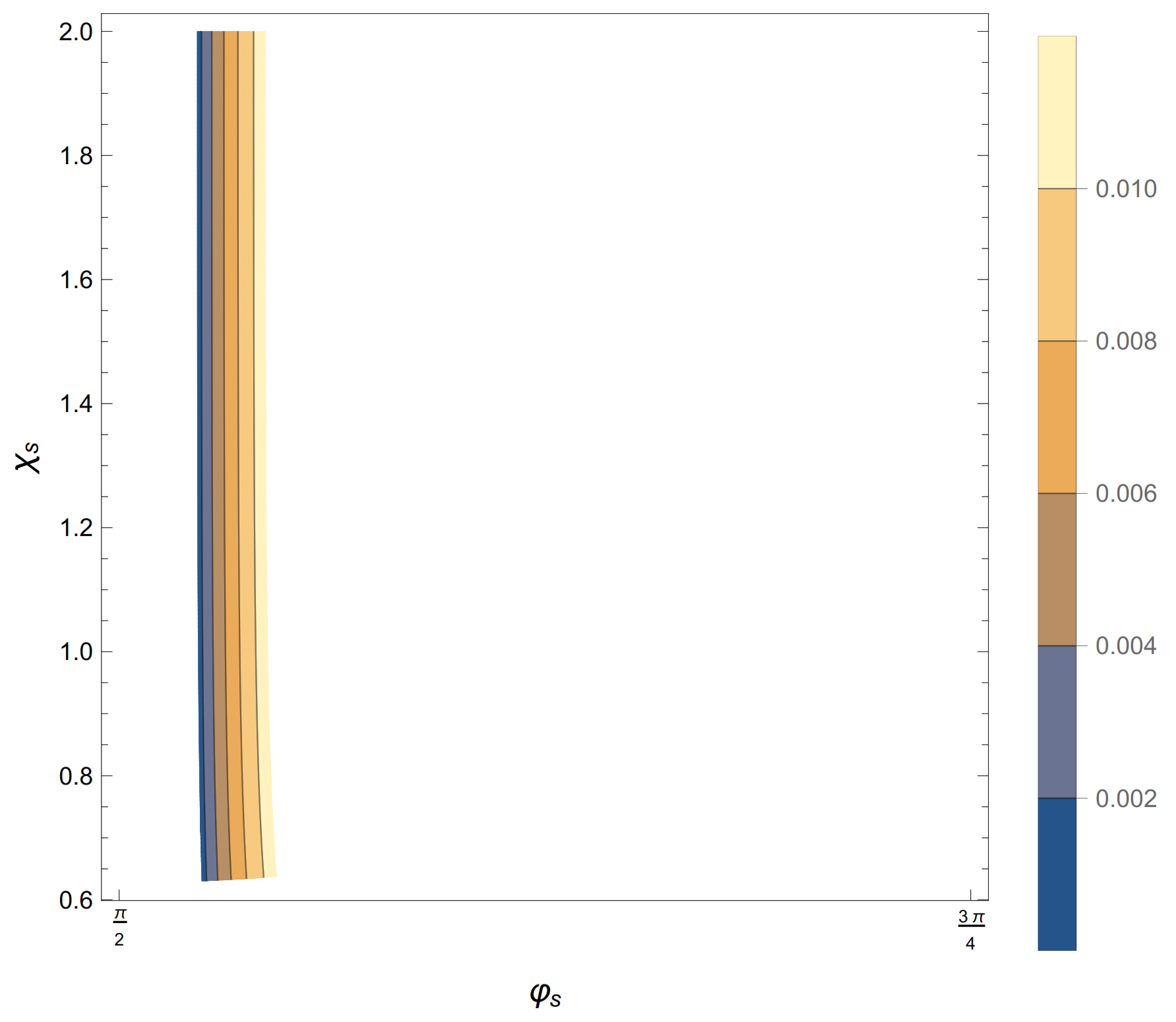}{$\ddk\varphi_{{}_S}=-0.05$}
  \caption{QC model: allowed region in $(\varphi_{{}_S},\chi_{{}_S})$ for fixed $\ddk\varphi_{{}_S}$.}
  \label{fig:A-QC-phiddot}
\end{figure}
so that $\dk\varphi_{{}_S}$ and $\ddk\varphi_{{}_S}$ respectively diverge like $\frac{1}{\varphis}$ and $-\frac{1}{\varphis^3}$ (or, close to $\varphis=\pi$, $\dk\varphi_{{}_S}$ and $\ddk\varphi_{{}_S}$ change sign and diverge like $\frac{1}{\pi-\varphis}$ and $\frac{1}{(\pi-\varphis)^3}$). For larger values of $\chis$, along the borderline $\sin^2\varphis\simeq 0.08772\,e^{-2 \chis}$, on which the corrections of relative order $e^{-4\chis}/\sin^2\varphis\simeq11.4\,e^{-2\chis}$ in eq.~\eqref{eq:phi-limit-large} are small for $\chis\gtrsim2.5$,
\begin{equation}\label{eq:phi-border-St}
\begin{split}
  \dk\varphi_{{}_S} &\simeq \frac{\dns{}}{2}\,\tan\varphi_{{}_S}\simeq 0.14808~\dns{}~e^{-\chis}\,,\\
  \ddk\varphi_{{}_S} &\simeq \frac{\das{}}{2}\,\tan\varphi_{{}_S}
  + \frac{\big(\dns{}\big)^2}{4}\,\tan\varphi_{{}_S}\,\sec^2\varphi_{{}_S}\simeq  \left( \das{}\,
  + \frac{\big(\dns{}\big)^2}{2}\right) 0.14808\,e^{-\chis}\,.
\end{split}
\end{equation}
As one reaches $\chi_{{}_{S,3}}$ given in eq.~\eqref{chis3-phis3-star}, $\varphis$ becomes equal to $\varphi_{{}_{S,3}}$, and for $\chis\geq\chi_{{}_{S,3}}$ the lower edge stays at $\varphis\simeq\varphi_{{}_{S,\infty}}\simeq\varphi_{{}_{S,3}}$, eq.~\eqref{phis-infty}, along which
\begin{equation}\label{eq:phi-chi3-St}
\begin{split}
  \dk\varphi_{{}_S} &\simeq 7.88\times 10^{-4}\,\dns{}\,,\\
  \ddk\varphi_{{}_S} &\simeq 3.94\times 10^{-4} \left(2 \das{}
  + \big(\dns{}\big)^2\right)\,.
\end{split}
\end{equation}
For the ACT DR6 mean values these are $\dk\varphi_{{}_S}\simeq6.0\times10^{-6}$ and $\ddk\varphi_{{}_S}\simeq5.3\times10^{-6}$. Hence, even though $\dk\varphi_{{}_S}$ and $\ddk\varphi_{{}_S}$ are quite small,
they are not zero. Of course along this edge, $\Lambda^{\rm max}_{{}_S}\gtrsim H$.

In the QC model also $\gamma_{{}_S}>R\simeq3.5$, so $\varphi_{{}_S}=0,\pi$ are excluded for all
$\chi_{{}_S}$ and eq.~\eqref{eq:phi-limit-small} is not relevant. The allowed region for $\chis\gtrsim 1$, lies at
$\sin^2\varphi_{{}_S}\gtrsim3.5\,e^{-2\chi_{{}_S}}$, where $v\gg1$ and the large-$\chi_{{}_S}$ result
\eqref{eq:phi-limit-large} holds; one therefore cannot approach $\varphi_{{}_S}=0,\pi$ closely enough
to see the blow-up of $\dk\varphi_{{}_S}$ and $\ddk\varphi_{{}_S}$.

As $\chi_{{}_S}\to\infty$, the lower edge of
For $1\lesssim\chi_{{}_S}<\chi_{{}_{S,3}}$ the lower edge of
the allowed region is $\varphi_{{}_S}\simeq1.87\,e^{-\chi_{{}_S}}$, and along this edge
$\dk\varphi_{{}_S}\simeq0.0071\,e^{-\chi_{{}_S}}$ and $\ddk\varphi_{{}_S}\simeq0.0064\,e^{-\chi_{{}_S}}$ for the ACT DR6
mean values, i.e.\ both decay like $e^{-\chi_{{}_S}}$, since $\tan\varphi_{{}_S}\simeq\varphi_{{}_S}$ there. For $\chi_{{}_S}\geq\chi_{{}_{S,3}}$ the edge is fixed by the EFT condition, Eq. \eqref{EFT-Validity}, at $\varphi_{{}_S}\simeq\varphi_{{}_{S,\infty}}\simeq0.0423$, where $\dk\varphi_{{}_S}\simeq2.12\times10^{-2}\,\dns{}\simeq1.6\times10^{-4}$ and $\ddk\varphi_{{}_S}\simeq1.06\times10^{-2}\big(2\das{}+(\dns{})^2\big)\simeq1.4\times10^{-4}$. This
edge gives the smallest required $|\dk\varphi_{{}_S}|$ and $|\ddk\varphi_{{}_S}|$ at a given $\chi_{{}_S}$; at
fixed $\varphi_{{}_S}$ they tend to the $\chi_{{}_S}$-independent values in eq.~\eqref{eq:phi-limit-large}.
\begin{figure}[t]
  \centering
  \panel{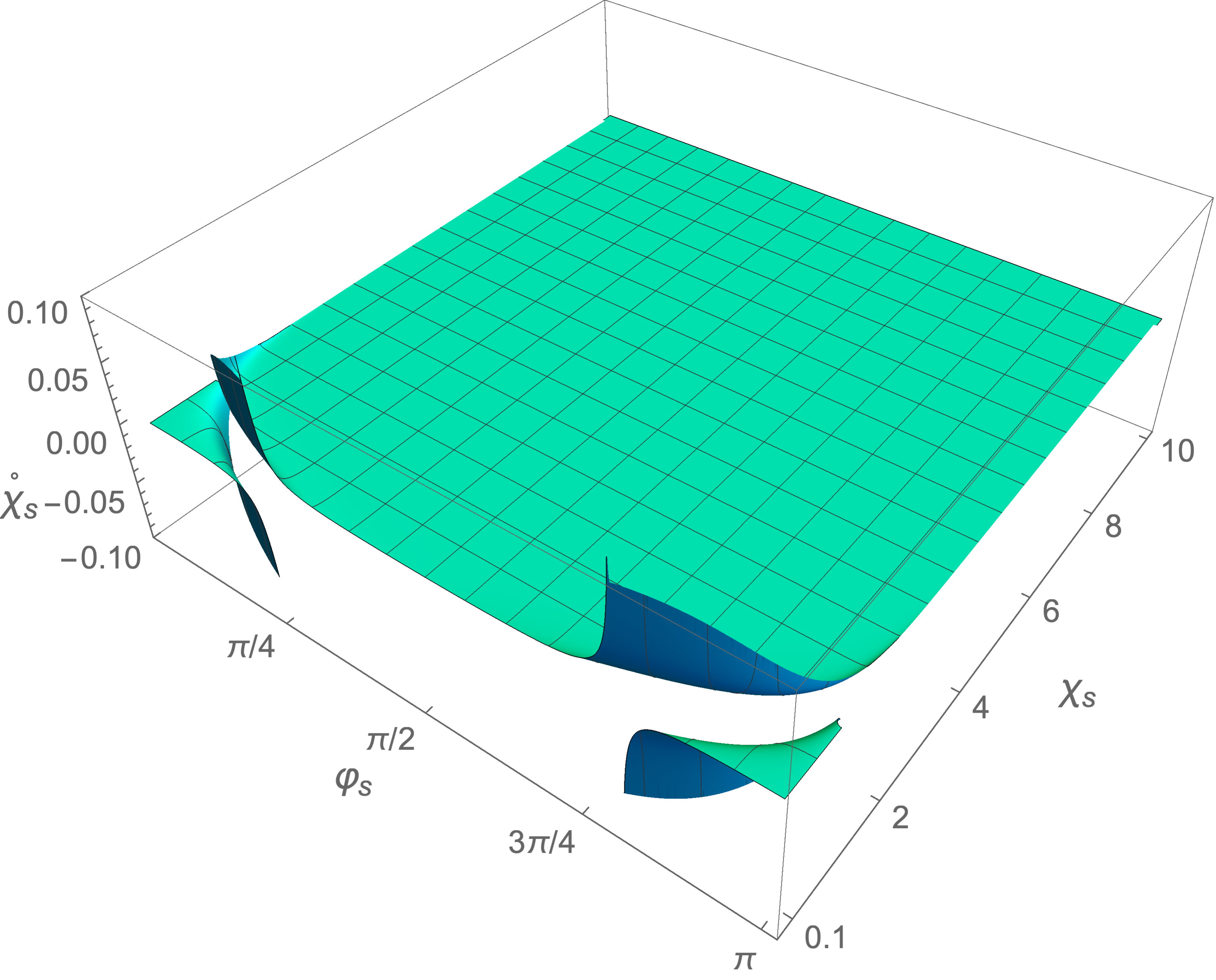}{Starobinsky}\hfill\panel{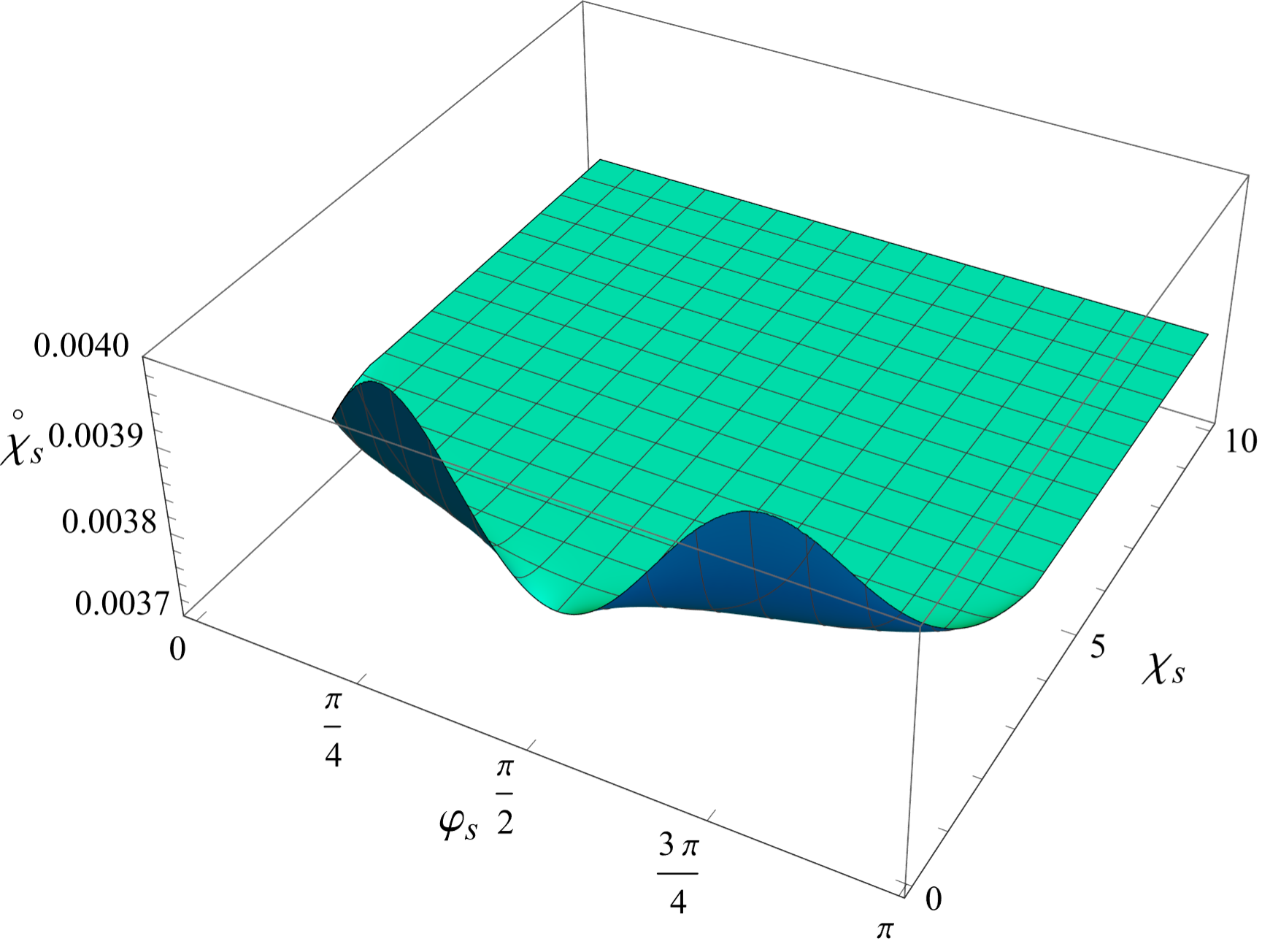}{QC}
  \caption{Plot of $\dk\chi_{{}_S}$ as a function of $(\varphi_{{}_S},\chi_{{}_S})$, when $\varphi_{{}_S}$ is scale-independent. In this plot, we have assumed that $n_{\rm s}$ has taken
  the mean value of $n_{\rm s}^{\Ob}$ from ACT DR6.  The domains are snipped from requiring $r<r_{{}_{\rm max}}$, Eq. \eqref{eq:r0.038}, and the validity of the EFT, Eq. \eqref{EFT-Validity}.}
  \label{fig:chidot-mean}
\end{figure}
\begin{figure}[t]
  \centering
  \panel{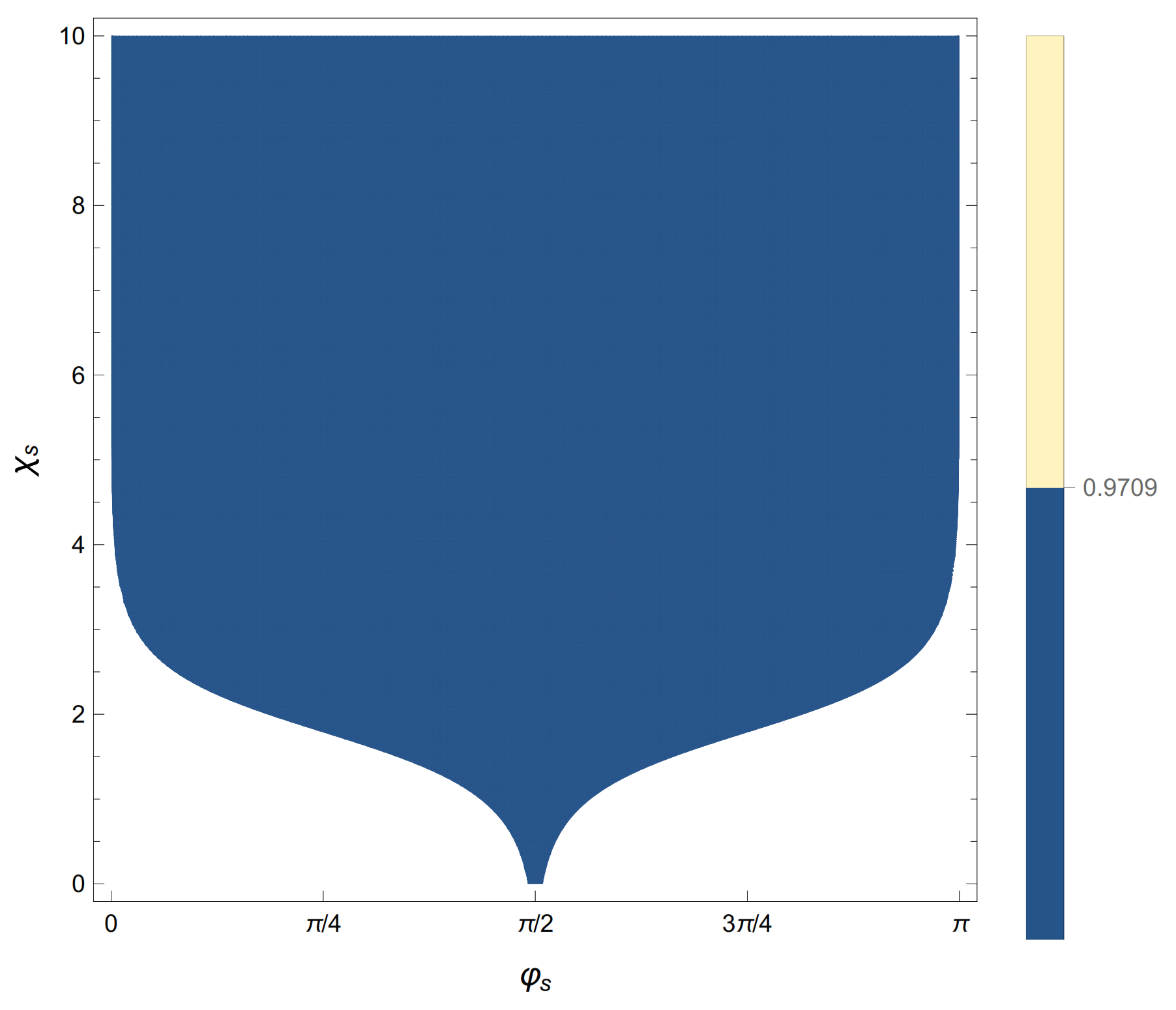}{$\dk\chi_{{}_S}=0.00212$}\hfill\panel{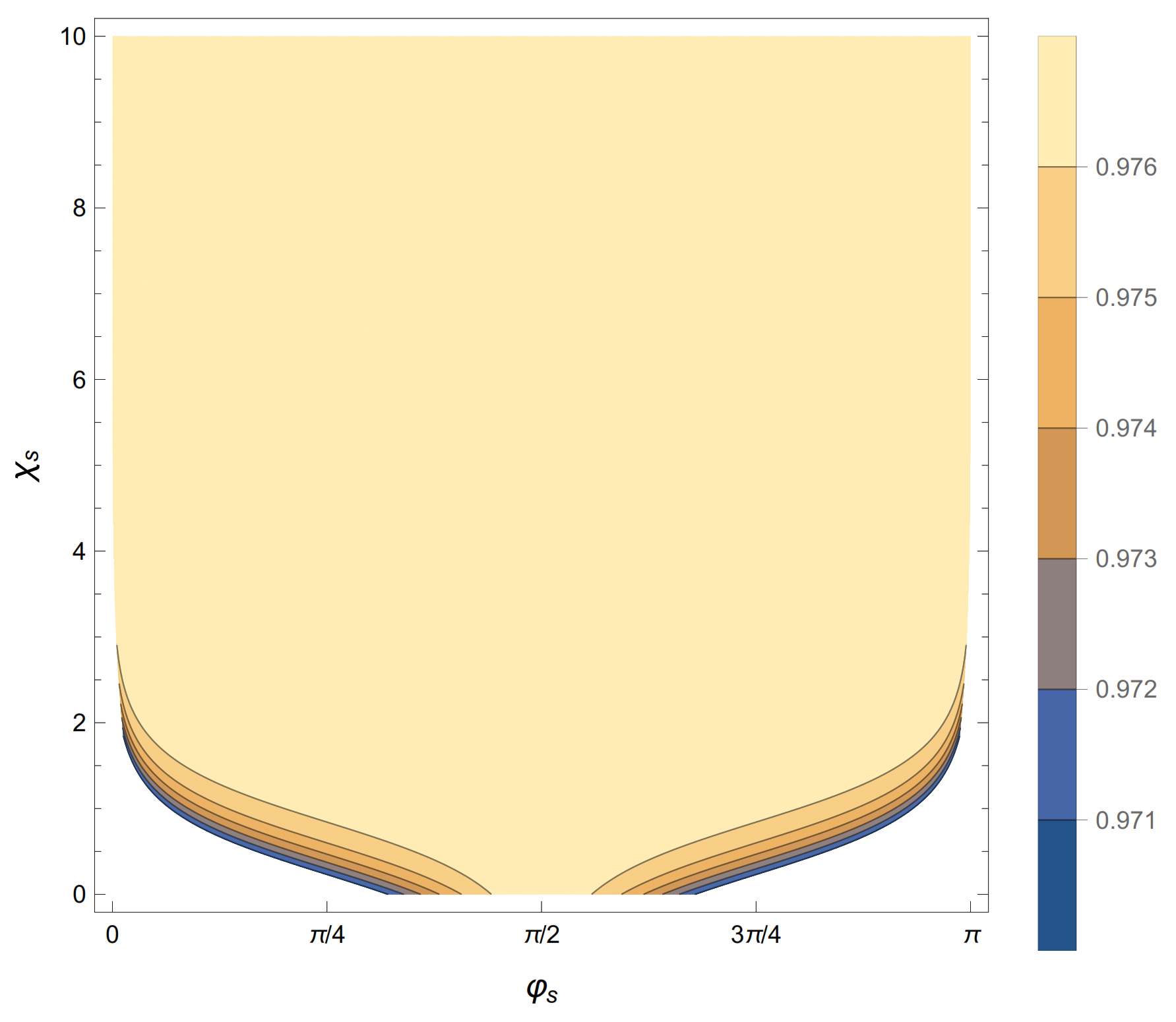}{$\dk\chi_{{}_S}=0.005$}\\[2mm]
  \panel{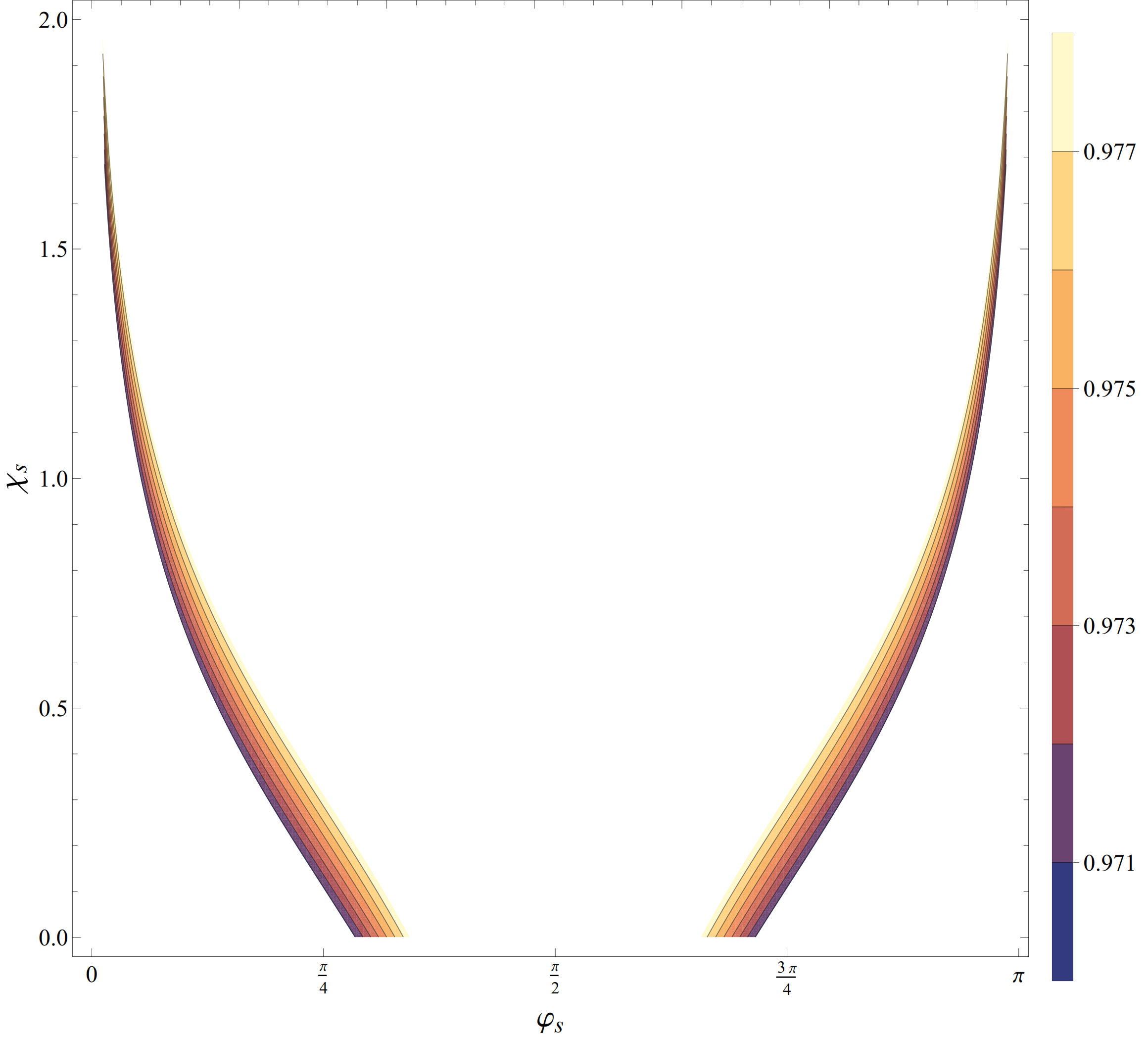}{$\dk\chi_{{}_S}=0.01$}\hfill\panel{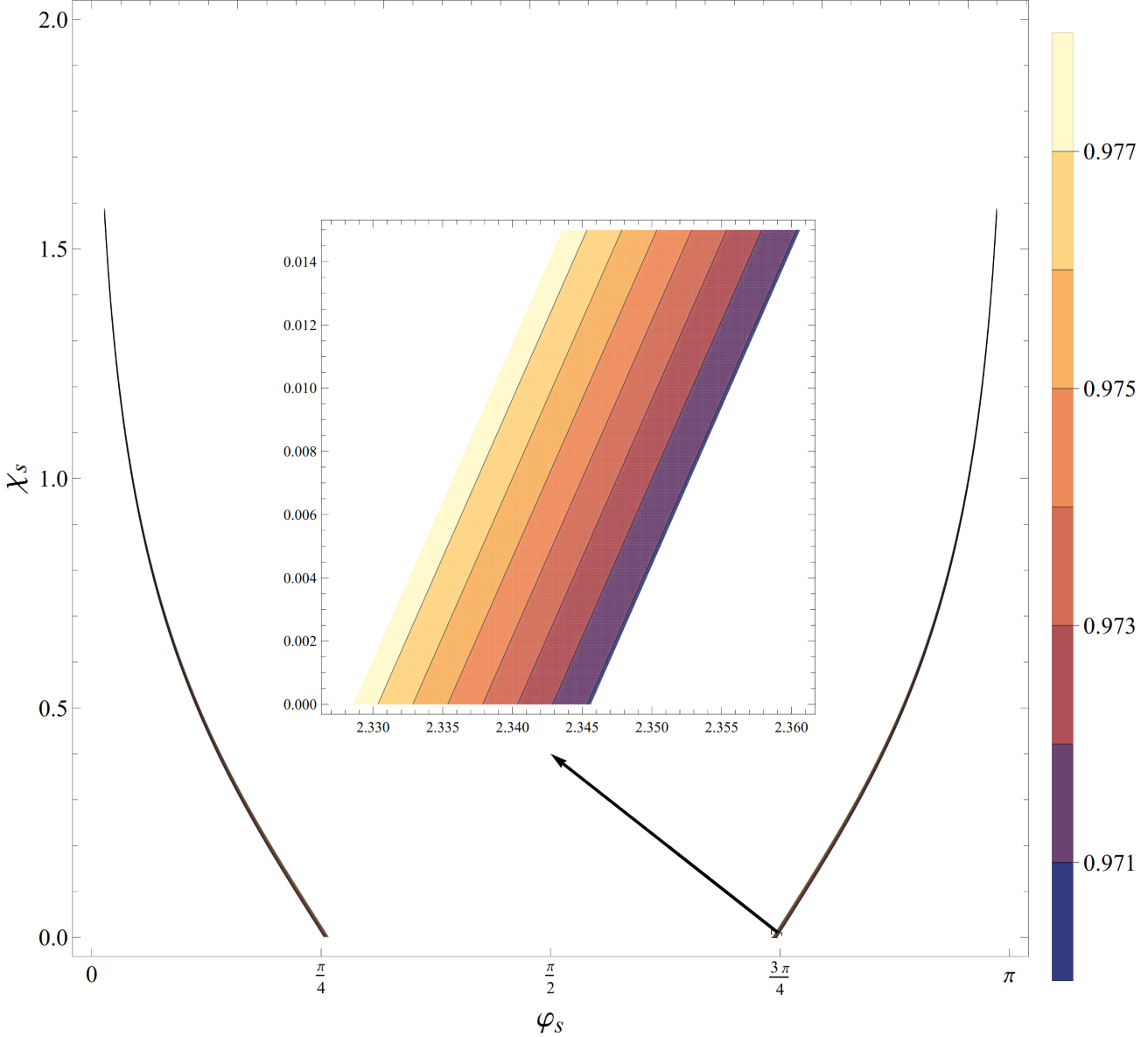}{$\dk\chi_{{}_S}=0.1$}\\[2mm]
  \caption{Starobinsky model: allowed region in $(\varphi_{{}_S},\chi_{{}_S})$ for fixed $\dk\chi_{{}_S}>0$.}
  \label{fig:A-St-chidot}
\end{figure}

\begin{figure}[t]
  \centering
  \panel{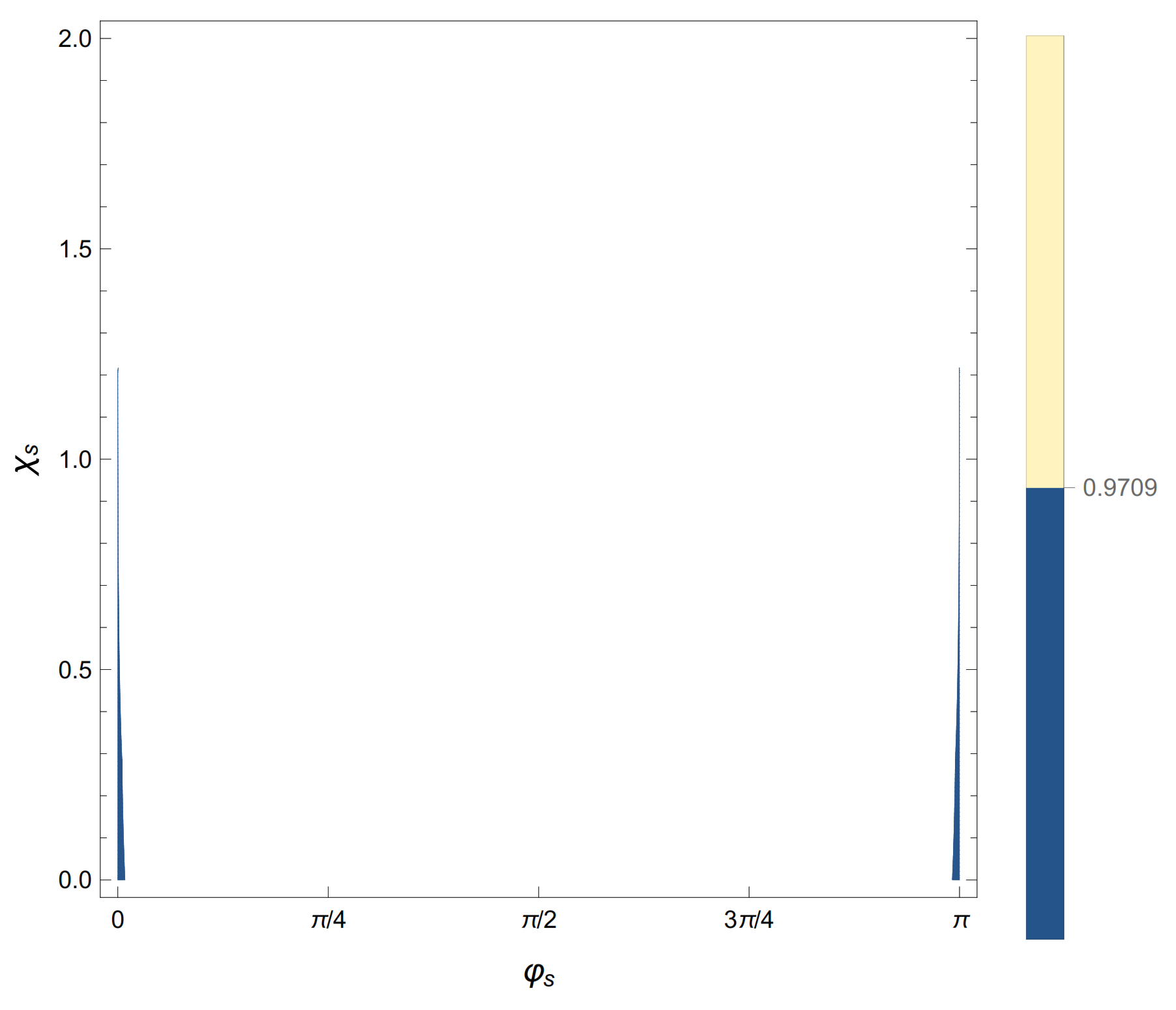}{$\dk\chi_{{}_S}=-0.00212$}\hfill\panel{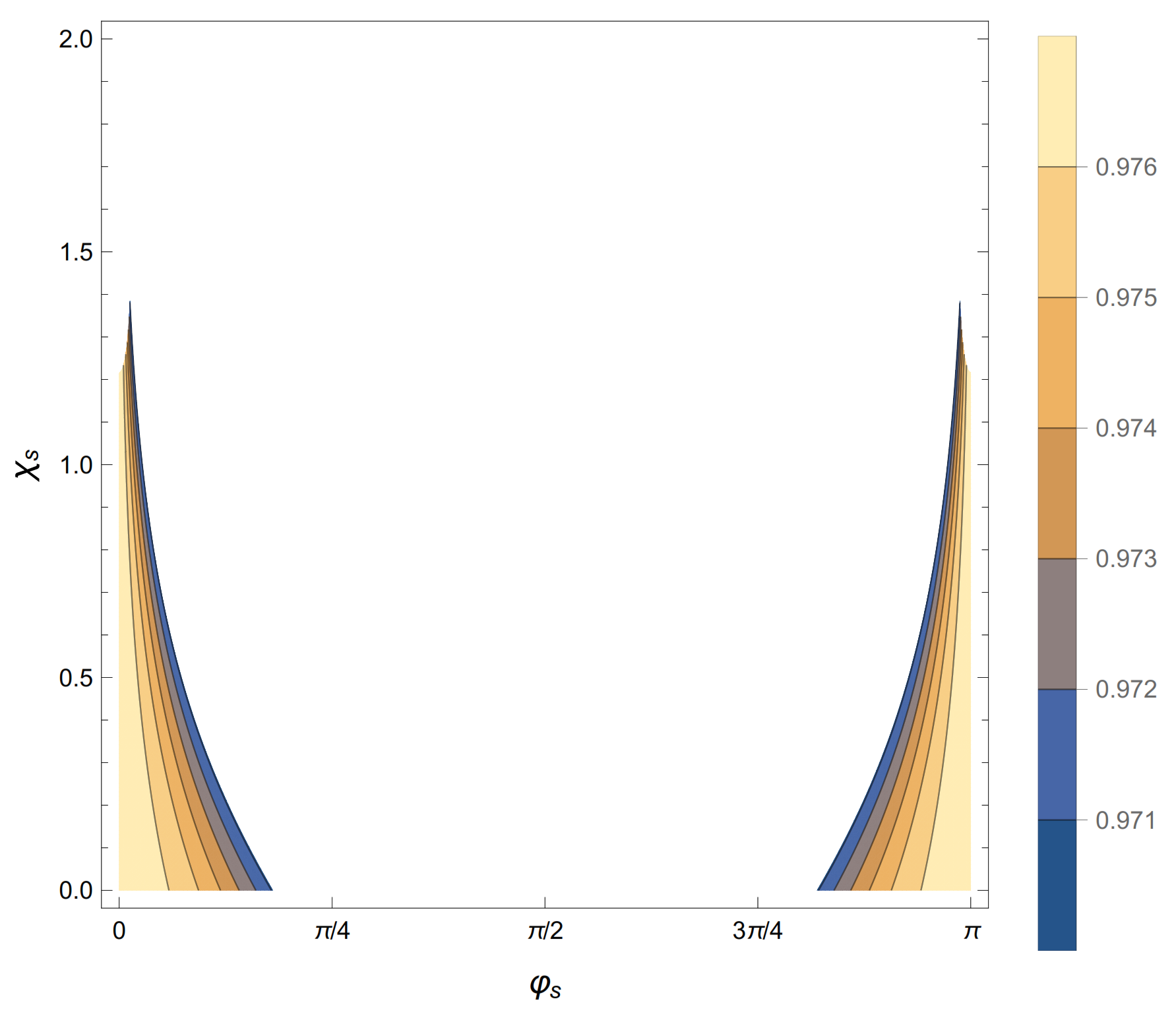}{$\dk\chi_{{}_S}=-0.005$}\\[2mm]
  \panel{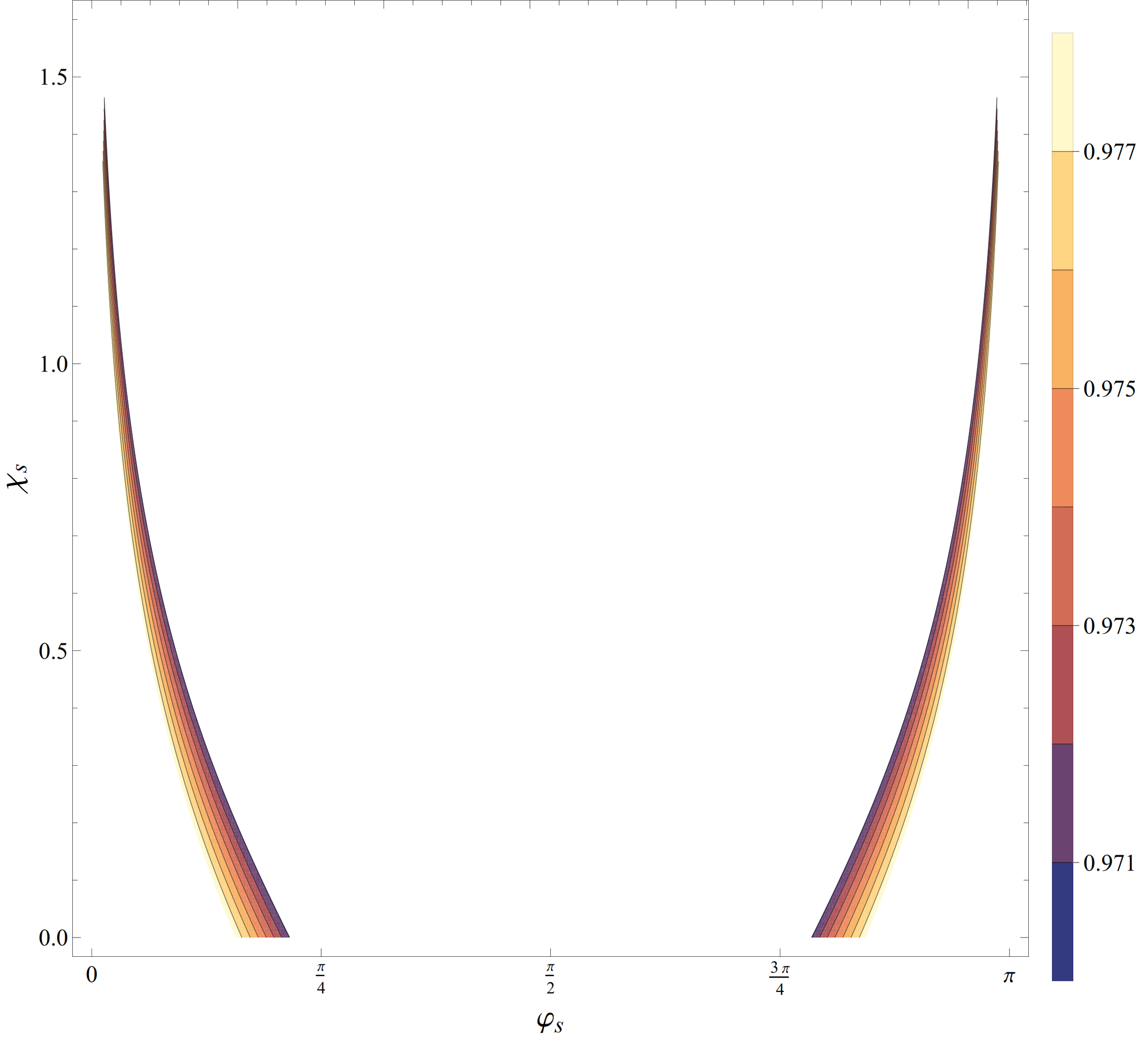}{$\dk\chi_{{}_S}=-0.01$}\hfill\panel{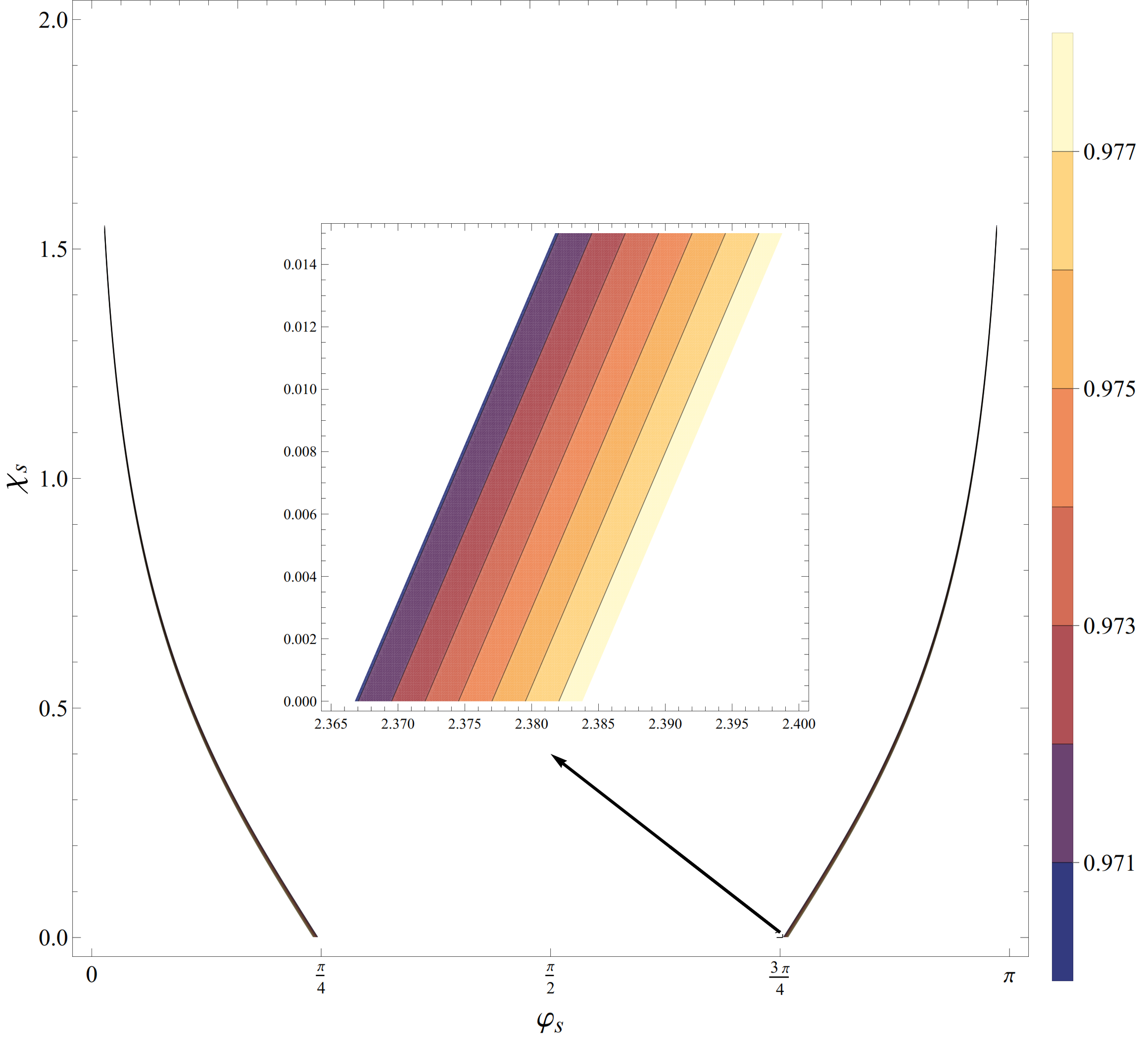}{$\dk\chi_{{}_S}=-0.1$}\\[2mm]
  \caption{Starobinsky model: allowed region in $(\varphi_{{}_S},\chi_{{}_S})$ for fixed $\dk\chi_{{}_S}<0$.}
  \label{fig:B-St-chidot}
\end{figure}

\begin{figure}[t]
  \centering
  \panel{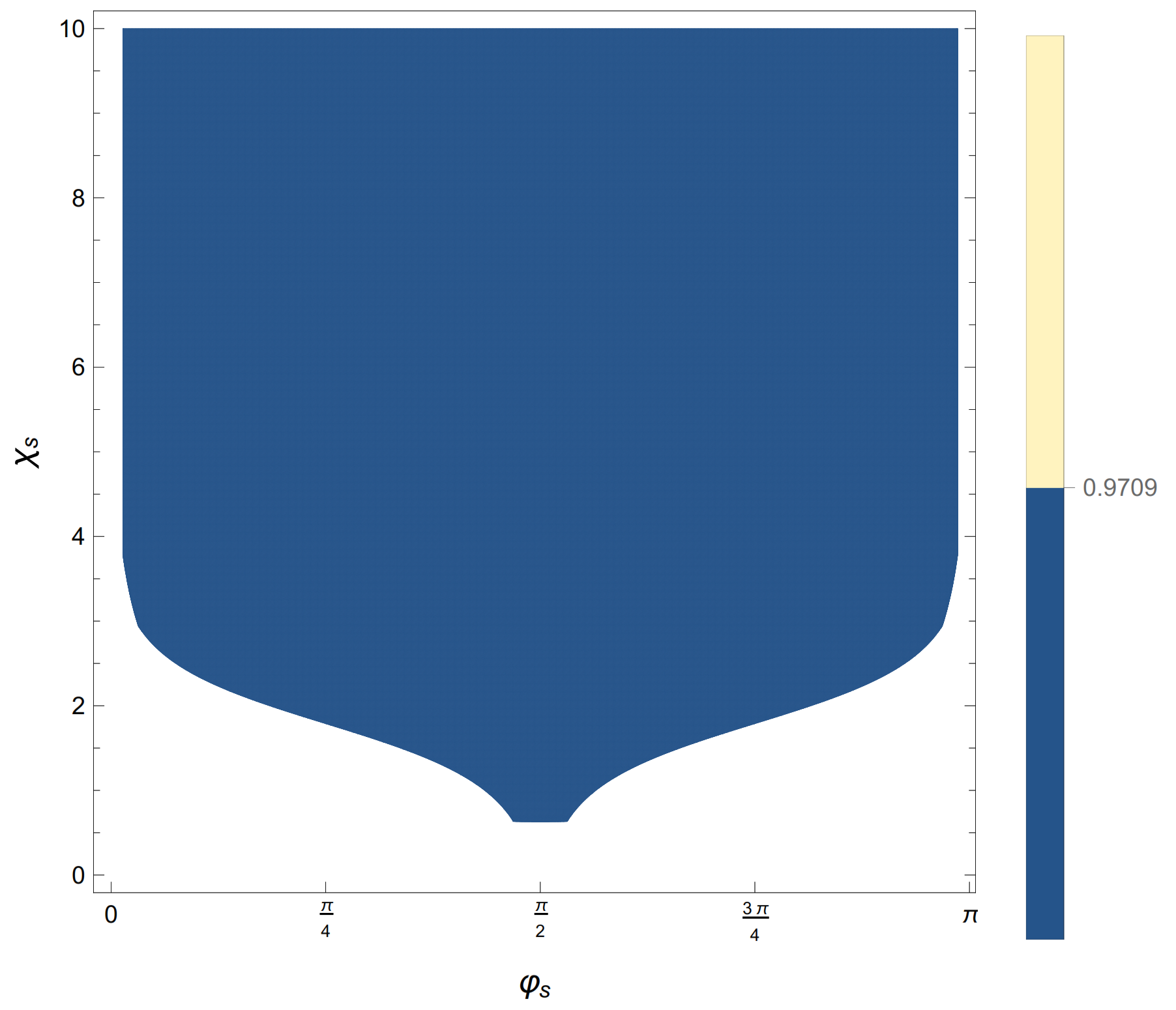}{$\dk\chi_{{}_S}=0.00212$}\hfill\panel{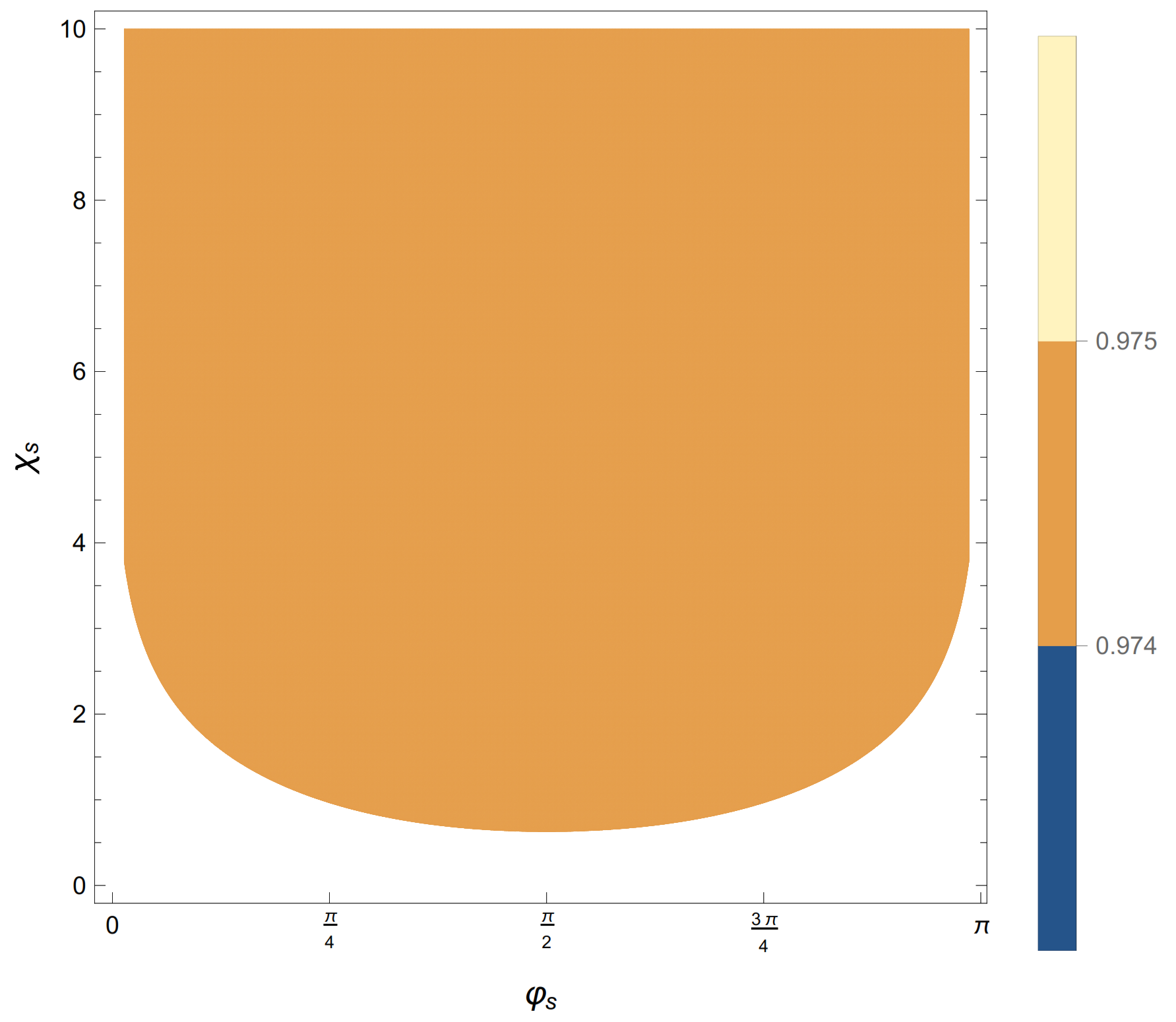}{$\dk\chi_{{}_S}=0.004$}\\[2mm]
  \panel{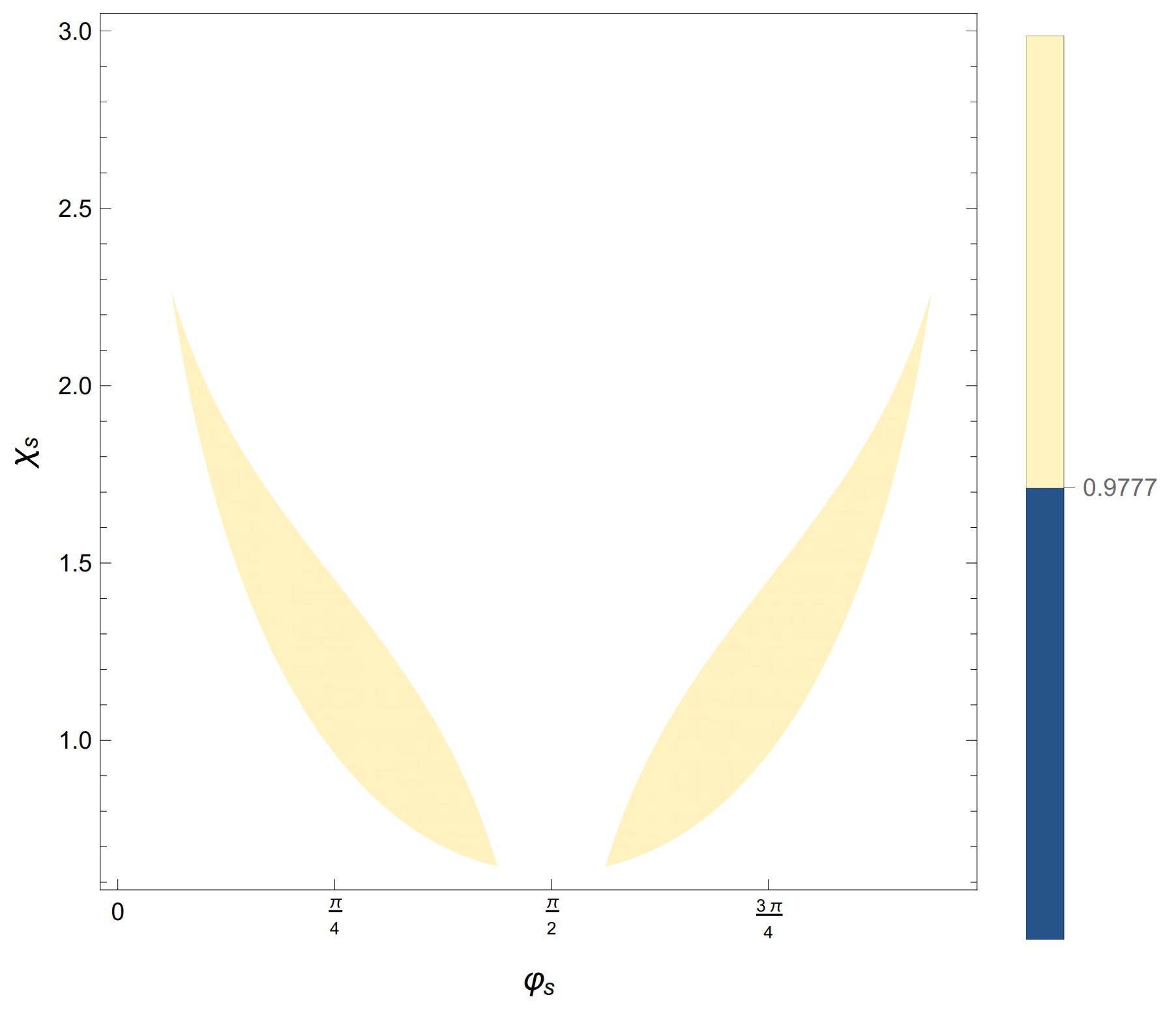}{$\dk\chi_{{}_S}=0.00555$}\hfill\panel{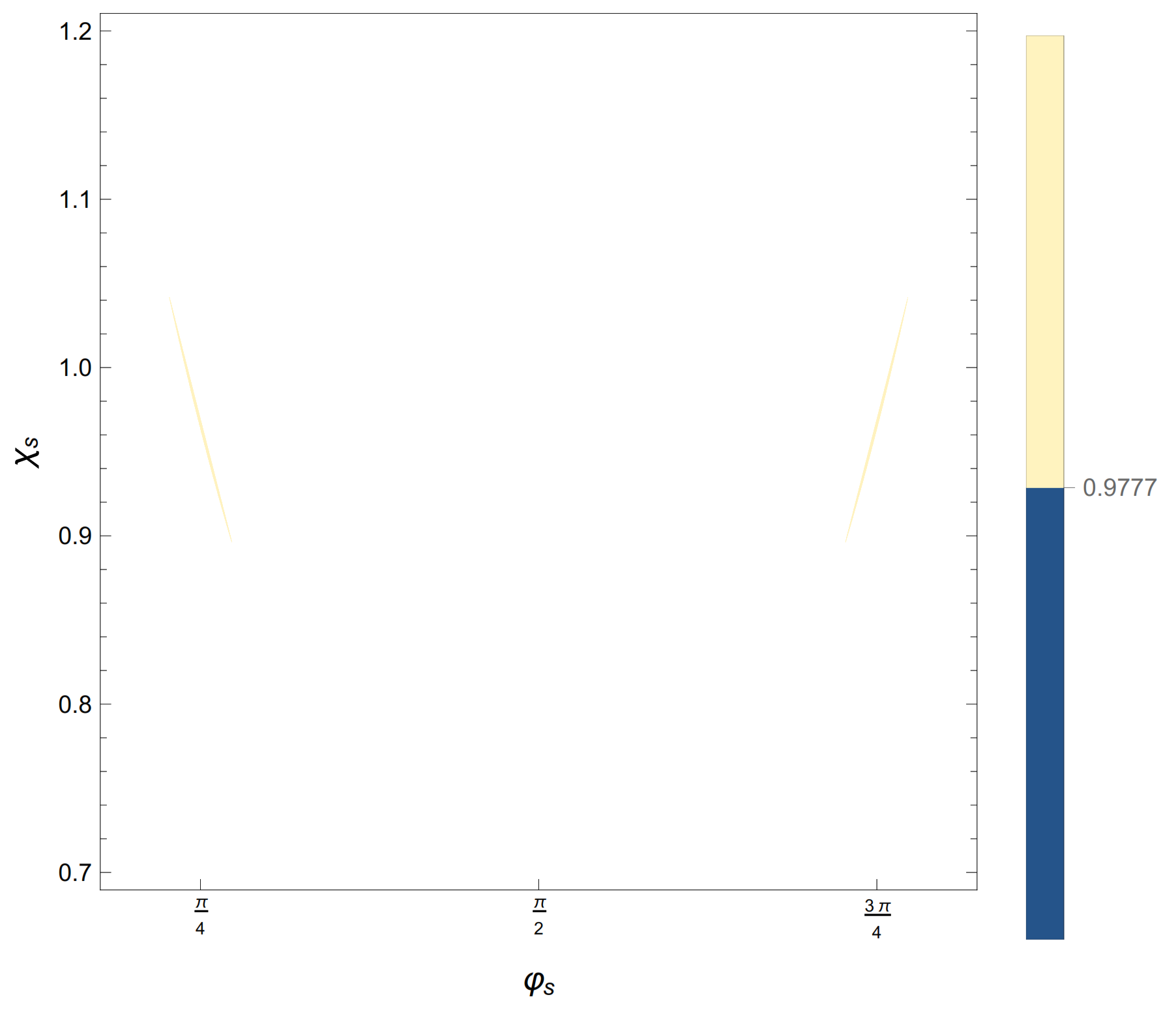}{$\dk\chi_{{}_S}=0.00575$}
  \caption{QC model: allowed region in $(\varphi_{{}_S},\chi_{{}_S})$ for fixed $\dk\chi_{{}_S}$.}
  \label{fig:A-QC-chidot}
\end{figure}

\begin{figure}[t]
  \centering
  \panel{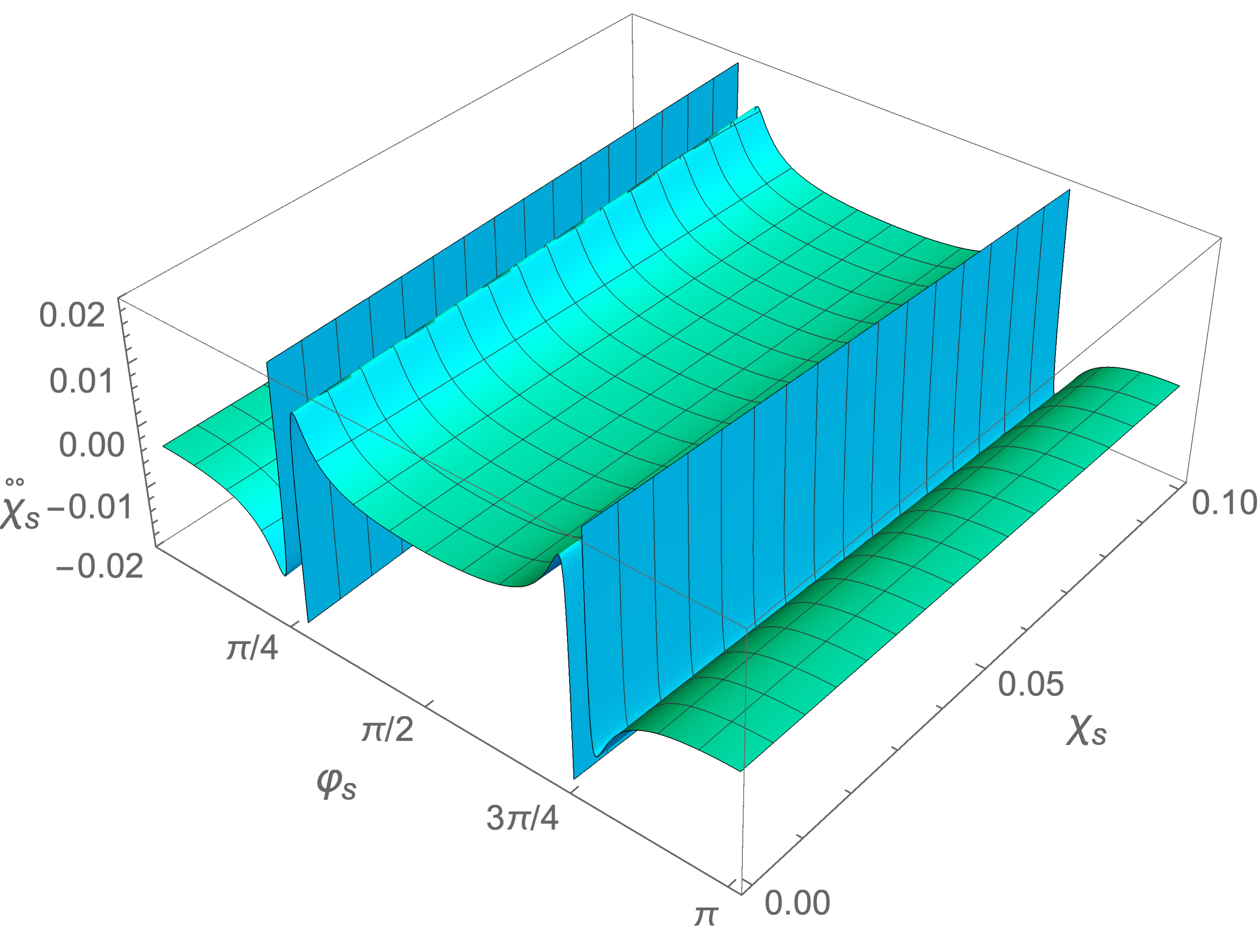}{Starobinsky, $0\leq \chi_{{}_S}\leq 0.1$}\hfill\panel{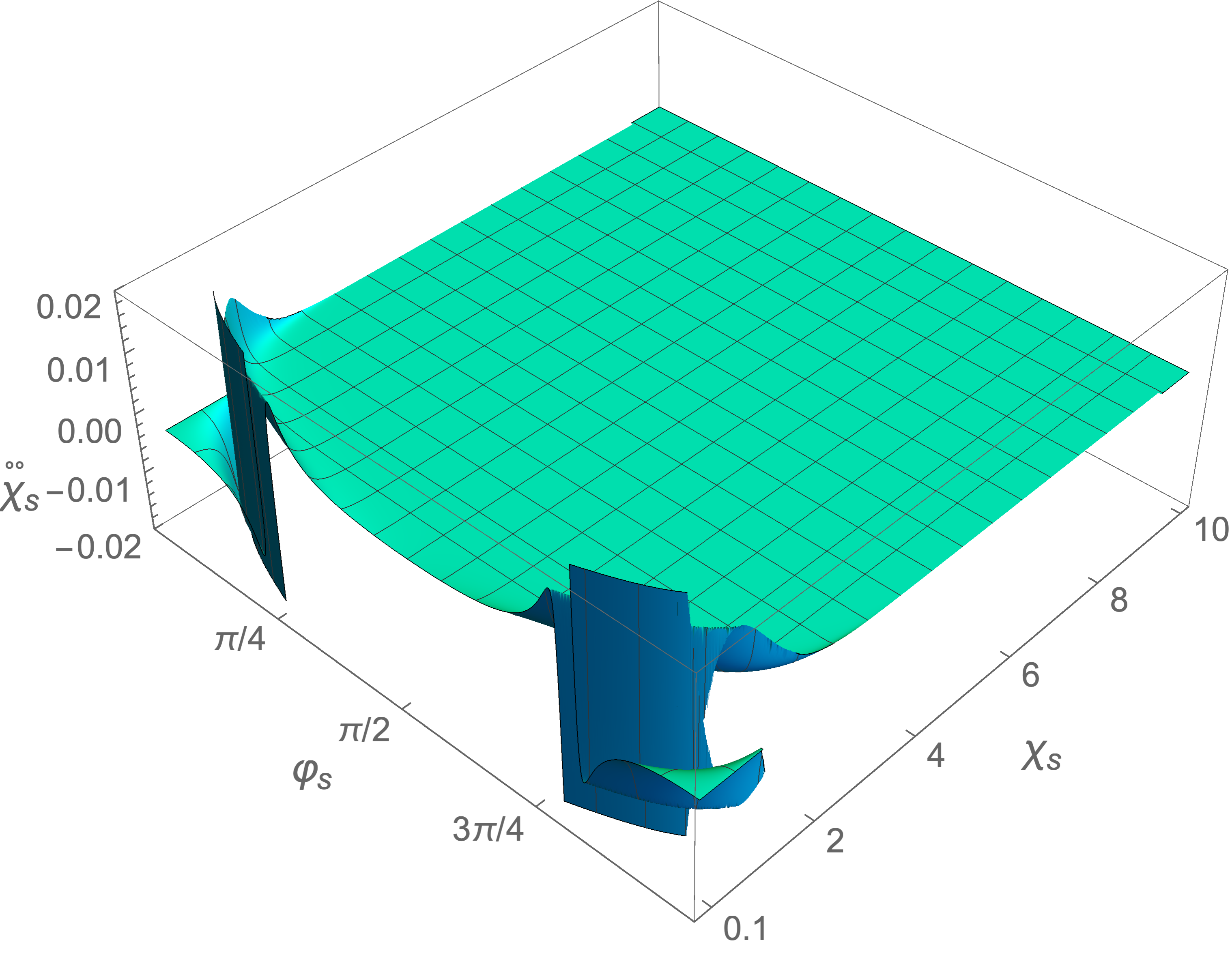}{Starobinsky, $0.1\leq \chi_{{}_S}\leq 10$}
  \caption{Plots of  $\ddk\chi_{{}_S}$ as a function of $(\varphi_{{}_S},\chi_{{}_S})$ for the Starobinsky inflation when $\varphi_{{}_S}$ is scale-independent, assuming the mean values of $n_{\rm s}^{\Ob}$ and  $\alpha_{{}_S}^{\Ob}$ from the ACT DR6.}
  \label{fig:chiddot-mean-Starobinsky}
\end{figure}

\begin{figure}[t]
  \centering
 \panel{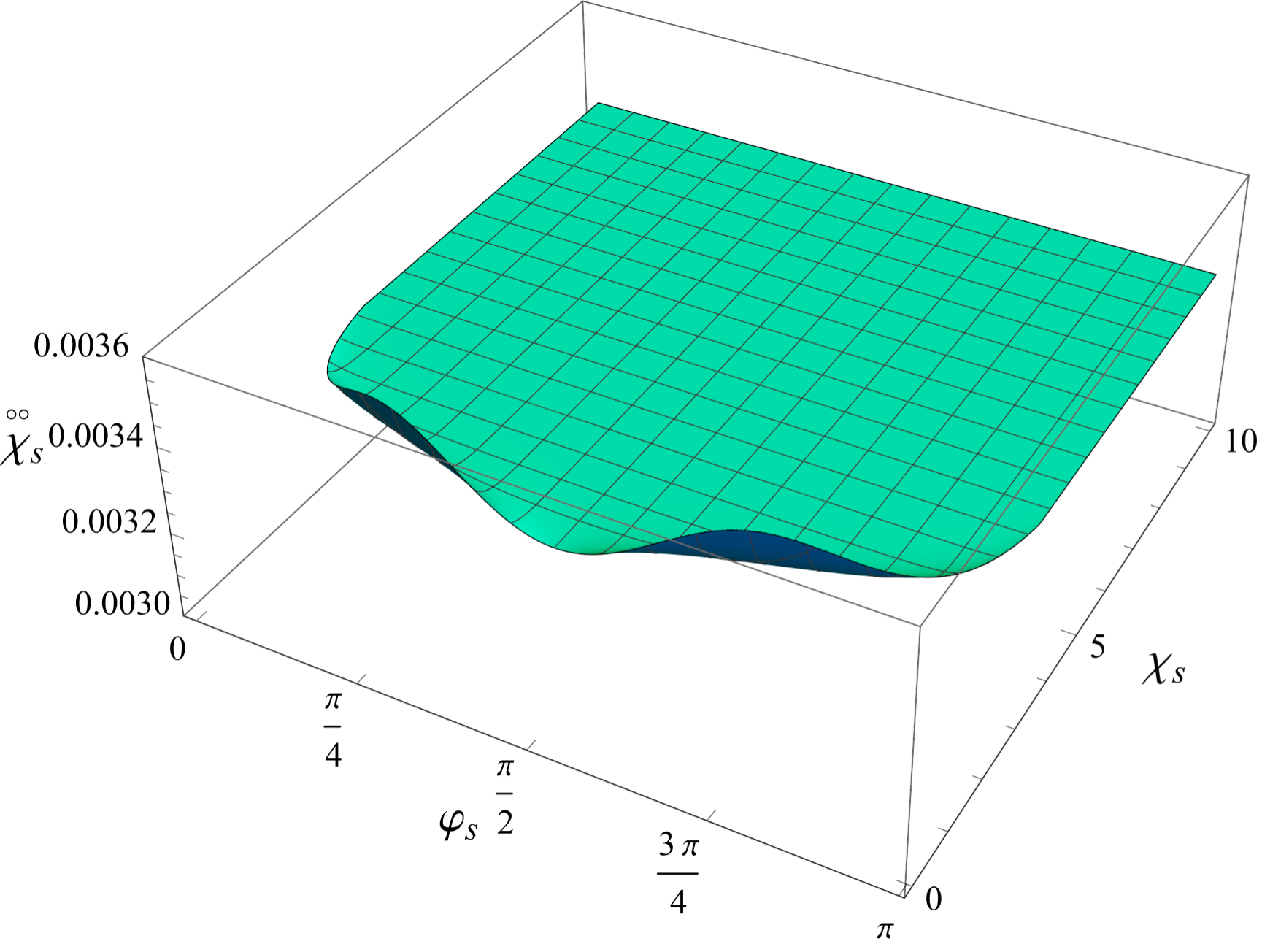}{QC}
  \caption{Plots of  $\ddk\chi_{{}_S}$ as a function of $(\varphi_{{}_S},\chi_{{}_S})$ for the QC model when $ \varphi_{{}_S}$  is scale-independent, assuming the mean values of $n_{\rm s}^{\Ob}$ and  $\alpha_{{}_S}^{\Ob}$ from the ACT DR6.}
  \label{fig:chiddot-mean-QC}
\end{figure}

\begin{figure}[p]
  \centering

  \begin{subfigure}{0.48\textwidth}
    \centering
    \includegraphics[width=\linewidth]{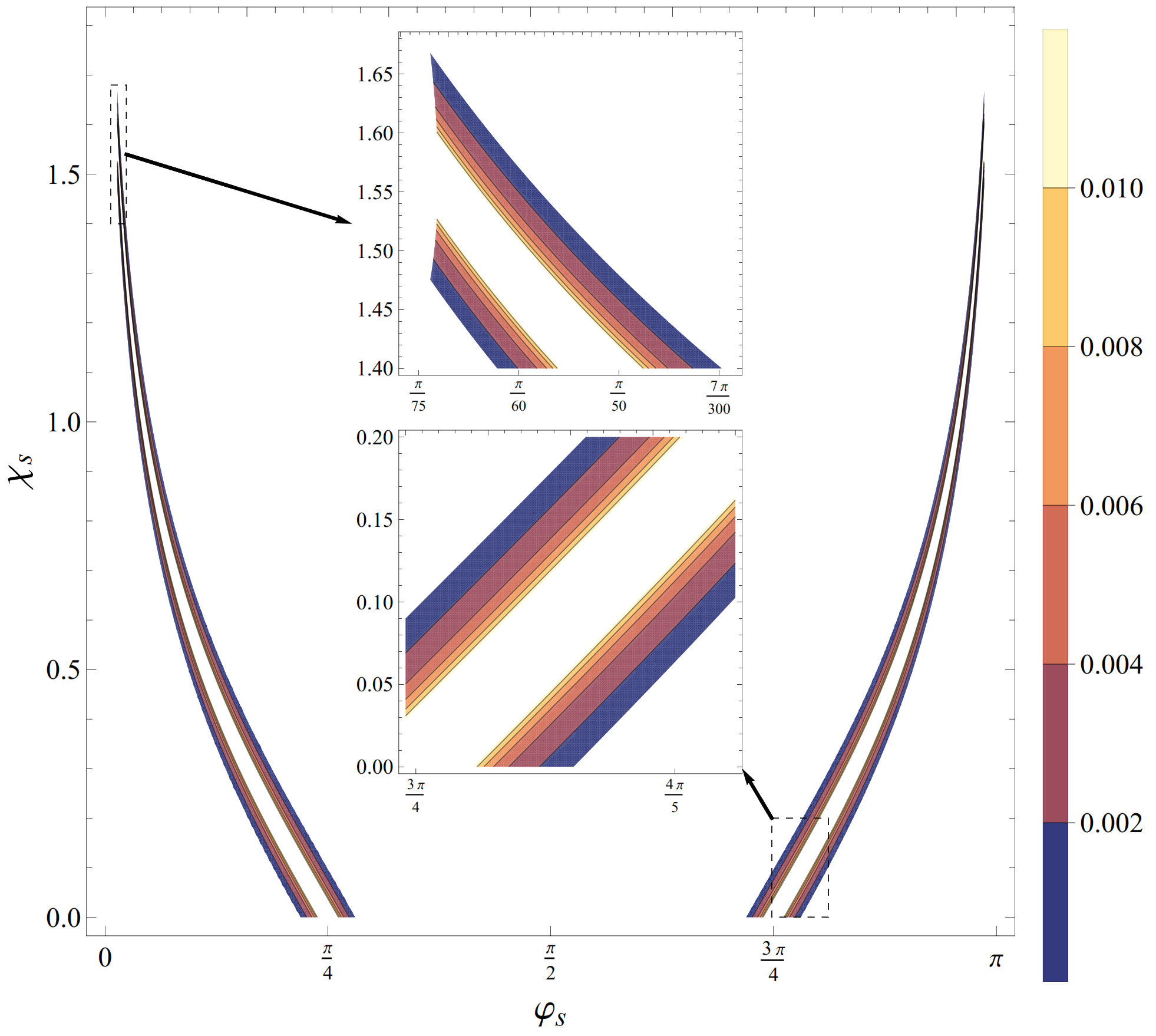}
    \caption{$\ddk\chi_{{}_S}=0$}
    \label{fig:A-St-chiddot-pos-a}
  \end{subfigure}
  \hfill
  \begin{subfigure}{0.48\textwidth}
    \centering
    \includegraphics[width=\linewidth]{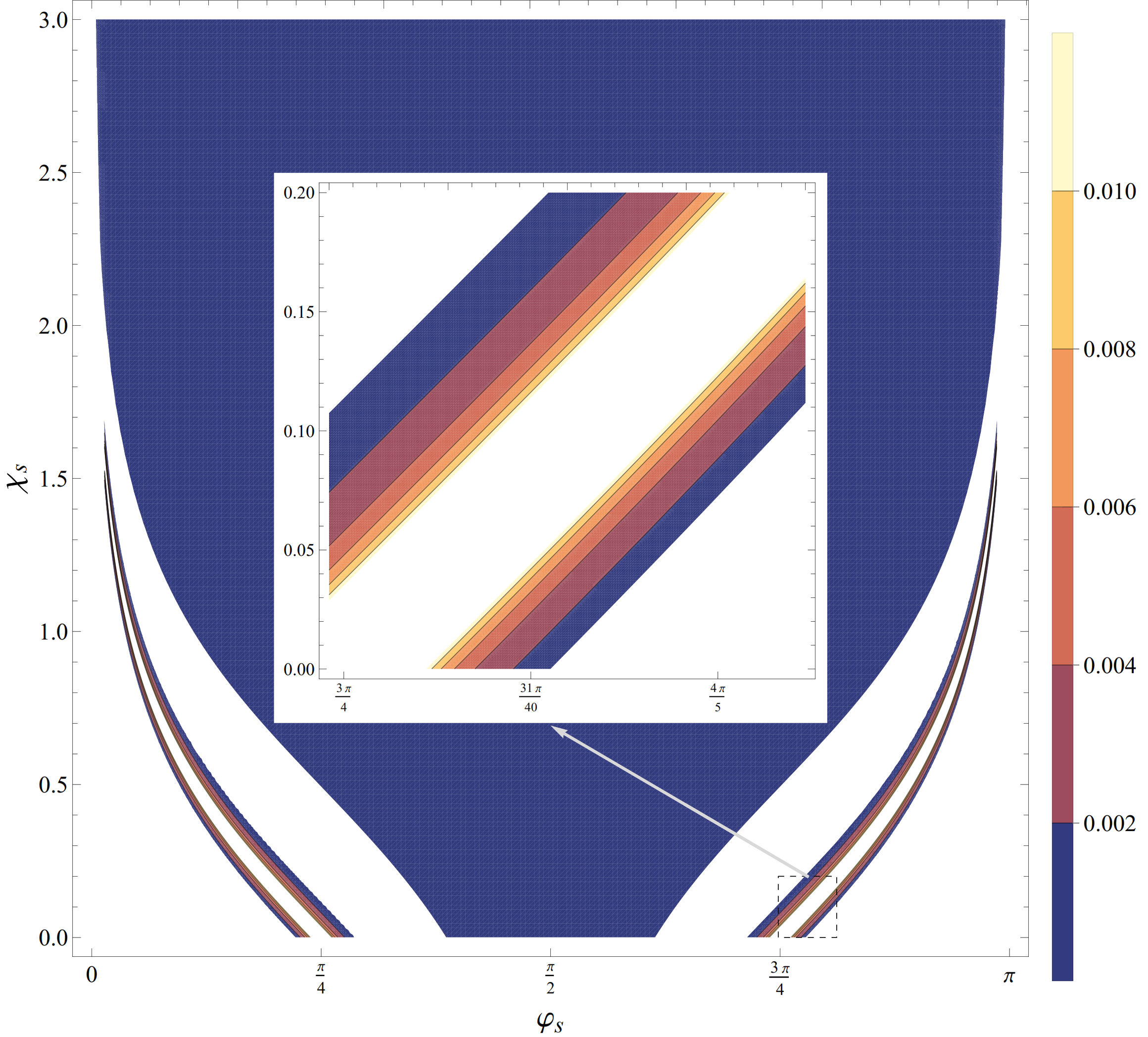}
    \caption{$\ddk\chi_{{}_S}=0.001$}
    \label{fig:A-St-chiddot-pos-b}
  \end{subfigure}

  \vspace{2mm}

  \begin{subfigure}{0.48\textwidth}
    \centering
    \includegraphics[width=\linewidth]{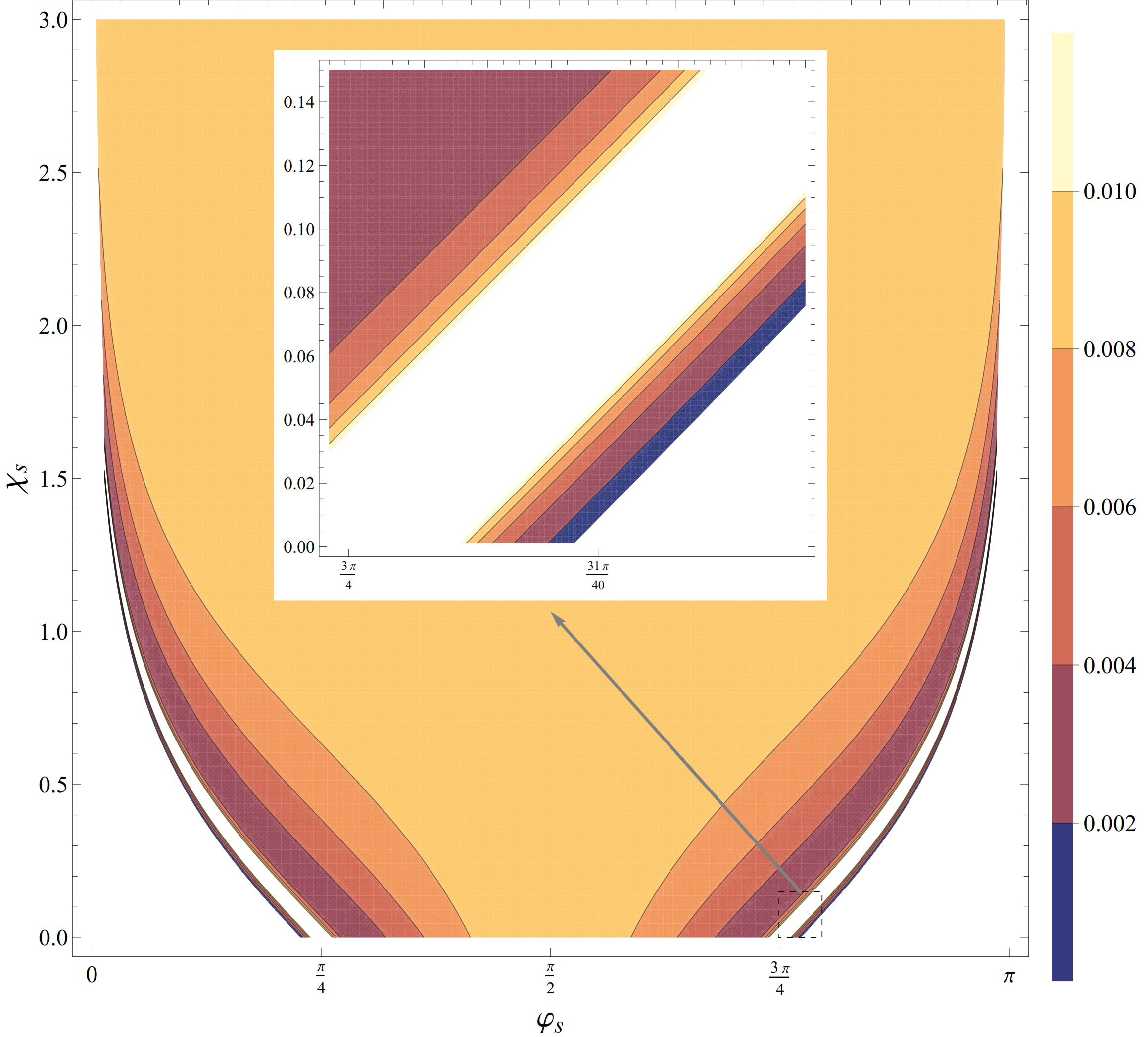}
    \caption{$\ddk\chi_{{}_S}=0.005$}
    \label{fig:A-St-chiddot-pos-c}
  \end{subfigure}
  \hfill
  \begin{subfigure}{0.48\textwidth}
    \centering
    \includegraphics[width=\linewidth]{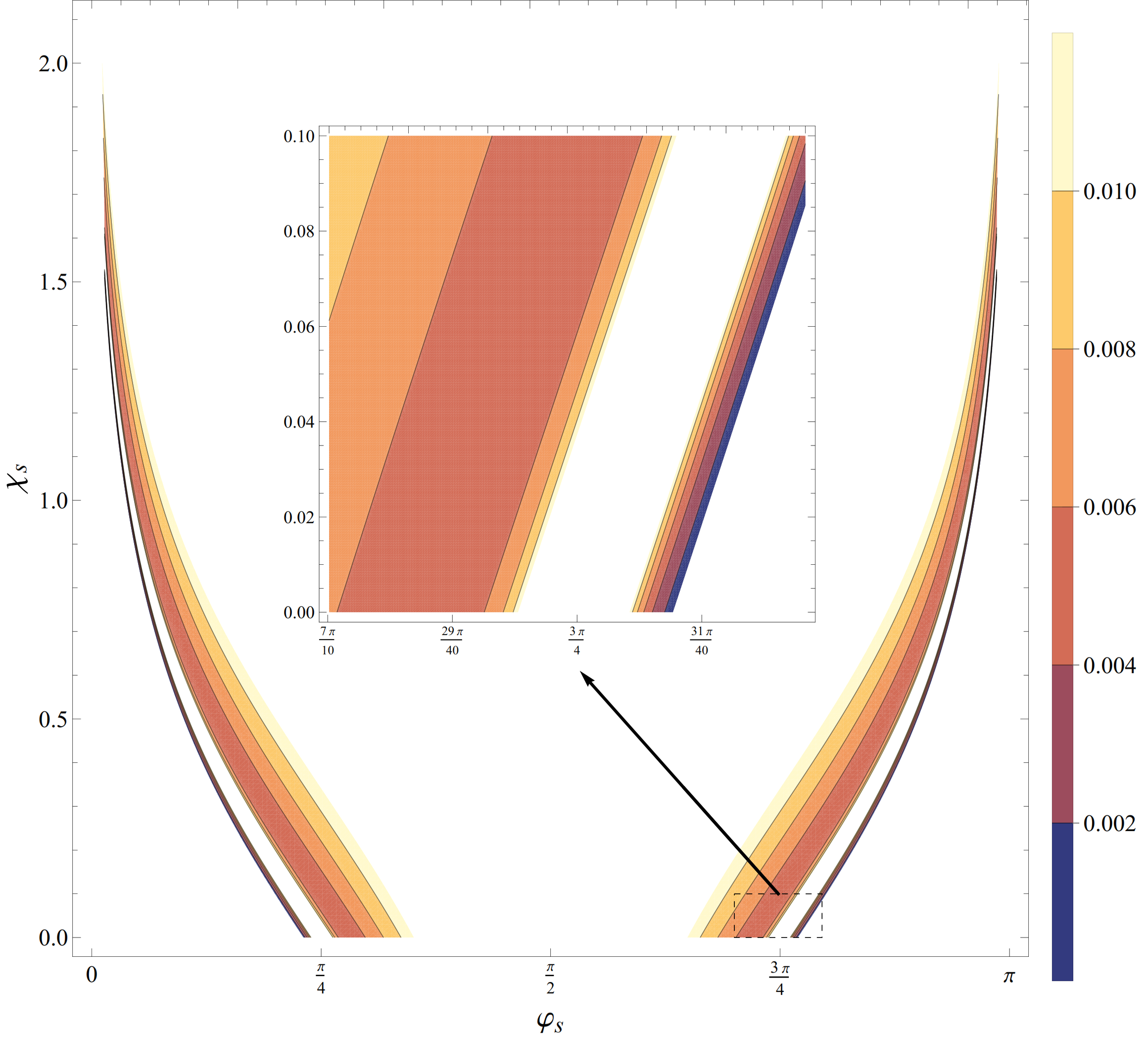}
    \caption{$\ddk\chi_{{}_S}=0.01$}
    \label{fig:A-St-chiddot-pos-d}
  \end{subfigure}

  \vspace{2mm}

  \begin{subfigure}{0.48\textwidth}
    \centering
    \includegraphics[width=\linewidth]{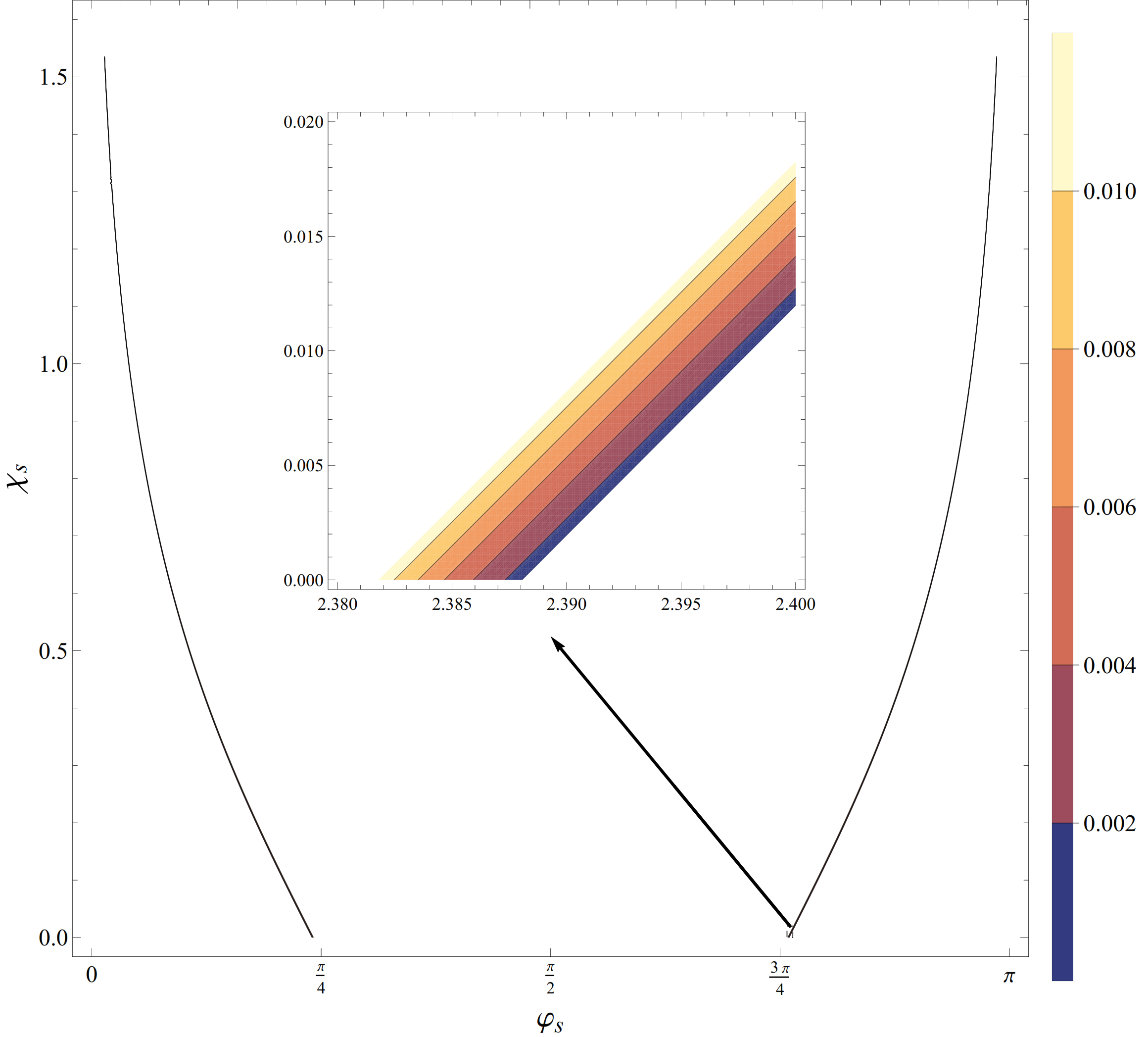}
    \caption{$\ddk\chi_{{}_S}=0.1$}
    \label{fig:A-St-chiddot-pos-e}
  \end{subfigure}

  \caption{Starobinsky model: allowed region in
  $(\varphi_{{}_S},\chi_{{}_S})$ for fixed
  $\ddk\chi_{{}_S}\geq0$.}
  \label{fig:A-St-chiddot-pos}
\end{figure}

\begin{figure}[htbp]
  \centering
  \panel{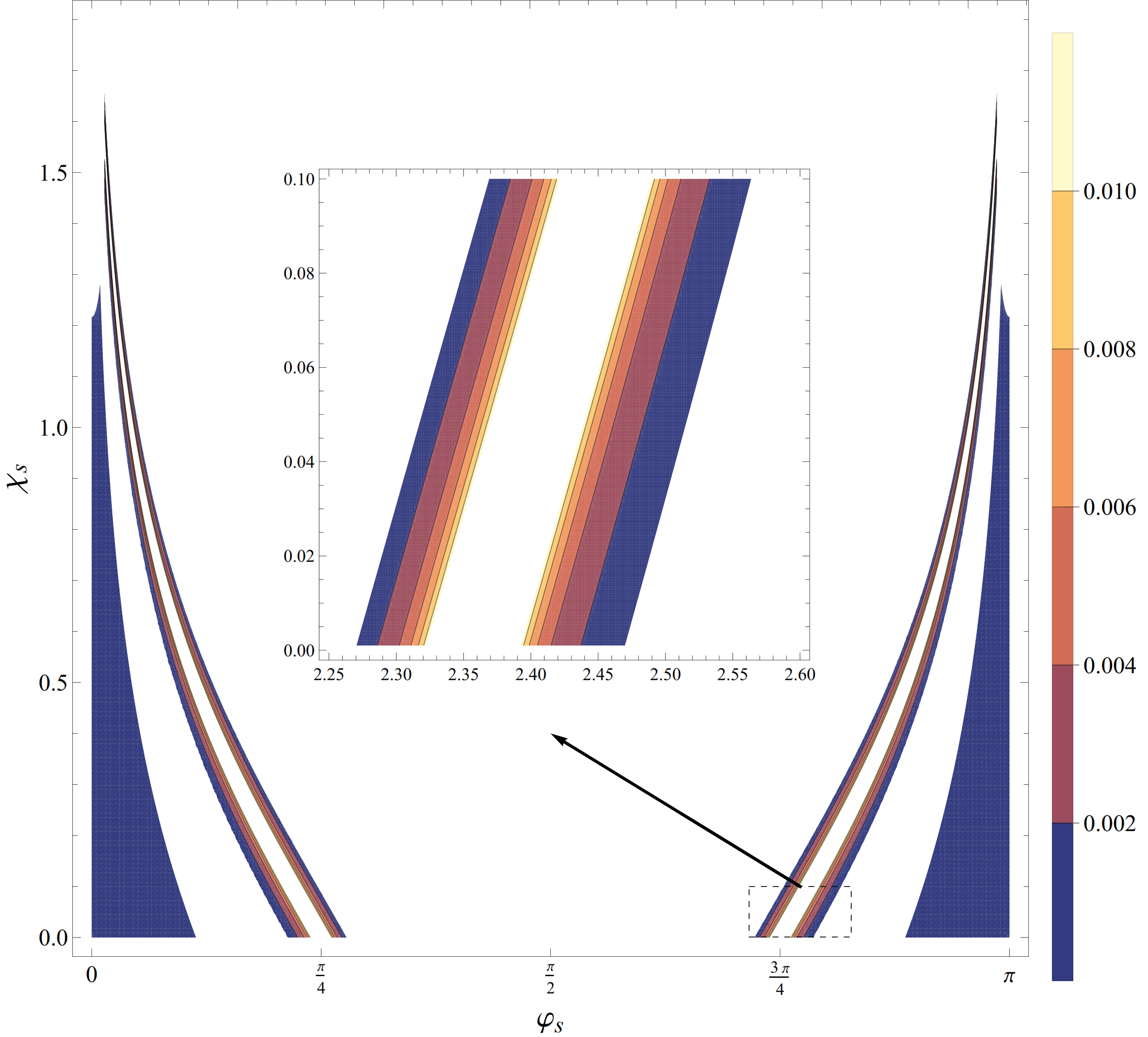}{$\ddk\chi_{{}_S}=-0.001$}\hfill\panel{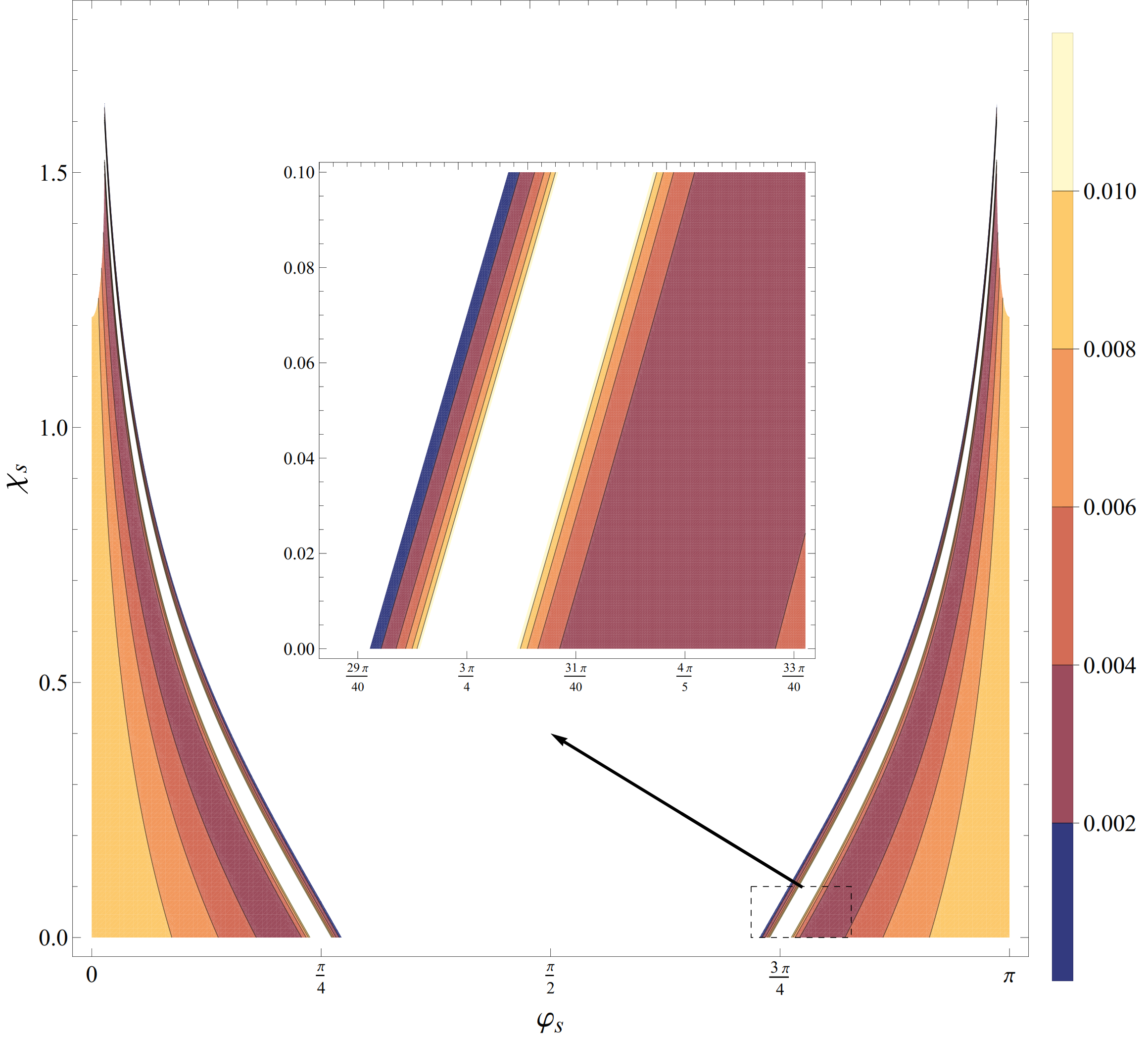}{$\ddk\chi_{{}_S}=-0.005$}\\[2mm]
  \panel{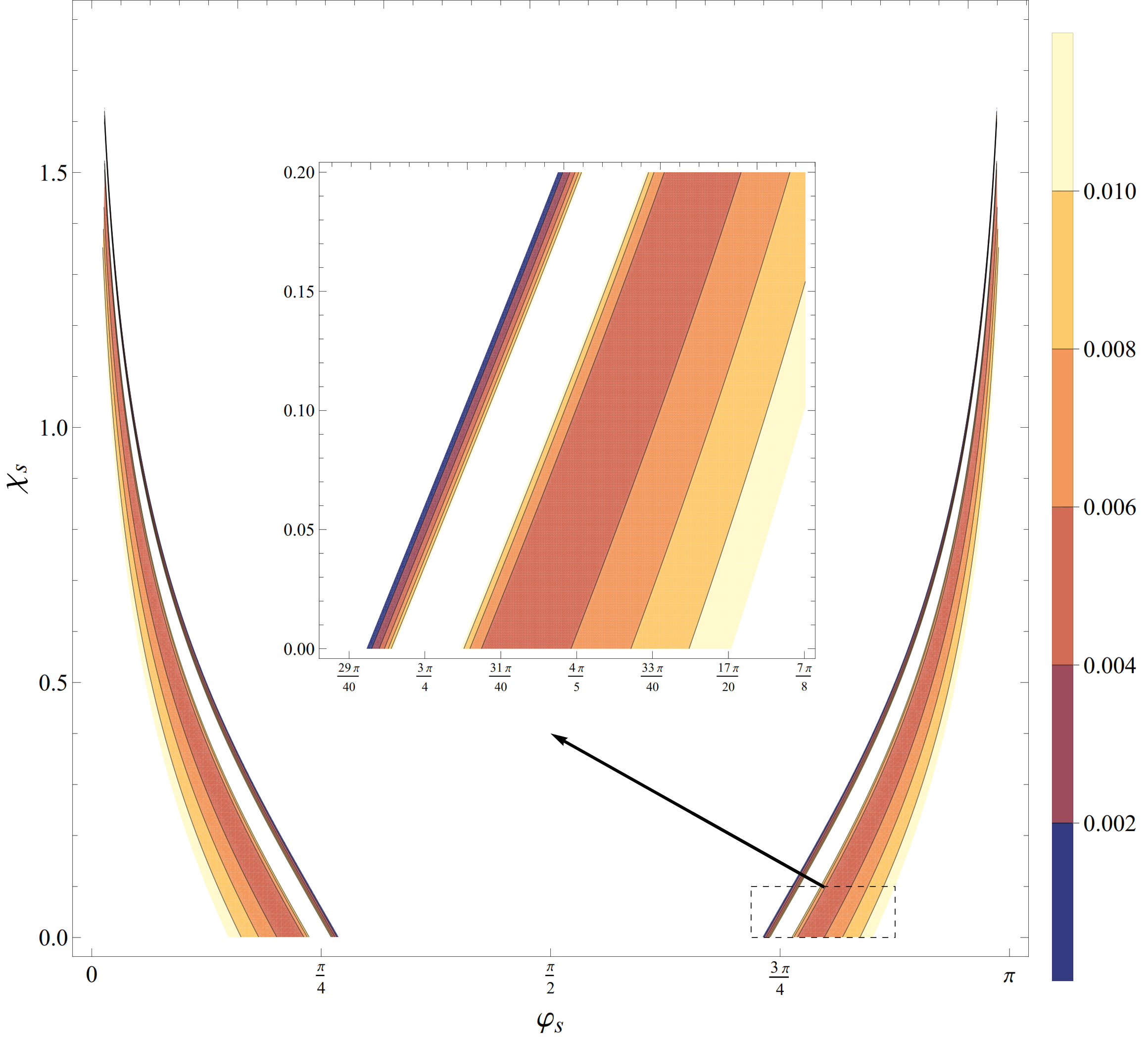}{$\ddk\chi_{{}_S}=-0.01$}\hfill\panel{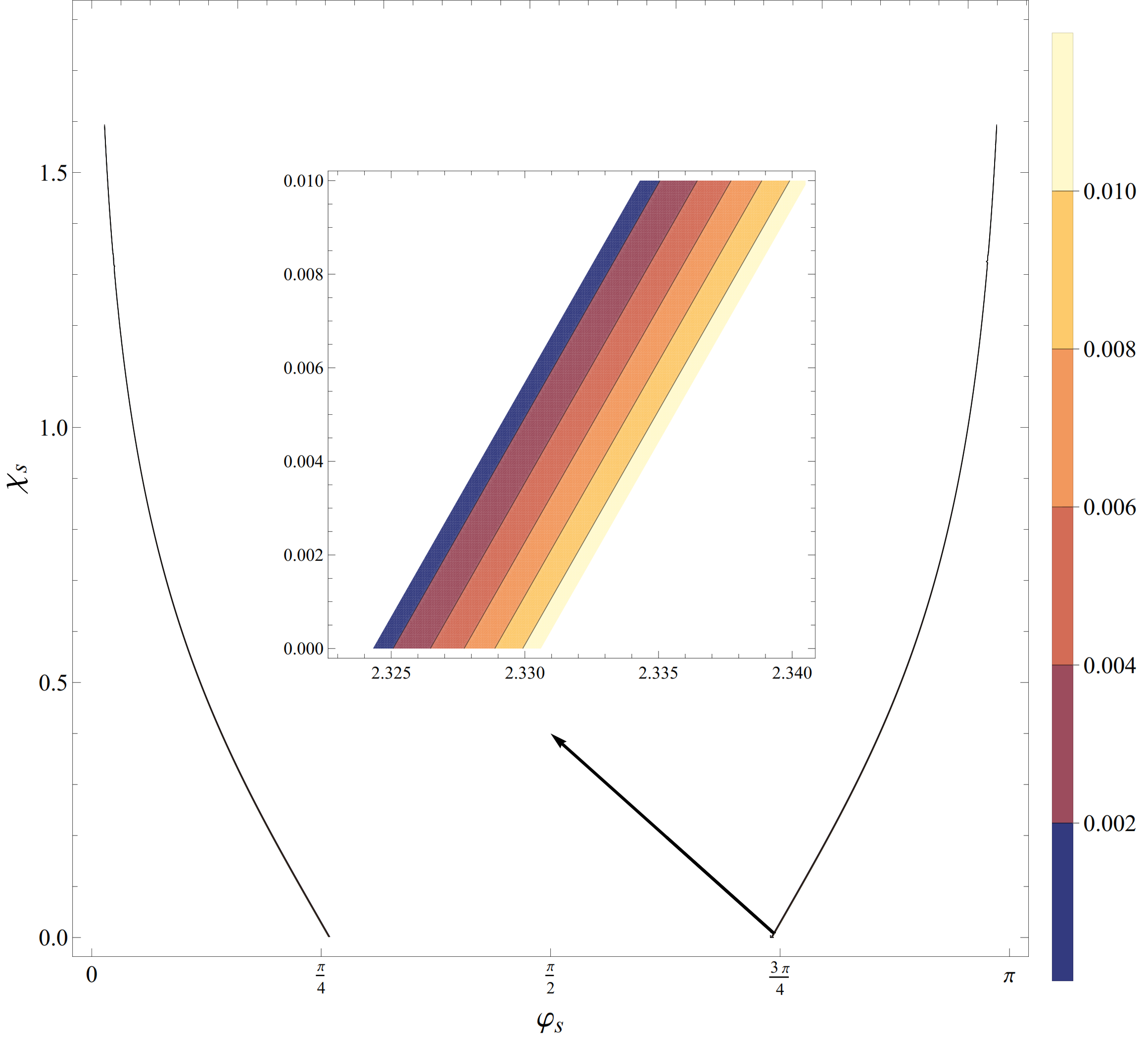}{$\ddk\chi_{{}_S}=-0.1$}
  \caption{Starobinsky model: allowed region in $(\varphi_{{}_S},\chi_{{}_S})$ for fixed $\ddk\chi_{{}_S}<0$.}
  \label{fig:A-St-chiddot-neg}
\end{figure}

\begin{figure}[htbp]
  \centering
  \panel{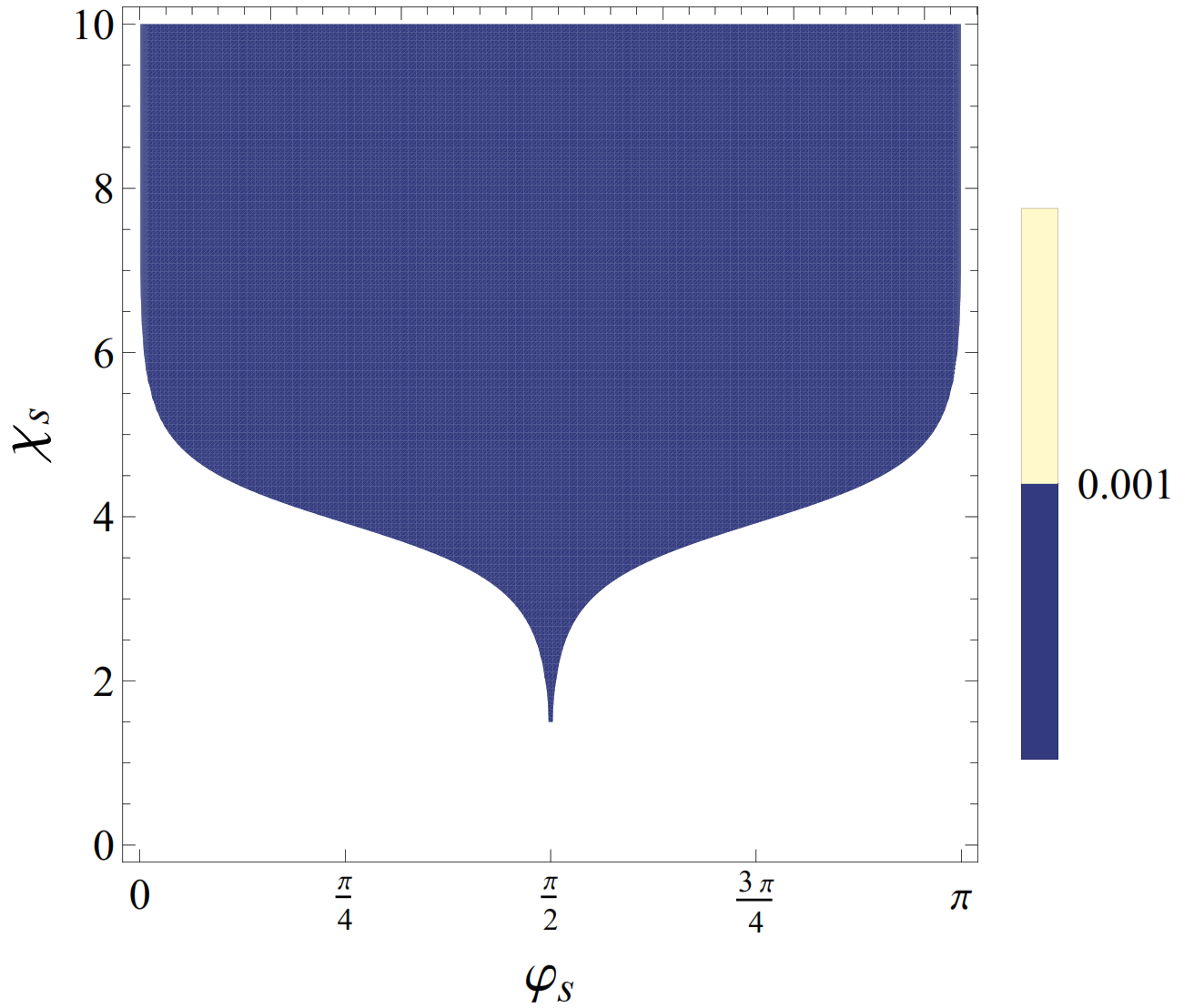}{$\ddk\chi_{{}_S}=0.00078$}\hfill\panel{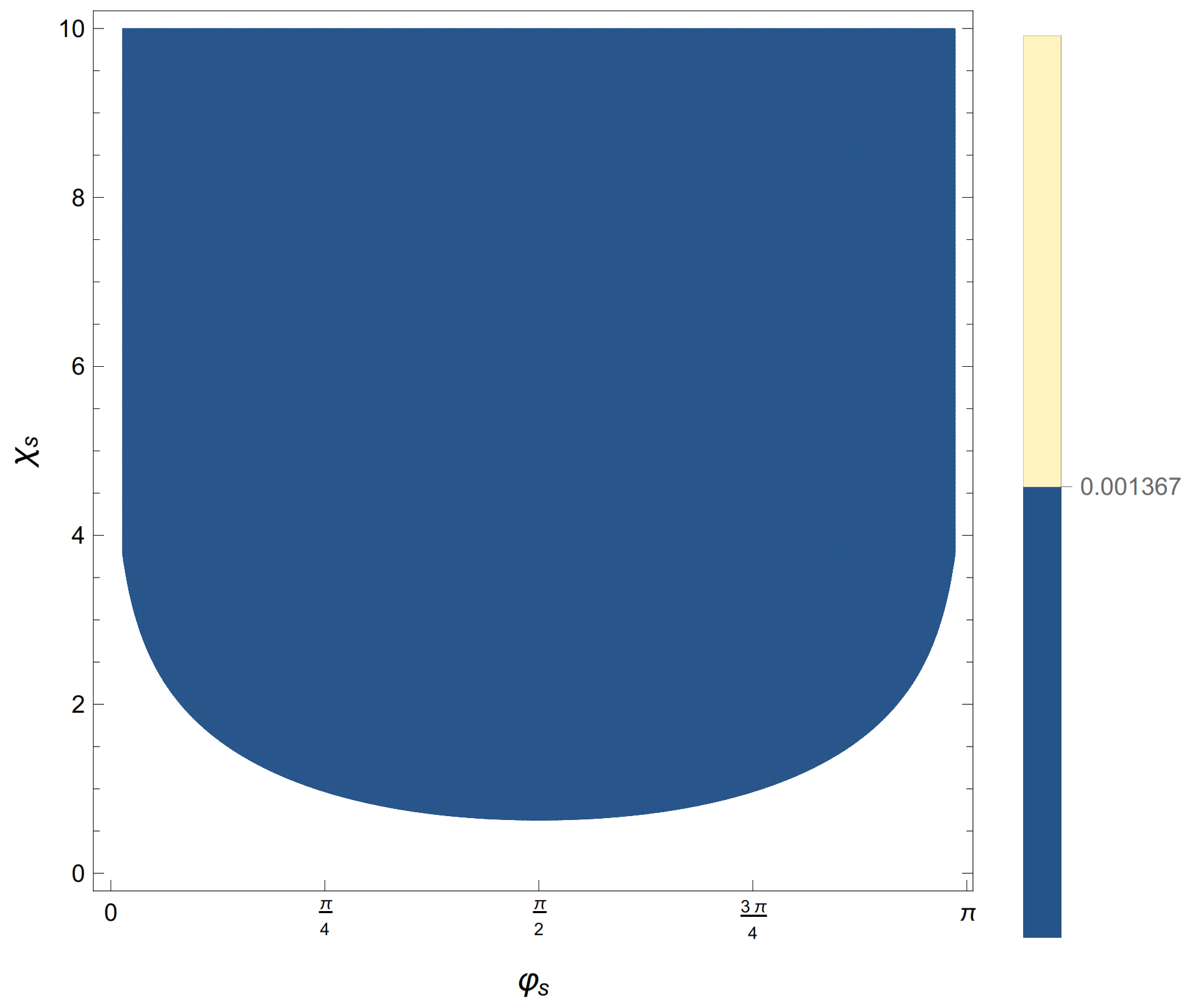}{$\ddk\chi_{{}_S}=0.001$}\\[2mm]
  \panel{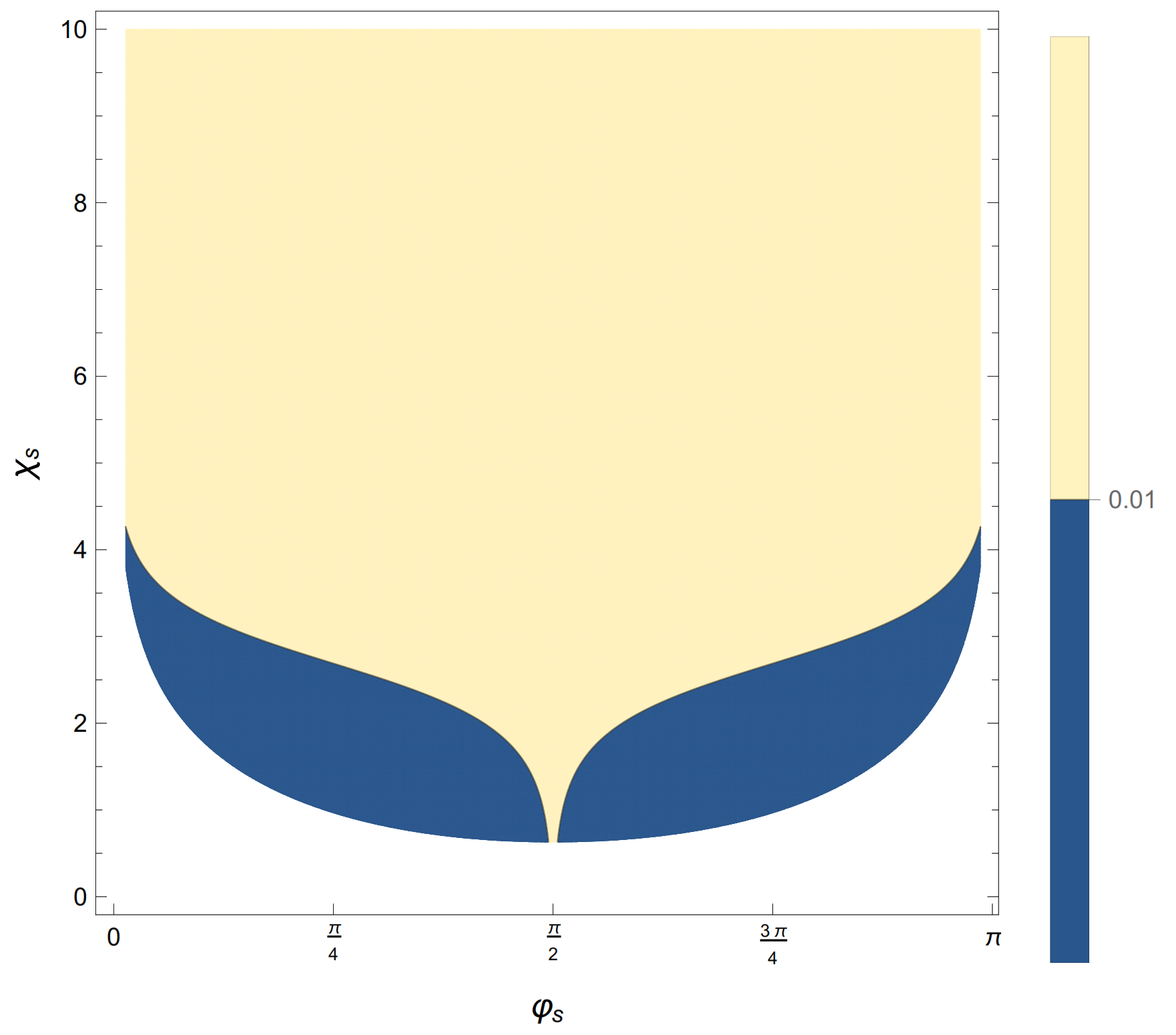}{$\ddk\chi_{{}_S}=0.005278$}\hfill\panel{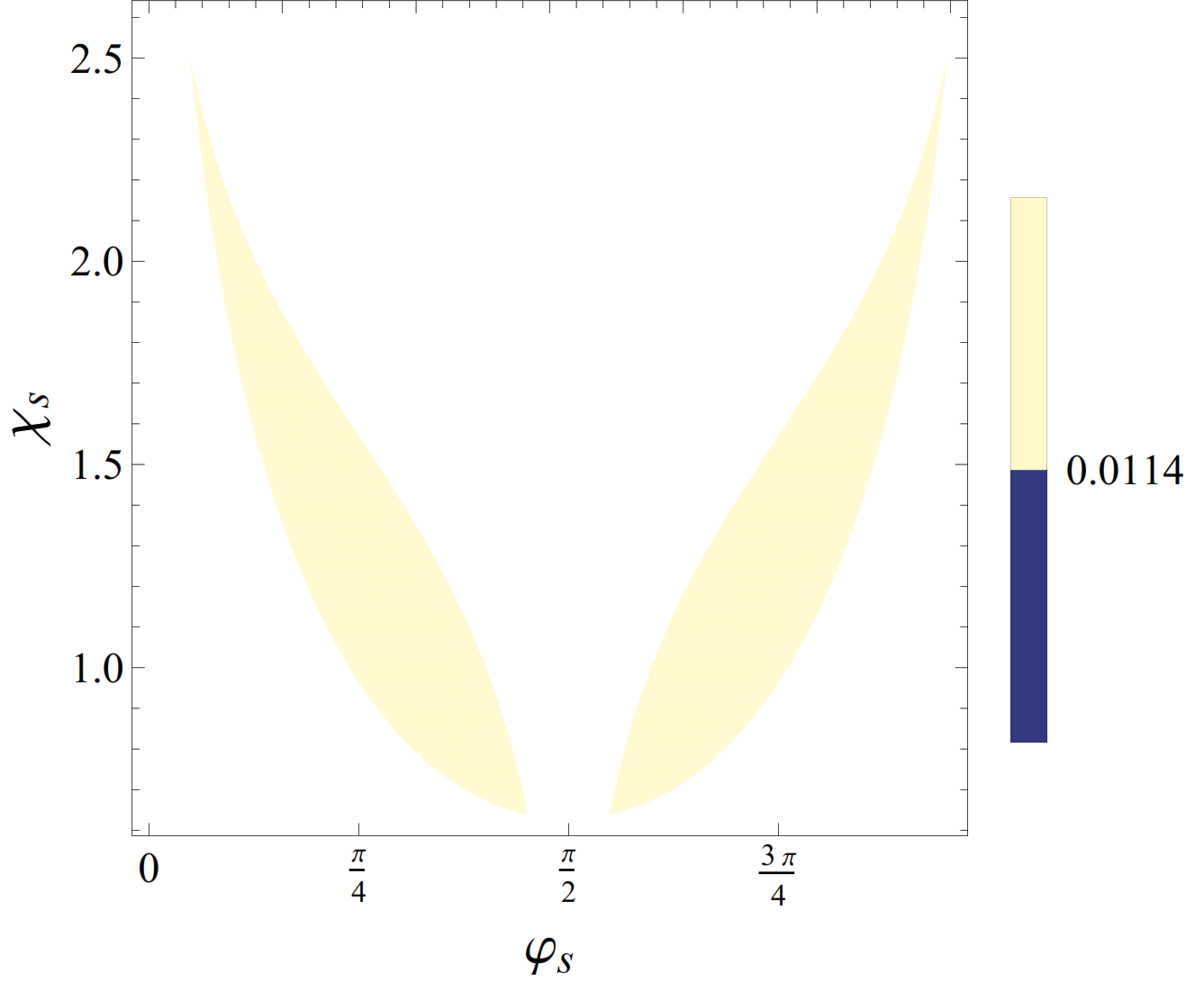}{$\ddk\chi_{{}_S}=0.006$}
   \panel{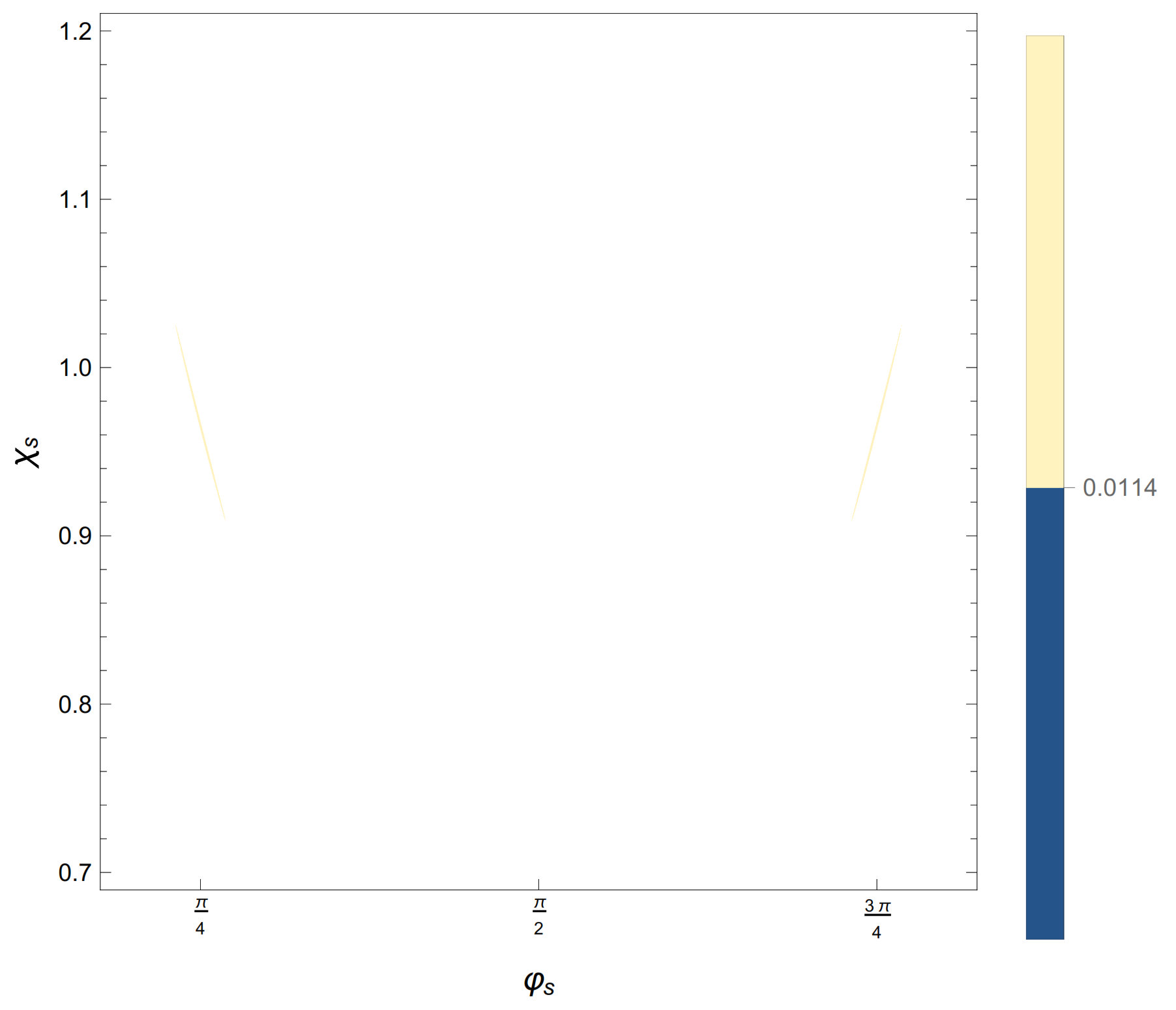}{$\ddk\chi_{{}_S}=0.00623$}
  \caption{QC model: allowed region in $(\varphi_{{}_S},\chi_{{}_S})$ plane for fixed $\ddk\chi_{{}_S}$.}
  \label{fig:A-QC-chiddot}
\end{figure}

\paragraph{Case B: $\dk\chi_{{}_S}\neq0$, $\ddk\chi_{{}_S}\neq0$, $k$-independent $\varphi_{{}_S}$: }
\begin{align}
  \dk\chi_{{}_S} &= \frac{\dns{}}{2}\,\frac{\gamma_{{}_S}}{\Bs}\,, \label{eq:chidot}\\
  \ddk\chi_{{}_S} &= \frac{\das{}}{2}\,\frac{\gamma_{{}_S}}{\Bs}
  - \big(\dns{}\big)^2\,\frac{\sin^2 2\varphi_{{}_S}\,\gamma_{{}_S}}{2\,\Bs^3}\,. \label{eq:chiddot}
\end{align}
For $\dk\chi_{{}_S}\neq0$, and scale-independent $\varphi_{{}_S}$, the surface plots of $\dk\chi_{{}_S}$
in the $(\varphi_{{}_S},\chi_{{}_S})$ is shown in figure~\ref{fig:chidot-mean}, for the ACT mean
value of $n_{\rm s}^{\Ob}$. Contour plots in the $(\varphi_{{}_S},\chi_{{}_S})$ plane for a fixed value of $\dk\chi_{{}_S}$, with
$n_{\rm s}^{\Ob}$ ranging over all viable values in the $1\sigma$ region of the scalar spectral index of the ACT DR6 results, are presented in figures~\ref{fig:A-St-chidot}, \ref{fig:B-St-chidot} and
\ref{fig:A-QC-chidot}. For $\dk\chi_{{}_S}\neq0$, $\ddk\chi_{{}_S}\neq0$ and scale-independent $\varphis$, we have plotted
$\ddk\chi_{{}_S}$ in the allowed region of $\varphi_{{}_S},\chi_{{}_S})$, for the mean values of $n_{\rm s}^{\Ob}$ $\alpha_{\rm s}^{\Ob}$  in
figures~\ref{fig:chiddot-mean-Starobinsky} and \ref{fig:chiddot-mean-QC}. The constraints on $(\varphi_{{}_S},\chi_{{}_S})$ for different values of
$\ddk\chi_{{}_S}$, evaluated over the ACT DR6 range of $\alpha_{\rm s}^{\Ob}$, for the mean value of $n_{\rm s}^{\Ob}$, are displayed in
figures~\ref{fig:A-St-chiddot-pos}--\ref{fig:A-QC-chiddot} for the Starobinsky and QC models.


It is also instructive to analyse the behaviour of eqs.~\eqref{eq:chidot} and \eqref{eq:chiddot} in the limits of large $\chi_{{}_S}$
and of $\varphi_{{}_S}\to0,\pi$. Defining $x\equiv e^{2\chi_{{}_S}}\sin^2\varphi_{{}_S}$ and
$y\equiv e^{-2\chi_{{}_S}}\cos^2\varphi_{{}_S}$, one can write
$\gamma_{{}_S}=x+y$ and $\Bs=x-y$. Noting that using $xy=\frac{\sin^2 2\varphi_{{}_S}}{4}$, the ratios appearing in these equations depend only on
$u\equiv y/x=e^{-4\chi_{{}_S}}\cot^2\varphi_{{}_S}$, 
\begin{equation}
  \frac{\gamma_{{}_S}}{\Bs}=\frac{1+u}{1-u}\,,
  \qquad
  \frac{\sin^2 2\varphi_{{}_S}\,\gamma_{{}_S}}{2\Bs^3}=2u\,\frac{1+u}{(1-u)^3}\,.
\end{equation}
For $\chi_{{}_S}\to\infty$ at fixed $\sin\varphi_{{}_S}\neq0$, i.e.\ for $u\ll1$, we obtain
\begin{equation}\label{eq:chi-limit-large}
\begin{split}
  \dk\chi_{{}_S} &\simeq \frac{\dns{}}{2}\left(1+2e^{-4\chi_{{}_S}}\cot^2\varphi_{{}_S}\right),\\
  \ddk\chi_{{}_S} &\simeq \frac{\das{}}{2}
  + e^{-4\chi_{{}_S}}\cot^2\varphi_{{}_S}\left[\das{}-2\big(\dns{}\big)^2\right]
  + \mathcal{O}\!\left(e^{-8\chi_{{}_S}}\right).
\end{split}
\end{equation}
Hence $\dk\chi_{{}_S}\to\dns{}/2$ and $\ddk\chi_{{}_S}\to\das{}/2$
exponentially fast, independently of $\varphi_{{}_S}$. This is consistent with
$\gamma_{{}_S}\simeq e^{2\chi_{{}_S}}\sin^2\varphi_{{}_S}$ in eq.~\eqref{eq:gamma-large}, for which
$\ln\gamma_{{}_S}\simeq2\chi_{{}_S}+\text{const}$. For the ACT DR6 mean values,
eq.~\eqref{eq:delta-mean}, the limiting values are $\dk\chi_{{}_S}\to0.0038$ and
$\ddk\chi_{{}_S}\to0.0034$. The term proportional to $(\dns{})^2$ in $  \ddk\chi_{{}_S}$ is of order
$10^{-4}$ and enters only at relative order $e^{-4\chi_{{}_S}}$, so it is negligible.

The behaviour near $\varphi_{{}_S}\to0,\pi$ is different, since $\Bs$ changes sign at
$\tan^2\varphi_{{}_S}=e^{-4\chi_{{}_S}}$, where $\dk\chi_{{}_S}$ and $\ddk\chi_{{}_S}$ diverge. Using
$\gamma_{{}_S}=\cosh2\chi_{{}_S}-\sinh2\chi_{{}_S}\cos2\varphi_{{}_S}$ and
$\Bs=\sinh2\chi_{{}_S}-\cosh2\chi_{{}_S}\cos2\varphi_{{}_S}$, one finds that $\Bs<0$ is equivalent to
$\cos2\varphi_{{}_S}>\tanh2\chi_{{}_S}$, which implies
\begin{equation}
  \gamma_{{}_S}<\cosh2\chi_{{}_S}-\sinh2\chi_{{}_S}\tanh2\chi_{{}_S}=\frac{1}{\cosh2\chi_{{}_S}}\le1\,.
\end{equation}
In the QC model the bound \eqref{eq:r0.038} requires $\gamma_{{}_S}>r_\BD/r_{{}_{\rm max}}\simeq3.5$, so
$\Bs>0$ throughout the allowed region. There $\dk\chi_{{}_S}$ and $\ddk\chi_{{}_S}$ never diverge or
change sign, and the values $\varphi_{{}_S}=0,\pi$ are excluded for all $\chi_{{}_S}$, see figure \ref{fig:chiddot-mean-QC}.

In the Starobinsky model, on the other hand, the bound \eqref{eq:r0.038} only requires $\gamma_{{}_S}>0.0877$, and part
of the region with $\Bs<0$ survives. This requires $1/\cosh2\chi_{{}_S}>0.0877$, i.e.\
$\chi_{{}_S}<\chi_{{}_{S,2}}\simeq1.562$, eq.~\eqref{chis2-phis2-star}, so the region near $\varphi_{{}_S}\to0,\pi$ is retained only for bounded $\chi_{{}_S}$.
At $\varphi_{{}_S}=0,\pi$ one has exactly $\gamma_{{}_S}=e^{-2\chi_{{}_S}}$, so that
$\dns{}=-2\dk\chi_{{}_S}$ and $\das{}=-2\ddk\chi_{{}_S}$, and
\eqref{eq:r0.038} limits $\chi_{{}_S}<1.2168$. For $v\equiv e^{4\chi_{{}_S}}\tan^2\varphi_{{}_S}\ll1$,
\begin{equation}\label{eq:chi-limit-small}
\begin{split}
  \dk\chi_{{}_S} &\simeq -\frac{\dns{}}{2}\left(1+2v\right),\\
  \ddk\chi_{{}_S} &\simeq -\frac{\das{}}{2}
  - v\left[\das{}-2\big(\dns{}\big)^2\right]
  + \mathcal{O}\!\left(v^2\right).
\end{split}
\end{equation}
Compared with the large-$\chi_{{}_S}$ regime, the required scale dependence has the opposite sign:
for the ACT DR6 mean values $\dk\chi_{{}_S}\to-0.0038$ and $\ddk\chi_{{}_S}\to-0.0034$ as
$\varphi_{{}_S}\to0,\pi$. The two regimes are connected through the curve $\Bs=0$, on which the
required $\dk\chi_{{}_S}$ and $\ddk\chi_{{}_S}$ diverge, as seen in figure~\ref{fig:chiddot-mean-Starobinsky}.

We now present the contour plots for each choice of fixed $\dk\chi_{{}_S}$ and $\ddk\chi_{{}_S}$. For $\Delta\dns{}$ and $\Delta \das{}$ within the 1$\sigma$ region of the ACT DR6 results, the corresponding contour regions in the $(\varphi_{{}_S},\chi_{{}_S})$ plane are displayed in figures \ref{fig:A-St-chidot}--\ref{fig:A-QC-chiddot}, for both the Starobinsky and the QC models.
One can understand the behavior of the plots from the exact properties of eqs.~\eqref{eq:chidot} and \eqref{eq:chiddot}. Since
$\gamma_{{}_S}^2-\Bs^2=4xy=\sin^22\varphi_{{}_S}$, both equations depend on $(\varphi_{{}_S},\chi_{{}_S})$ only through the single variable
\begin{equation}\label{eq:q-def}
q\equiv\frac{\gamma_{{}_S}}{\Bs}=\frac{1+u}{1-u}\,,\qquad u=\frac{q-1}{q+1}=e^{-4\chi_{{}_S}}\cot^2\varphi_{{}_S}\,,
\end{equation}
and take the exact form
\begin{align}
\dk\chi_{{}_S}&=\frac{q\,\dns{}}{2}\,,\label{dkchis-q}\\
\ddk\chi_{{}_S}&=\frac{q}{2}\Big[\das{}-\big(q^2-1\big)\big(\dns{}\big)^2\Big]\,, \label{ddkchis-q}
\end{align}
with $|q|\geq1$, and $q>0$ ($q<0$) for $\Bs>0$ ($\Bs<0$). The contours of fixed $\dk\chi_{{}_S}$ or $\ddk\chi_{{}_S}$ are therefore curves of constant $u$,
\begin{equation}\label{eq:u-curves}
\chi_{{}_S}=\frac{1}{2}\ln|\cot\varphi_{{}_S}|-\frac{1}{4}\ln u\,,
\end{equation}
which are vertical translates of each other in the $(\varphi_{{}_S},\chi_{{}_S})$ plane.

For a fixed $\dk\chi_{{}_S}$, letting $\dns{}$ vary over its $1\sigma$ interval $[\delta_1,\delta_2]=[0.00423,\,0.01103]$ selects
$q\in[2\dk\chi_{{}_S}/\delta_2,\,2\dk\chi_{{}_S}/\delta_1]$. The allowed region is the band between the two curves \eqref{eq:u-curves} with $u_i=(2\dk\chi_{{}_S}-\delta_i)/(2\dk\chi_{{}_S}+\delta_i)$. At fixed $\varphi_{{}_S}$ its width in $\chi_{{}_S}$ is independent of $\varphi_{{}_S}$,
\begin{equation}\label{eq:width-chidot}
\Delta\chi_{{}_S}=\frac{1}{4}\left|\ln\frac{u_1}{u_2}\right|=\frac{1}{4}\left|\ln\frac{\big(2\dk\chi_{{}_S}-\delta_1\big)\big(2\dk\chi_{{}_S}+\delta_2\big)}{\big(2\dk\chi_{{}_S}+\delta_1\big)\big(2\dk\chi_{{}_S}-\delta_2\big)}\right|
\simeq\frac{\delta_2-\delta_1}{4|\dk\chi_{{}_S}|}\quad\text{for }|\dk\chi_{{}_S}|\gg\frac{\delta_2}{2}\,,
\end{equation}
and at fixed $\chi_{{}_S}$ its width in $\varphi_{{}_S}$ is $\Delta\varphi_{{}_S}\simeq\sin2\varphi_{{}_S}\,\Delta\chi_{{}_S}$, which is largest at $\varphi_{{}_S}=\pi/4,3\pi/4$, as in Case A. Eq.~\eqref{eq:width-chidot} holds for $|\dk\chi_{{}_S}|>\delta_2/2$. For $\delta_1/2<\dk\chi_{{}_S}\leq\delta_2/2$, the interval of $q$ contains $q=1$ ($u=0$), and the band is bounded only from below, by the curve $u=u_1$, extending to $\chi_{{}_S}\to\infty$ and reaching down to the $r$ bound along $\varphi_{{}_S}=\pi/2$. Since $|q|\geq1$, eq.~\eqref{dkchis-q} also implies $|\dk\chi_{{}_S}|\geq\dns{}/2\geq\delta_1/2\simeq0.00212$ in both models. As $\dk\chi_{{}_S}\to\delta_1/2$, $u_1\to0$ and the band shrinks onto the line $\varphi_{{}_S}=\pi/2$. Fixed values of $\ddk\chi_{{}_S}$ behave in the same way. For $(q^2-1)(\dns{})^2\ll\das{}$, inverting eq.~\eqref{ddkchis-q} at the mean value of $\dns{}$ gives $q\simeq2\ddk\chi_{{}_S}/\das{}$, so eq.~\eqref{eq:width-chidot} applies with $\dk\chi_{{}_S}\to\ddk\chi_{{}_S}$ and $\delta_i\to\delta^{(\alpha)}_i$, where $[\delta^{(\alpha)}_1,\delta^{(\alpha)}_2]=[0.00156,\,0.01196]$. Close to the curve $\Bs=0$, where $|q|\gg1$, the cubic term dominates instead and $|\ddk\chi_{{}_S}|\simeq|q|^3(\dns{})^2/2$. In the Starobinsky model this cubic term also allows $\ddk\chi_{{}_S}=0$ at $q^2=1+\das{}/(\dns{})^2$. Therefore for the Starobinsky inflationary model, there is a non-empty region of parameter space for which $\ddk\chi_{{}_S}=0$, but yield a positive $\alpha_{\rm s}$ with just  $\dk\chi_{{}_S}\neq0$, see Fig.~\ref{fig:A-St-chiddot-pos-a}.

In the QC model, the bound \eqref{eq:r0.038} confines $q$ to a narrow interval. With $R\equiv r_{{}_\BD}/r_{{}_{\rm max}}\simeq3.5088$, one has $\gamma_{{}_S}>R$ and $\Bs^2=\gamma_{{}_S}^2-\sin^22\varphi_{{}_S}\geq\gamma_{{}_S}^2-1$, so
\begin{equation}\label{eq:qmax}
1\leq q\leq\frac{\gamma_{{}_S}}{\sqrt{\gamma_{{}_S}^2-1}}<q_{\rm max}\equiv\frac{R}{\sqrt{R^2-1}}\simeq1.04327\,,
\end{equation}
or equivalently $0\leq u<u_{\rm max}=\big(R-\sqrt{R^2-1}\big)^2\simeq0.0212$. The value $q=1$ is reached along $\varphi_{{}_S}=\pi/2$ (and as $\chi_{{}_S}\to\infty$), while $q\to q_{\rm max}$ at the points $\varphi_{{}_S}=\pi/4,3\pi/4$, $\chi_{{}_S}=\frac12\operatorname{arccosh}R\simeq0.967$ on the boundary of the allowed region. Physically, the $r$ bound forces $e^{2\chi_{{}_S}}\sin^2\varphi_{{}_S}\gg e^{-2\chi_{{}_S}}\cos^2\varphi_{{}_S}$, so that $\ln\gamma_{{}_S}\simeq2\chi_{{}_S}+{\rm const}$ and $\dns{}\simeq2\dk\chi_{{}_S}$ throughout the allowed region. Since $q-1\leq0.044$, the cubic term in eq.~\eqref{ddkchis-q} is at most $(q_{\rm max}^2-1)(\dns{})^2\simeq10^{-5}$ and can never compensate $\das{}$, so $\ddk\chi_{{}_S}$ cannot vanish. Eqs.~\eqref{dkchis-q}, \eqref{ddkchis-q}, and \eqref{eq:qmax} then give the bounds
\begin{align}
\frac{\delta_1}{2}\leq\;&\dk\chi_{{}_S}\;\leq q_{\rm max}\,\frac{\delta_2}{2}\,, &&\text{i.e.}\quad 0.00212\lesssim\dk\chi_{{}_S}\lesssim0.00576\,,\label{eq:QC-chidot-bounds}\\
\frac{\delta^{(\alpha)}_1}{2}\leq\;&\ddk\chi_{{}_S}\;\leq \frac{q_{\rm max}}{2}\left[\delta^{(\alpha)}_2-\frac{\big(\dns{}\big)^2}{R^2-1}\right], &&\text{i.e.}\quad 0.00078\lesssim\ddk\chi_{{}_S}\lesssim0.00623\,,\label{eq:QC-chiddot-bounds}
\end{align}
where in eq.~\eqref{eq:QC-chiddot-bounds} $\dns{}$ takes its mean value. The lower bounds are attained for $q=1$ with the smallest allowed shifts. There the band of figures~\ref{fig:A-QC-chidot} and \ref{fig:A-QC-chiddot} collapses onto $\varphi_{{}_S}=\pi/2$. The upper bounds are attained for $q=q_{\rm max}$ with the largest shifts, where the band shrinks to the points $(\pi/4,0.967)$ and $(3\pi/4,0.967)$. This explains the narrow ranges of $\dk\chi_{{}_S}$ and $\ddk\chi_{{}_S}$ in figures~\ref{fig:A-QC-chidot} and \ref{fig:A-QC-chiddot}. For the ACT DR6 mean values these ranges shrink further to $0.00382\lesssim\dk\chi_{{}_S}\lesssim0.00398$ and $0.00338\lesssim\ddk\chi_{{}_S}\lesssim0.00352$, in agreement with figures~\ref{fig:chidot-mean} and \ref{fig:chiddot-mean-QC}. In the Starobinsky model, by contrast, the curve $\Bs=0$ crosses the allowed region, so $|q|$ and hence $|\dk\chi_{{}_S}|$ and $|\ddk\chi_{{}_S}|$ are unbounded from above, while the lower bound $|\dk\chi_{{}_S}|\geq\delta_1/2$ still holds.

%
%
%
%

\section{Conclusions}
\label{sec:conclusions}

We have studied whether a scale-dependent excited (non-BD) initial state for the scalar perturbations can reconcile single-field slow-roll models with the ACT DR6 measurements of $n_{\rm s}$ and $\alpha_{\rm s}$, without modifying the inflaton potential. We considered two benchmark models: the Starobinsky model, which agrees very well with Planck but is in tension with the higher $n_{\rm s}$ preferred by ACT DR6, and the QC model, which is in addition excluded by its large tensor-to-scalar ratio. For $N_e=60$, both models predict $n_{{\rm s},\BD}=0.9667$ and $\alpha_{{\rm s},\BD}=-0.00056$ in the BD vacuum, with $r_{{}_\BD}=0.0033$ and $r_{{}_\BD}=0.133$, respectively. Matching the ACT DR6 values requires $\dns{}\in[0.00423,0.01103]$ and $\das{}\in[0.00156,0.01196]$ at $1\sigma$, with $\dns{}=0.0076$ and $\das{}=0.0068$ for the mean values.

Parametrising the Bogoliubov coefficients of the scalar modes by $\chi_{{}_S}$ and $\varphi_{{}_S}$, the scalar power spectrum is rescaled by $\gamma_{{}_S}=e^{2\chi_{{}_S}}\sin^2\varphi_{{}_S}+e^{-2\chi_{{}_S}}\cos^2\varphi_{{}_S}$. A constant $\gamma_{{}_S}$ leaves $n_{\rm s}$ and $\alpha_{\rm s}$ unchanged, so the shifts must come from the scale dependence of $\chi_{{}_S}$ and/or $\varphi_{{}_S}$. We derived these shifts in closed form, eqs.~\eqref{eq:ns-shift-general} and \eqref{eq:alphas-shift-general}, and analysed two cases, in which either the phase (Case A) or the excitation parameter (Case B) carries the scale dependence. Keeping the tensor modes in the BD state, the same factor suppresses the tensor-to-scalar ratio, $r=r_{{}_\BD}/\gamma_{{}_S}$. Two conditions restrict the parameter space: the observational bound $r<r_{{}_{\rm max}}=0.038$, and the backreaction condition, which requires the scale of new physics $\Lambda$ at which the modes are excited to be larger than $H$. Following ref.~\cite{Holman:2007na}, the backreaction bound involves the slow-roll parameter $\eta=\ddot H/(2\epsilon H^3)$, which is $\simeq-1/N_e$ for the Starobinsky model but only $\simeq1/(12N_e^2)$ for the QC model. Its value for the QC model is therefore much smaller than assumed in earlier work \cite{Ashoorioon:2013eia}, which makes the EFT condition considerably more restrictive.

For the Starobinsky model, the bound on $r$ is irrelevant except near $\varphi_{{}_S}=0,\pi$, where $\gamma_{{}_S}<1$ enhances $r$. At $\varphi_{{}_S}=0,\pi$ it requires $\chi_{{}_S}<1.217$; the excluded region is widest at $(\varphi_{{}_S},\chi_{{}_S})\simeq(0.0439,1.562)$, and for larger $\chi_{{}_S}$ it narrows as $\varphi_{{}_S}\simeq0.296\,e^{-\chi_{{}_S}}$ until, for $\chi_{{}_S}\gtrsim5.24$, the EFT condition takes over and fixes the edge at $\varphi_{{}_S}\simeq0.0016$. Elsewhere, $\Lambda^{\rm max}_{{}_S}/H$ grows like $17.8\,\chi_{{}_S}^{-1/2}$ for $\chi_{{}_S}\ll1$ and approaches $25.2\sin^{1/2}\varphi_{{}_S}$ for $\chi_{{}_S}\gg1$, leaving ample room for a separation between $\Lambda$ and $H$. For the QC model, the excited state has to do more work: suppressing $r$ below $r_{{}_{\rm max}}$ requires $\gamma_{{}_S}>3.51$, so that $\chi_{{}_S}>0.628$ even at $\varphi_{{}_S}=\pi/2$, and the minimum allowed $\chi_{{}_S}$ grows as $\varphi_{{}_S}$ moves towards $0$ or $\pi$. With the correct value of $\eta$, the backreaction condition limits the scale of new physics to $\Lambda\lesssim5.75\,H$, with $\Lambda^{\rm max}_{{}_S}/H\to4.86$ at large $\chi_{{}_S}$, and excludes $\varphi_{{}_S}\lesssim0.042$ (and $\pi-\varphi_{{}_S}\lesssim0.042$) for $\chi_{{}_S}\gtrsim3.8$. A viable window therefore remains for the QC model, but the scale of new physics must lie within a few times the magnitude of $H$.

In Case A, the running phase must satisfy $\dk\varphi_{{}_S}\simeq(\dns{}/2)\tan\varphi_{{}_S}\simeq0.0038\tan\varphi_{{}_S}$ at large $\chi_{{}_S}$, independently of $\chi_{{}_S}$. For a fixed $\dk\varphi_{{}_S}$, the allowed region is a band in $\varphi_{{}_S}$ of width $\Delta\varphi_{{}_S}\simeq\frac12\sin2\varphi_{{}_S}\,\Delta\dns{}/\dns{}$, which narrows as $|\dk\varphi_{{}_S}|$ grows. The required $|\dk\varphi_{{}_S}|$ and $|\ddk\varphi_{{}_S}|$ diverge at $\varphi_{{}_S}=0,\pi/2,\pi$, where $\gamma_{{}_S}$ is stationary in $\varphi_{{}_S}$, and are smallest along the edges of the allowed region, where they can be as small as $\sim10^{-5}$ for the Starobinsky model and $\sim10^{-4}$ for the QC model.

In Case B, we showed that the required $\dk\chi_{{}_S}$ and $\ddk\chi_{{}_S}$ depend on $(\varphi_{{}_S},\chi_{{}_S})$ only through the single combination $q=\gamma_{{}_S}/\Bs$, eqs.~\eqref{dkchis-q} and \eqref{ddkchis-q}. The allowed regions are then bands between curves of constant $e^{-4\chi_{{}_S}}\cot^2\varphi_{{}_S}$, whose widths we obtained analytically, and $|\dk\chi_{{}_S}|\geq\dns{}/2\geq0.0021$ in both models. For the Starobinsky model, the curve $\Bs=0$ crosses the allowed region, so the required running of $\chi_{{}_S}$ is unbounded from above and changes sign near $\varphi_{{}_S}=0,\pi$. Moreover, a positive $\alpha_{\rm s}$ can be generated with $\ddk\chi_{{}_S}=0$, by $\dk\chi_{{}_S}$ alone. For the QC model, the bound on $r$ forces $1\leq q<1.043$, so that $\ln\gamma_{{}_S}\simeq2\chi_{{}_S}+{\rm const}$ throughout the allowed region and the running of $\chi_{{}_S}$ is tightly fixed by the observed shifts: $0.0021\lesssim\dk\chi_{{}_S}\lesssim0.0058$ and $0.0008\lesssim\ddk\chi_{{}_S}\lesssim0.0062$ over the $1\sigma$ ACT DR6 ranges, and $0.0038\lesssim\dk\chi_{{}_S}\lesssim0.0040$, $0.0034\lesssim\ddk\chi_{{}_S}\lesssim0.0035$ for the mean values.

Taken together, these results show that the tension between ACT DR6 and the Starobinsky model can be removed by a mild, scale-dependent excitation of the scalar modes over a large part of the parameter space, with the scale of new physics comfortably above $H$. The QC model can also be reconciled with ACT DR6 and with the bound on $r$, but only with a sizeable excitation, $\chi_{{}_S}\gtrsim0.6$, a narrowly determined running, and a scale of new physics within a factor of about six of $H$, which makes this possibility considerably more fine-tuned.

Our analysis has several limitations. We treated the cases in which only $\varphi_{{}_S}$ or only $\chi_{{}_S}$ runs; when both run, eqs.~\eqref{eq:ns-shift-general} and \eqref{eq:alphas-shift-general} open up a larger parameter space, including the mixed term $\dk\chi_{{}_S}\dk\varphi_{{}_S}$. We assumed BD tensor modes and fixed $N_e=60$; an excitation of the tensor modes would modify $r$ independently.  On the observational side, the scenario has distinctive signatures. With BD tensor modes, the tensor tilt remains $n_{{}_T}=-2\epsilon$ while $r=-8n_{{}_T}/\gamma_{{}_S}$, so the consistency relation is violated by the factor $\gamma_{{}_S}$, which exceeds $3.5$ in the QC model. Excited initial states are also known to enhance the bispectrum in the folded and local configurations \cite{Chen:2006nt, Holman:2007na, Ashoorioon:2010xg,Agullo:2010ws,Ashoorioon:2013eia}. Future measurements of $n_{\rm s}$, $\alpha_{\rm s}$, $r$, the tensor tilt, and primordial non-Gaussianity will therefore provide direct tests of this possibility.

\acknowledgments

We thank K. Rezazadeh for his collaboration at the beginning of this project. A.A. is thankful for the hospitality of the CERN theory group where this work was completed.

%

\bibliographystyle{JHEP}
\bibliography{ref}

\end{document}